\documentclass[twocolumn,amsmath,amssymb,aps,pra]{revtex4}
	\usepackage{amsmath}
	\usepackage{amssymb}
	\usepackage{graphicx} 
	\usepackage{cancel}
	\usepackage[breaklinks=true,colorlinks,citecolor=blue,linkcolor=blue,urlcolor=blue]{hyperref}
	\usepackage{bbold}
	\usepackage{soul}
   	\usepackage{enumerate}
    	\usepackage{color}
    	\usepackage{ulem}

\begin{document}
\title{Heat rectification through single and coupled quantum dots}

\author{Ludovico Tesser}
\affiliation{Department of Microtechnology and Nanoscience (MC2), Chalmers University of Technology, S-412 96 G\"oteborg, Sweden}
\author{Bibek Bhandari}
\affiliation{Department of Physics and Astronomy, University of Rochester, Rochester, NY 14627, USA}
\affiliation{Institute for Quantum Studies, Chapman University, Orange, CA 92866, USA}
\author{Paolo Andrea Erdman}
\affiliation{Freie Universit\"at Berlin, Department of Mathematics and Computer Science, Arnimallee 6, 14195 Berlin, Germany}
\author{Elisabetta Paladino}
\affiliation{Dipartimento di Fisica e Astronomia Ettore Majorana, Universit\`a di Catania, Via S. Sofia 64, 95123 Catania, Italy}
\affiliation{INFN, Sez. Catania, I-95123, Catania, Italy}
\affiliation{CNR-IMM, Via S. Sofia 64, I-95123, Catania, Italy}
\author{Rosario Fazio}
\affiliation{Abdus Salam ICTP, Strada Costiera 11, I-34151 Trieste, Italy}
\affiliation{Dipartimento di Fisica E. Pancini, Universit\`a di Napoli Federico II, , 80126 Napoli, Italy}
\affiliation{NEST, Istituto Nanoscienze-CNR, I-56126 Pisa, Italy}
\author{Fabio Taddei}
\affiliation{NEST, Istituto Nanoscienze-CNR and Scuola Normale Superiore, I-56126 Pisa, Italy}

\begin{abstract}
We study heat rectification through quantum dots in the Coulomb blockade regime using a master equation approach. We consider both cases of two-terminal and four-terminal 
devices. In the two-terminal configuration, we analyze the case of a single quantum dot with either a doubly-degenerate level or two non-degenerate levels. In the sequential tunneling regime we analyze the behaviour of heat currents and rectification as functions of the position of the energy levels and of the temperature bias.
In particular, we derive an upper bound for rectification in the closed-circuit setup with the doubly-degenerate level. We also prove the absence of a bound for the case of two non-degenerate levels and identify the ideal system parameters to achieve nearly perfect rectification.
The second part of the paper deals with the effect of second-order cotunneling contributions, including both elastic and inelastic processes.
In all cases we find that there exists ranges of values of parameters (such as the levels' position) where rectification is enhanced by cotunneling. 
In particular, in the doubly-degenerate level case we find that cotunneling corrections
can enhance rectification when they reduce the magnitude of the heat currents.
For the four-terminal configuration, we analyze the non-local situation of two Coulomb-coupled quantum dots, each connected to two terminals: the temperature bias is applied 
to the two terminals connected to one quantum dot, while the heat currents of interest are the ones flowing in the other quantum dot.
Remarkably, in this situation we find that non-local rectification can be perfect as a consequence of the fact that the heat currents vanish for properly tuned parameters.
\end{abstract}
\date{\today}

\maketitle

\section{Introduction}
Rectification is the phenomenon for which the magnitude of a current flowing in a system depends on the sign of the bias applied.
In other words, by reversing the bias the current not only changes sign but also its magnitude.
Perfect rectification is obtained when the current can flow only in one direction.
The most familiar example of rectification is the one occurring in diodes, a two-terminal electronic component which allows the flow of charge current primarily in 
one direction, i.e.~presenting low resistance in one direction and high resistance in the other.

Thermal rectification, i.e.~the rectification of heat currents, occurs in a two-terminal system when the absolute value of the heat flux changes by reversing the 
sign of temperature bias applied to the two leads.
This phenomenon has recently attracted increasing interest as a mean to improve thermal management in nanoscale systems, for example by blocking the flow 
of heat in certain areas of an electronic circuit to prevent overheating.
Such interest is fueled by recent advancements in the experimental realization of nanostructured devices where thermal fluxes can be measured~\cite{giazotto2006,giazotto2012,pekola2015,ronzani2018,maillet2019,maillet2020}.
Thermal rectification was first observed experimentally a long time ago in Ref.~\onlinecite{starr1935}.
More recently, in solid-state quantum systems it has been theoretically studied in Refs.~\onlinecite{terraneo2002,li2004,segal2005,eckmann2006,zeng2008,ojanen2009,ruokola2009,wu2009,wu2009b,otey2010,kuo2010,kuo2010b,ruokola2011,
gunawardana2012,martinez2013,giazotto2013,landi2014,liu2014,jiang2015,sanchez2015,joulain2016,agarwalla2017,vicioso2018,goury2019,giazotto2020,bhandari2021,iorio2021,upadhyay2021}.
In electronic nanoscale systems thermal rectification has been studied theoretically in hybrid quantum devices \cite{wu2009,wu2009b,martinez2013,goury2019,giazotto2020,iorio2021} 
and Quantum Dots (QDs)~\cite{xueou2008,kuo2010,ruokola2011,lopez2013,vicioso2018,kuo2020,aligia2020}, and experimentally measured in Refs.~\onlinecite{chang2006,scheibner2008,senior2019,schmotz2011,martinez2015}.
In Ref.~\onlinecite{bhandari2021} thermal rectification has been calculated for a multi-level bosonic quantum system consisting of a nonlinear resonator attached to two baths.

\begin{figure}[!tb]
	\centering
	\includegraphics[width=0.99\columnwidth]{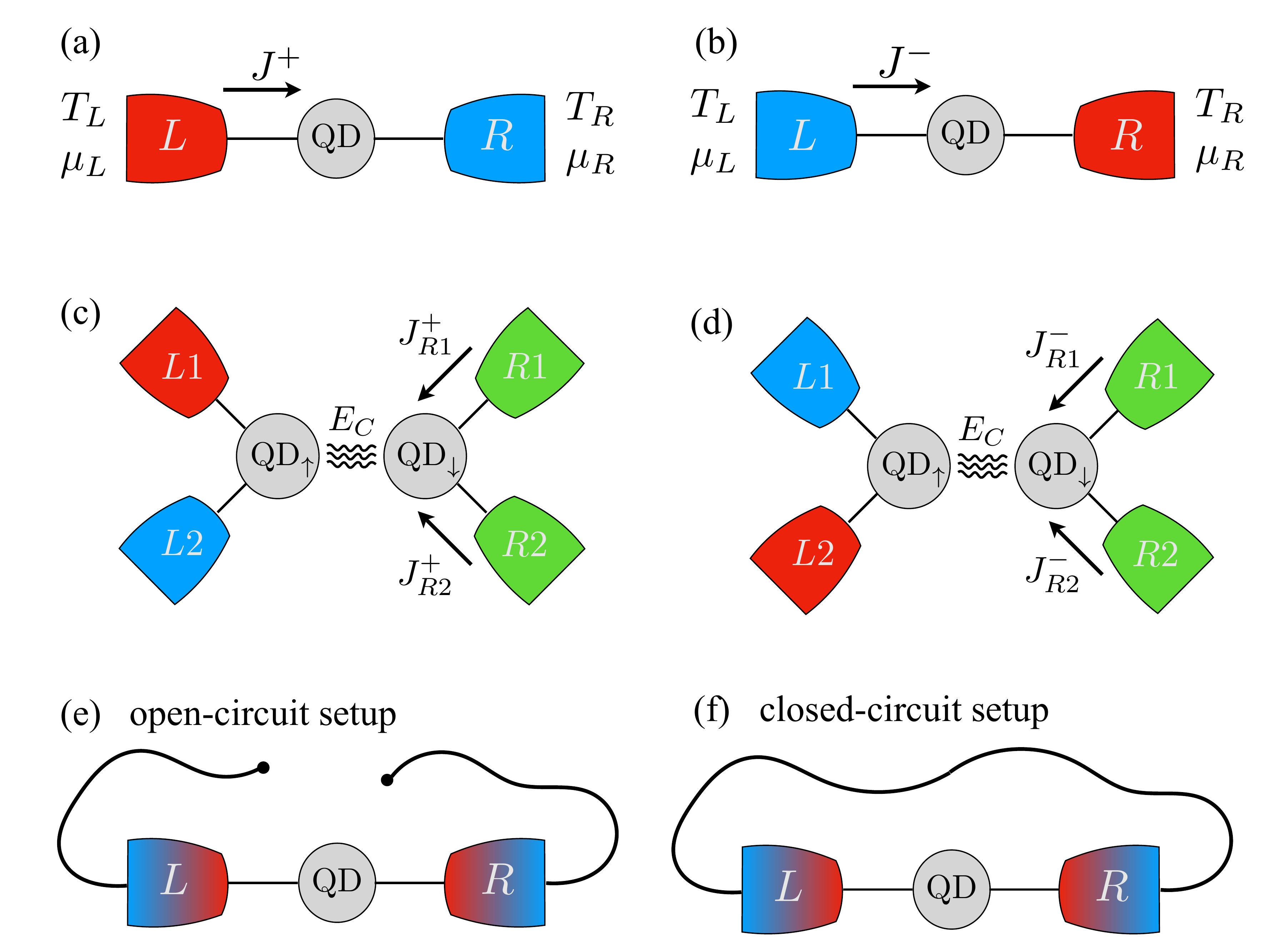}
	\caption{Sketch of the systems considered: grey circles represent QDs, while red, blue and green objects represent the reservoirs. Panel (a) and (b) refer to the two-terminal setup, while panel (c) and (d) refer to the four-terminal setup. For the two-terminal setup, the left (right) reservoir is characterized by a temperature $T_L=T+\Delta T/2$ ($T_R=T-\Delta T/2$) and a chemical potential $\mu_L=\Delta\mu/2$ ($\mu_R=-\Delta\mu/2$). Panel (a) represents the forward bias configuration, where $\Delta T>0$ (the left reservoir is hot and the right reservoir is cold), while panel (b) represents the backward bias configuration, where $\Delta T<0$ (the left reservoir is cold and the right reservoir is hot). $J^\pm$ is the heat current flowing in the left lead in the forward (backward) bias configuration. In the four-terminal setup, the two QDs (identified by the arrows $\uparrow$ and $\downarrow$) are each connected to two reservoirs and are Coulomb-coupled to each other (with charging energy $E_C$). We refer to $L1$, $L2$ and QD$_\uparrow$ as the {\it drive} circuit, while to $R1$, $R2$ and QD$_\downarrow$ to the {\it drag} circuit and we set $T_{L1}=T+\Delta T/2$, $T_{L2}=T-\Delta T/2$, $T_{R1}=T$ and $T_{R2}=T$. All reservoirs are kept at the same chemical potential. Panel (c) depicts the forward bias configuration, where $\Delta T>0$, while panel (d) depicts the backward bias configuration, where $\Delta T<0$. We are interested in the heat currents $J^\pm_{R1}$ and $J^\pm_{R2}$ flowing in the lead $R1$ and $R2$, respectively. Panel (e) and (f) illustrate the open-circuit and the closed-circuit setups, respectively. In the former case no charge current flows through the system, while in the latter the two reservoirs are electrically connected ($\mu_L=\mu_R$).}
	\label{setup}
\end{figure}

Focus of the present work is to investigate how heat can be rectified using QD-based devices. As mentioned above, few papers on heat rectification in QDs are available in the literature.
In Ref.~\onlinecite{xueou2008}, motivated by the experiment reported in Ref.~\onlinecite{scheibner2008}, thermal rectification has been investigated for a two-level QD using a nonequilibrium Green function method and focusing on the role of the energy-dependence of the tunnel couplings between QD and leads.
In Refs.~\onlinecite{kuo2010,ruokola2011,aligia2020} the case of multiple capacitively-coupled QDs was considered: while in Ref.~\onlinecite{kuo2010} all QDs were connected to two leads, in Refs.~\onlinecite{ruokola2011,aligia2020} each of the two QDs considered were attached to one lead only, so that heat can be transported only by electronic fluctuations.
In Ref.~\onlinecite{lopez2013} a mean-field approximation was used to calculate self-consistently the heat rectification of a single QD.
The role of interference, quantum superposition and level degeneracy on the heat rectification was studied in Ref.~\onlinecite{vicioso2018} for various systems of coupled QDs, using the master equation approach up to sequential tunneling processes.
Finally, the case of an array of QDs was studied in Ref.~\onlinecite{kuo2020} using the Keldysh Green's function technique.

An important ingredient to rectify heat is the presence of non-linearities in the spectrum (e.g. due to the combined effect of confinement and electron-electron interaction).
This fact can be easily understood by noticing that, in the absence of interactions, the heat current can be calculated using the 
Landauer-B\"uttiker approach, i.e.~by an energy integral of the transmission probability of the QD multiplied by the difference of the Fermi distribution 
functions of the two terminals. Since the temperatures enter only the distribution functions (indeed the transmission probability consists of a set of 
narrow Lorentzian functions of energy, one for each QD level), an inversion of the temperature bias simply gives rise to a change of sign of the current, thus no rectification.
Non-linearities, however, are not enough. Indeed, to obtain rectification a necessary condition is to break the mirror-symmetry of the system, for example by coupling the system to the left and to the right terminals by a different extent.

In this paper we study heat rectification for QDs in the Coulomb blockade regime using the master equation approach~\cite{ruokola2011,vicioso2018}, with a particular emphasis on the role of second-order cotunneling contributions.
We consider both the two-terminal and the four-terminal configurations [see Figs.~\ref{setup}(a-b) and (c-d), respectively].
For the former, we analyze the case of a single QD with either a doubly-degenerate level or two non-degenerate levels.
The two reservoirs, labelled with $L$ and $R$ in Figs.~\ref{setup}(a-b), are characterized by their temperatures ($T_L=T+\Delta T/2$ and $T_R=T-\Delta T/2$) and their chemical potentials ($\mu_L=\Delta\mu/2$ and $\mu_R=-\Delta\mu/2$).
We are interested in studying the heat current that flows through the system when a temperature bias is applied between the reservoirs.
Furthermore, we assume that no work is performed on the system and consider both the {\it open-circuit setup} [where the charge current vanishes, see Fig.~\ref{setup}(e)] and the {\it closed-circuit setup} [where the bias voltage is set to zero,  see Fig.~\ref{setup}(f)].
For the sake of definiteness, we focus on the heat current flowing between the left terminal and the QD.
Referring to Fig.~\ref{setup}(a-b), we define the forward heat current $J^+$ as the one relative to $\Delta T>0$ and the backward heat current $J^-$ the one relative to $\Delta T<0$.
The laws of thermodynamics assure that heat will flow from left to right if $\Delta T > 0$ (forward bias), or from right to left if $\Delta T < 0$ (backward bias).
When $|J^+|$, induced by a forward bias, is different from $|J^-|$, induced by a backward bias, we define the heat rectification coefficient as
\begin{equation}
R=\frac{\left|J^+\right|-\left|J^-\right|}{\left|J^+\right|+\left|J^-\right|}.
\label{erre}
\end{equation}
The definition is such that $|R| \leq 1$.
In particular, $R = 0$ means that no rectification takes place, while $R = \pm1$ means that we have perfect rectification (i.e. the heat current is finite in one direction, and null in the other).

We first analyse the sequential tunneling regime (assuming up to single occupancy of the QD) deriving, when possible, analytical expressions for the heat currents.
Remarkably, we could derive an upper bound for rectification in the closed-circuit setup with a doubly-degenerate level, and prove the absence of a bound for the case of two non-degenerate levels.
We also analyze the behaviour of currents and rectification as functions of the levels' position and the temperature bias.
The most important part of the paper deals with the effect of cotunneling contributions, including both elastic and inelastic processes, in all setups of the two-terminal configuration.
The most remarkable results are the following:
i) in the open-circuit setup of the doubly-degenerate level case, cotunneling yields a finite (though little) rectification, contrary to what happens when only sequential processes are considered;
ii) in the closed-circuit setup of a doubly-degenerate level, cotunneling corrections to the forward heat current are opposite to the corrections to the backward heat current, thus yielding rectification enhancement when cotunneling lowers the magnitude of the two currents (an analogous result was reported in Ref.~\onlinecite{bhandari2021});
iii) in the case of two non-degenerate levels, in the open-circuit setup, cotunneling always increases the currents with respect to the sequential regime;
iv) in all cases there exists ranges of values of the levels' position where rectification is enhanced by cotunneling. 

For the four-terminal configuration, we analyze the case of two Coulomb-coupled QDs, each connected to two terminals [see Fig.~\ref{setup}(c-d)].
Such a setup has been actually realized in Refs.~\onlinecite{mcclure2007,shinkai2009,shinkai2009b,bischoff2015,hartmann2015,thierschmann2015,volk2015,koski2015,keller2016,singh2019,mu2021}.
This is a non-local configuration, where the temperature bias is applied to the terminals $L1$ and $L2$ on the left ({\it drive} circuit), while the heat currents of interest are the ones flowing in terminals $R1$ and $R2$ on the right ({\it drag} circuit).
Remarkably, in this situation we find that non-local rectification (defined for the currents in the drag circuit) can reach the ideal value as a consequence of the fact that the heat currents in the drag circuit can change sign (thus going to zero) as a function, for example, of the energy level of one of the QDs.
The absolute value of the heat currents in the drag circuit, however, is small when compared to the heat currents in the drive circuit.

In addition, we consider the case where the tunnel couplings between QDs and leads depend on energy, since this situation occurs in experimental realizations~\cite{keller2016}.
We find that the heat currents in the drag circuit have similar amplitude but opposite signs, meaning that if heat is extracted from one reservoir, a similar amount of heat is deposited into the other.

The paper is organized as follows: in Sec.~\ref{sec:model} we define the systems under investigation and we describe the model Hamiltonian, while in Sec.~\ref{ress} we discuss the results we obtain.
We first consider the sequential tunneling regime for a single QD with a doubly-degenerate level in Sec.~\ref{sec2deg}, with two non-degenerate levels in Sec.~\ref{ndegSeq}.
Then we discuss the results obtained when cotunneling contributions are accounted for in Sec.~\ref{secCot} (with a doubly-degenerate level in Sec.~\ref{degCot}, and with two non-degenerate levels in Sec.~\ref{ndegCot}).
Sec.~\ref{secNL} is devoted to the results obtained with the four-terminal (non-local) configuration, with two QDs each coupled to two reservoirs.
The summary can be found in Sec.~\ref{secConc}.
Details of the calculations relative to the master equation in the sequential tunneling regime are reported in Appendix~\ref{ME}, while cotunneling contributions are reported in Appendix~\ref{appQD} and \ref{AppIntegrals}.

\section{System and model}
\label{sec:model}
We consider a system consisting of a QD with two levels (relative to spin up and spin down), whose Hamiltonian reads
\begin{equation}
\label{Hqd}
	H_\text{QD} =  (\epsilon_\uparrow \hat{n}_{\uparrow} + \epsilon_\downarrow \hat{n}_{\downarrow})
	+ E_\text{C} \hat{n}_{\uparrow} \hat{n}_{\downarrow} ,
\end{equation}
where $\epsilon_\uparrow=\epsilon-\Delta\epsilon/2$ and $\epsilon_\downarrow=\epsilon+\Delta\epsilon/2$ are the energy of the two levels.
Here $\hat{n}_{\sigma}=\hat{c}^\dagger_{\sigma}\hat{c}_{\sigma}$ is a number operator, while $\hat{c}^\dagger_{\sigma}$ and $\hat{c}_{\sigma}$ are creation and destruction fermionic operators, respectively, for an electron with spin $\sigma$ in the QD.
The first two terms describe the discrete levels of the QD, while the last one accounts for the Coulomb repulsion between the electrons within the QD ($E_C$ represents the charging energy).
We assume the spacing $\Delta\epsilon$ to be much smaller than $E_{\rm C}$, so that the electrostatic interaction plays a fundamental role in the transport properties of the system.

The QD is tunnel-coupled to two electronic reservoirs characterized by a well defined temperature $T_\alpha$ and chemical potential $\mu_\alpha$,
whose Hamiltonians are given by
\begin{equation}\label{hamres}
H_\alpha=\sum_{k,\sigma} (\epsilon_{\alpha k\sigma} -\mu_\alpha) \hat{b}^\dagger_{\alpha k\sigma} \hat{b}_{\alpha k\sigma} ,
\end{equation}
where $\hat{b}_{\alpha k\sigma}$ and $\hat{b}^\dagger_{\alpha k\sigma}$ are, respectively, the destruction and creation operators for electrons in lead $\alpha={L, R}$ with energy $\epsilon_{\alpha k\sigma}$, spin $\sigma$ and momentum $k$.
The coupling Hamiltonian reads
\begin{equation}
\label{ht}
H_{\rm T}=\sum_{k,\sigma} (t_{ L} \hat{b}^\dagger_{L k\sigma} \hat{c}_{\sigma} + t_{R} \hat{b}^\dagger_{R k\sigma} \hat{c}_{\sigma})  + {\rm h.~c.},
\end{equation}
where $t_\alpha$ is the tunneling amplitude
between the QD and lead $\alpha$.
The Hamiltonian $H_{\rm T}$ is not symmetric in the coupling, namely $t_{L}\ne t_{R}$.
Indeed, this is the condition needed to obtain a finite rectification.

In the final part of the paper we explicitly consider a four-terminal (non-local) configuration consisting of two single-level QDs, each tunnel-coupled to two reservoirs as sketched in Fig.~\ref{setup}(c-d), whose Hamiltonian is given by Eq.~(\ref{Hqd}) where $\sigma$ specifies the QD ($\uparrow$/$\downarrow$ for the QD on the left/right).
In this case the coupling Hamiltonian reads
\begin{align}
\label{ht2}
H_{\rm T}=\sum_{k} (t_{L1} \hat{b}^\dagger_{L1 k\uparrow} \hat{c}_{\uparrow} + t_{L2} \hat{b}^\dagger_{L2 k\uparrow} \hat{c}_{\uparrow} +\nonumber\\
+t_{R1} \hat{b}^\dagger_{R1 k\downarrow} \hat{c}_{\downarrow} + t_{R2} \hat{b}^\dagger_{R2 k\downarrow} \hat{c}_{\downarrow}
)  + {\rm h.~c.},
\end{align}
while the Hamiltonian for the reservoirs is given by Eq.~(\ref{hamres}), with the index $\alpha$ now taking the following values: $L1$, $L2$, $R1$, $R2$. 

The state of the QD (or QDs) is specified through the probability $P(\{ \sigma\})$ of finding the QD in the electronic configuration described by the set of occupancies $\{n_{\sigma}\}$ of its levels (with $n_{\sigma}=0,1$).
Tunneling processes change the state of the QD thus modifying the occupancies of the levels from one configuration to another.
In what follows we shall first consider the sequential tunneling regime which accounts for tunneling processes involving single-electron hopping through the tunnel barriers representing the coupling between the QD and the leads.
Therefore, the transition rates of such processes are obtained from the Fermi golden rule up to the leading order in the coupling Hamiltonian $H_{\rm T}$ (see Appendix~\ref{appQD}).
The tunneling constants $\Gamma_\alpha$ characterizing the interaction between the QD and lead $\alpha$ are defined as
\begin{equation}
\Gamma_\alpha=\frac{2\pi}{\hbar} D_\alpha|t_\alpha|^2, 
\label{tunc}
\end{equation}
where $D_\alpha$ is the density of states of lead $\alpha=L,R$ at the Fermi energy.
In part of our analysis, we will further assume that the charging energy $E_C$ is the largest energy scale.
This allows us to neglect all electronic configurations in which the total number of electrons in the QD exceeds one.
Therefore we can describe the state of the QD by specifying the probability of finding the QD in the state with zero electrons $P_0$, with one electron in a level with spin up ($P_\uparrow$), and with one electron in a level with spin down ($P_\downarrow$).
The master equation needed to determine such probabilities and the expressions of the currents are reported in App.~\ref{ME}.

\section{Results}
\label{ress}
\begin{figure}[!thb]
	\centering	\includegraphics[width=0.9 \columnwidth]{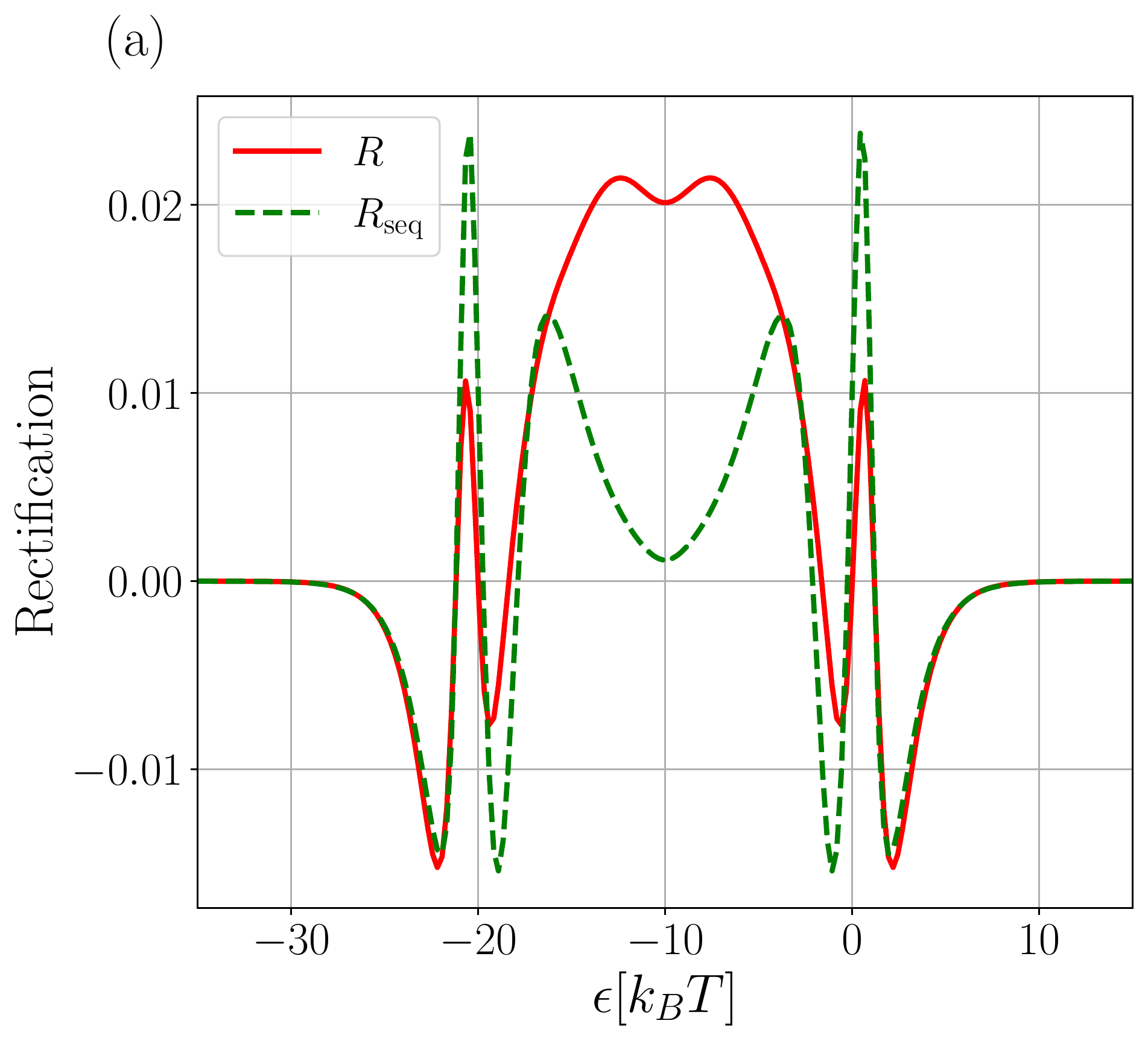}
	\centering	\includegraphics[width=0.9 \columnwidth]{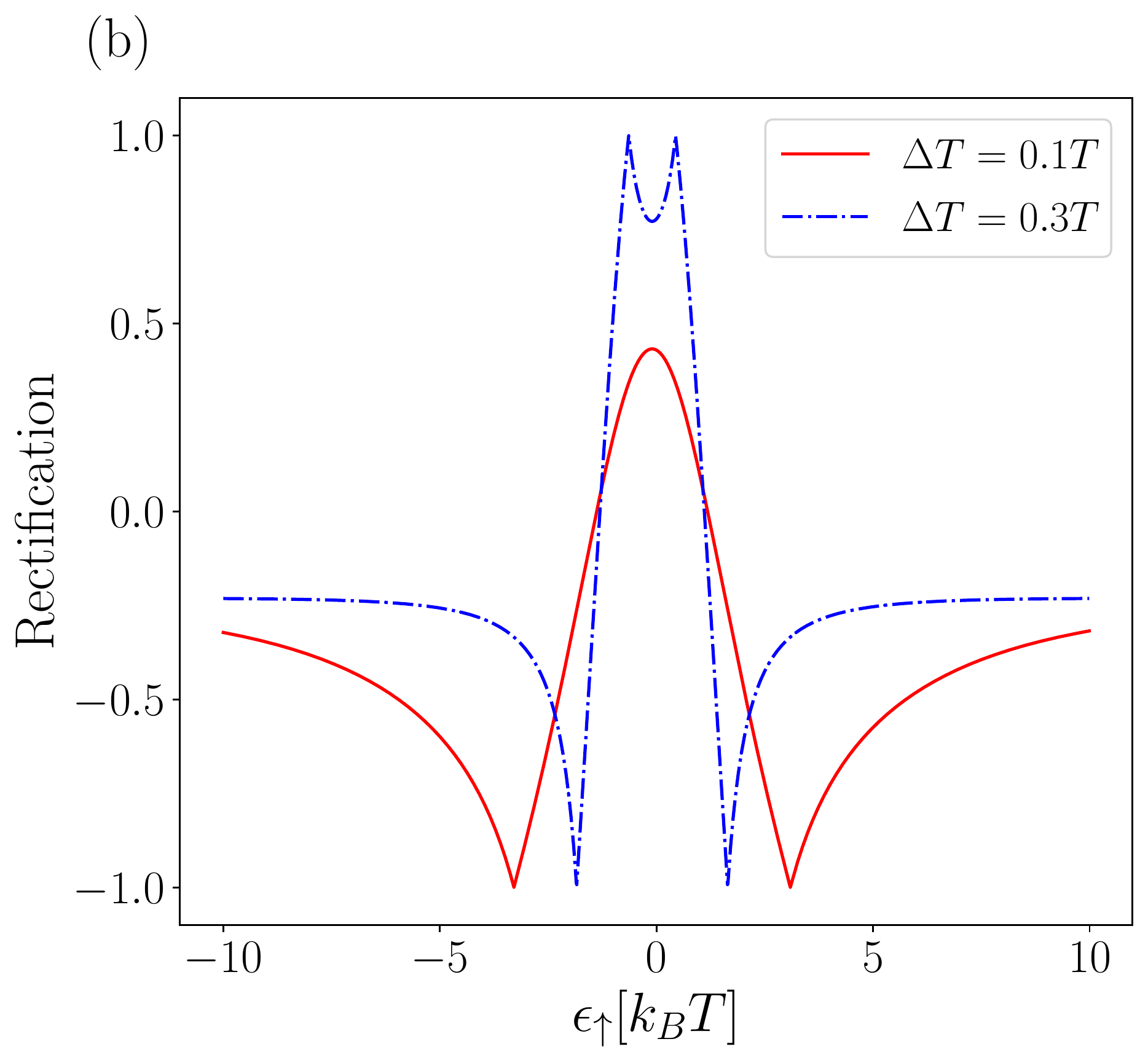}
	\caption{Rectification coefficient for two representative cases: (a) a single QD with two non-degenerate levels in the closed-circuit setup, and (b) two QDs in the four-terminal setup.
	The rectification coefficient is plotted as a function of the average energy of the levels $\epsilon$ in (a) and as a function of the level $\epsilon_\uparrow$ of the QD on the left-hand-side in (b).
The parameters used to obtain the curves in panel (a) and (b) are specified in the caption of Figs.~\ref{Cotunneling_c&r:QDnondeg:CCcotcurr} and \ref{figrecEnIn}, respectively. 
}
	\label{reprR}
\end{figure}
Let us first fix the notation: we denote by $I^c$ and $I$, respectively, the charge and energy current entering the QD from the left lead.
The heat current $J$ flowing through the left lead is thus expressed as
\begin{equation}
J=I-\frac{\mu_L}{(-e)} I^c ,
\end{equation}
where $-e$ is the electronic charge.
Note that both in the open-circuit setup, where $I^c=0$ [Fig.~\ref{setup}(e)], and in the closed-circuit setup, where $\mu_L=\mu_R=0$ [Fig.~\ref{setup}(f)], we have that heat and energy current coincide ($J=I$).

Before discussing in details our results in the various regimes and configurations, we first show the rectification coefficient obtained for two representative cases.
Namely, for a single QD [see Fig.~\ref{setup}(a) and (b)] with two non-degenerate levels in the closed-circuit setup and for a pair of QDs in the four-terminal setup [see Fig.~\ref{setup}(c) and (d)].
In Fig.~\ref{reprR}(a) the rectification coefficient is plotted for the former case as a function of the energy of the level $\epsilon$.
While $R$ includes cotunneling contribution (solid red curve), $R_{\rm seq}$ accounts for sequential tunneling processes only (green dashed curve).
Fig.~\ref{reprR}(a) shows that $R$ is typically not very large (for the parameters used here, the maximum value of $R$ is of the order of 2\%) and can take both negative and positive values depending on the position of the levels of the QD.
Remarkably, we find that cotunneling contributions can increase the rectification in a wide range of values of $\epsilon$.
All details will be discussed in Sec.~\ref{ndegCot}.

In Fig.~\ref{reprR}(b) the rectification coefficient for the four-terminal (non-local) configuration is plotted as a function of the level $\epsilon_\uparrow$ of the QD on the left-hand-side.
We consider two values of temperature bias: $\Delta T=0.1T$ (red solid curve) and $\Delta T=0.3T$ (blue dashed-dotted curve).
Remarkably, the blue curve spans the whole range of values of $R$  ($[-1,1]$), while the red curve takes values in the range $[-1, 0.4]$.
Overall, rectification is large in quite wide ranges of values of $\epsilon_\uparrow$.
As we will see in more details in Sec.~\ref{secNL}, however, the heat currents are rather small when compared with the single QD setup.

In the following sections we will describe the results obtained within the sequential tunneling regime (Sec.~\ref{secSeq}) and the results obtained accounting for the cotunneling contributions (Sec.~\ref{secCot}).

\subsection{Sequential tunneling regime}
\label{secSeq}
In this section we will assume that the charging energy $E_C$ is so large that we can neglect all electronic configurations in which the total number of electrons in the QD exceeds one.
This assumption, which will be lifted in Sec.~\ref{secCot}, allows us to obtain analytical results.
The results shown in the following Secs.~\ref{sec2deg} and \ref{ndegSeq} are in agreement and largely extend the results presented in Ref.~\onlinecite{vicioso2018}.
In particular, on the one hand, we will identify upper bounds for rectification and, on the other, we will discuss the relevant mechanisms allowing for the optimization of rectification both for degenerate and non-degenerate levels.

\subsubsection{Degenerate level}
\label{sec2deg}
Let us consider the degenerate case $\Delta\epsilon=0$ in which the charge current can be written as
\begin{equation}\label{Sequential:1:J^c}
		I^c=-2e \Gamma_L\Gamma_R \frac{f_L(\epsilon)-f_R(\epsilon)}{\Gamma_L[1+f_L(\epsilon)]+\Gamma_R[1+f_R(\epsilon)]},
\end{equation}
while the heat current, in accordance with Ref.~\onlinecite{vicioso2018}, takes the form
	\begin{equation}\label{Sequential:1:J^h}
		J=2(\epsilon-\mu_L)\Gamma_L\Gamma_R \frac{f_L(\epsilon)-f_R(\epsilon)}{\Gamma_L[1+f_L(\epsilon)]+\Gamma_R[1+f_R(\epsilon)]},
	\end{equation}
thus showing that the heat current is proportional to the charge current.
A direct consequence of this is that in the open-circuit setup (where there is no charge flow) the heat current is zero.

\begin{figure}[!t] 
\centering
	\includegraphics[width=0.95\columnwidth]{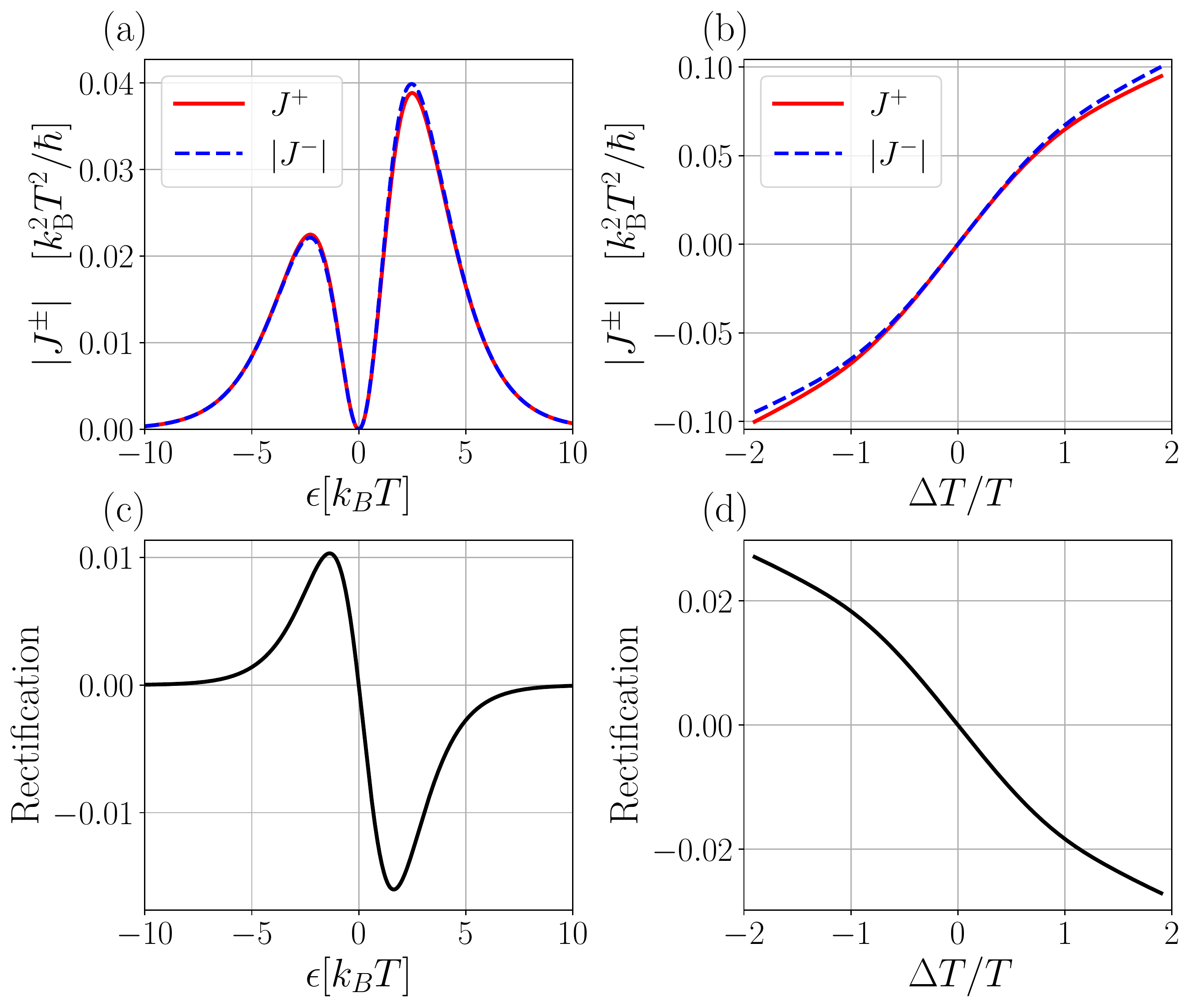}
	\caption{Closed-circuit setup for a doubly-degenerate level of energy $\epsilon$, measured with respect to $\mu_L=\mu_R=0$.
	Absolute value of the forward (red solid curve) and backward (blue dashed curve) heat currents (a), and the corresponding rectification $R$ (c). Panels (a) and (c) are in agreement with the results of Ref.~\onlinecite{vicioso2018} (see Fig.~3). The tunneling constants of the barriers are $\Gamma_L=2\Gamma_R=0.3k_BT/\hbar$.
In panels (a) and (c), we set $k_B\Delta T=0.5k_BT$
and heat currents and rectification are plotted as functions of $\epsilon$. In panels (b) and (d), the levels' energy is fixed at $\epsilon=2k_BT$,
and heat currents and the rectification are plotted as functions of the temperature bias $\Delta T$.}
\label{Sequential:1:grafici}
\end{figure}

In the closed-circuit setup, however, the heat currents are finite and the rectification can be written as~\cite{vicioso2018}
\begin{equation}\label{Sequential:1:rect}
	R=\frac{\Gamma_L-\Gamma_R}{\Gamma_L+\Gamma_R}\frac{f_R(\epsilon)-f_L(\epsilon)}{[2+f_L(\epsilon)+f_R(\epsilon)]},
\end{equation}
showing that a necessary condition to obtain rectification is that $\Gamma_L\ne \Gamma_R$ (this condition reflects the necessity to break the mirror-symmetry of the system).
In Fig.~\ref{Sequential:1:grafici} we plot the absolute values of the forward and backward heat currents and the resulting rectification.
In particular, in panel (a) $J^+$ and $|J^-|$ are plotted as functions of the QD levels' energy $\epsilon$.
When $\epsilon$ is zero (i.e.~when $\epsilon$ is aligned with the common chemical potential of the leads), both heat currents vanish because in this symmetric situation the sequential processes relative to the two leads cancel out.
When $|\epsilon|\gg 0$, the heat currents decrease exponentially because the QD is locked in the same state and the electrons cannot tunnel.
Indeed, $P_0$ goes quickly to 1 (and $P_\uparrow=P_\downarrow$ goes to zero) when $\epsilon$ increases to positive values over the scale set by $k_BT$, since electrons do not have enough energy to enter the QD, while $P_0$ goes quickly to zero (and the QD gets occupied) when $\epsilon$ decreases to negative values, as the energy level of the QD goes well below the chemical potential of the leads.
Notice that the currents display two asymmetric maxima at $|\epsilon|\approx 2.5k_BT$, and the maximum at $\epsilon\approx -2.5k_BT$ is lower than the one at $\epsilon\approx 2.5k_BT$.
The reason for this is that the probability of having one electron in the QD is higher when $\epsilon$ is negative (and of the order of $k_BT$), as compared to when $\epsilon$ is positive, so that fewer electrons can enter the QD (thus contributing to the current) because the charging energy does not allow any other electron to tunnel in the QD.

In panel (c) the rectification is plotted as a function of the QD levels' energy $\epsilon$.
Like the currents, the rectification goes to zero when $\epsilon=0$ and when $|\epsilon|\gg0$.
Intuitively, we can understand that $R$ gets suppressed for large positive values of $\epsilon$ by noting that in this situation Coulomb interaction plays little role (the QD is essentially unoccupied).
Thus the QD virtually behaves as a non-interacting one where rectification does not occur.
It turns out that the magnitude of rectification has two asymmetric maxima at $|\epsilon|\approx1.6k_BT$, where $|R|\approx 0.010$ and $|R|\approx 0.015$.
It is worth stressing that both heat currents and rectification are close to their maximum when $\epsilon$ is within the interval 1.6-2.5 $k_BT$, so that the QD operates as a heat rectifier to the best of its capabilities.
In panels (b) and (d), we plot the heat currents ($J^+$ and $|J^-|$) and the rectification, respectively, as functions of the temperature bias $\Delta T$.
Increasing $|\Delta T|$, the currents and their separation ($J^++J^-$) grow.
Notice that when $\Delta T/T\ll1$ and the currents are in the linear-response regime, the rectification vanishes ($R=0$ at $\Delta T=0$).
Increasing $\Delta T$, the currents go beyond the linear-response regime and $R$ varies linearly with $\Delta T$.
For values of $\Delta T/T$ larger than 0.5 the rectification is sublinear, but monotonous, thus reaching its maximum at $|\Delta T|=2T$.

Interestingly, in this configuration it is possible to find an upper bound that limits the rectification in this system.
Let us consider the rectification parameter $R$ as a function of $f_L$ and $f_R$, see Eq.~(\ref{Sequential:1:rect}), and look for its maximum over the possible values that the Fermi distribution functions can take.
It is important to notice that since $f_L$ and $f_R$ are evaluated at the same energy $\epsilon$, the quantities $f_L(\epsilon)$ and $f_R(\epsilon)$ cannot take arbitrary values between 0 and 1.
Actually, it is easy to see that both $f_{L}, f_{R}\leq 1/2$, when $\epsilon\geq 0$, or both $f_{L}, f_{R}\geq 1/2$, when $\epsilon\leq 0$ (recall that in the closed-circuit setup $\mu_L=\mu_R=0$), regardless of the temperatures $T_L$ and $T_R$.
With this constraint taken into account, it is possible to prove that the maxima of the function $|(f_L-f_R)/(2+f_L+f_R)|$, appearing in Eq.~(\ref{Sequential:1:rect}), occur when $f_L=0$ and $f_R =1/2$, or when $f_L=1/2$ and $f_R =0$.
In particular, $f_{L,R}=0$ corresponds to $\epsilon/(k_BT_{L,R})\gg 1$, i.e.~$T_{L,R}\ll\epsilon/k_B$, and $f_{L,R}=1/2$ corresponds to $\epsilon/(k_BT_{L,R})\simeq 0$, i.e.~$T_{L,R}\gg\epsilon/k_B$.
By substituting the above values of $f_L$ and $f_R$, we obtain the following upper bound
\begin{equation}
|R| \leq R_{\rm max}=\frac{1}{5}\left| \frac{\Gamma_L-\Gamma_R}{\Gamma_L+\Gamma_R} \right| .
\label{r15}
\end{equation}

\subsubsection{Non-degenerate levels}
\label{ndegSeq}
Let us now consider the non-degenerate case $\Delta\epsilon\ne 0$ in which the spin degeneracy of the level of the QD is broken, for example, through the Zeeman effect by applying a magnetic field.

\begin{figure}[h!]
\centering
	\includegraphics[width=\columnwidth]{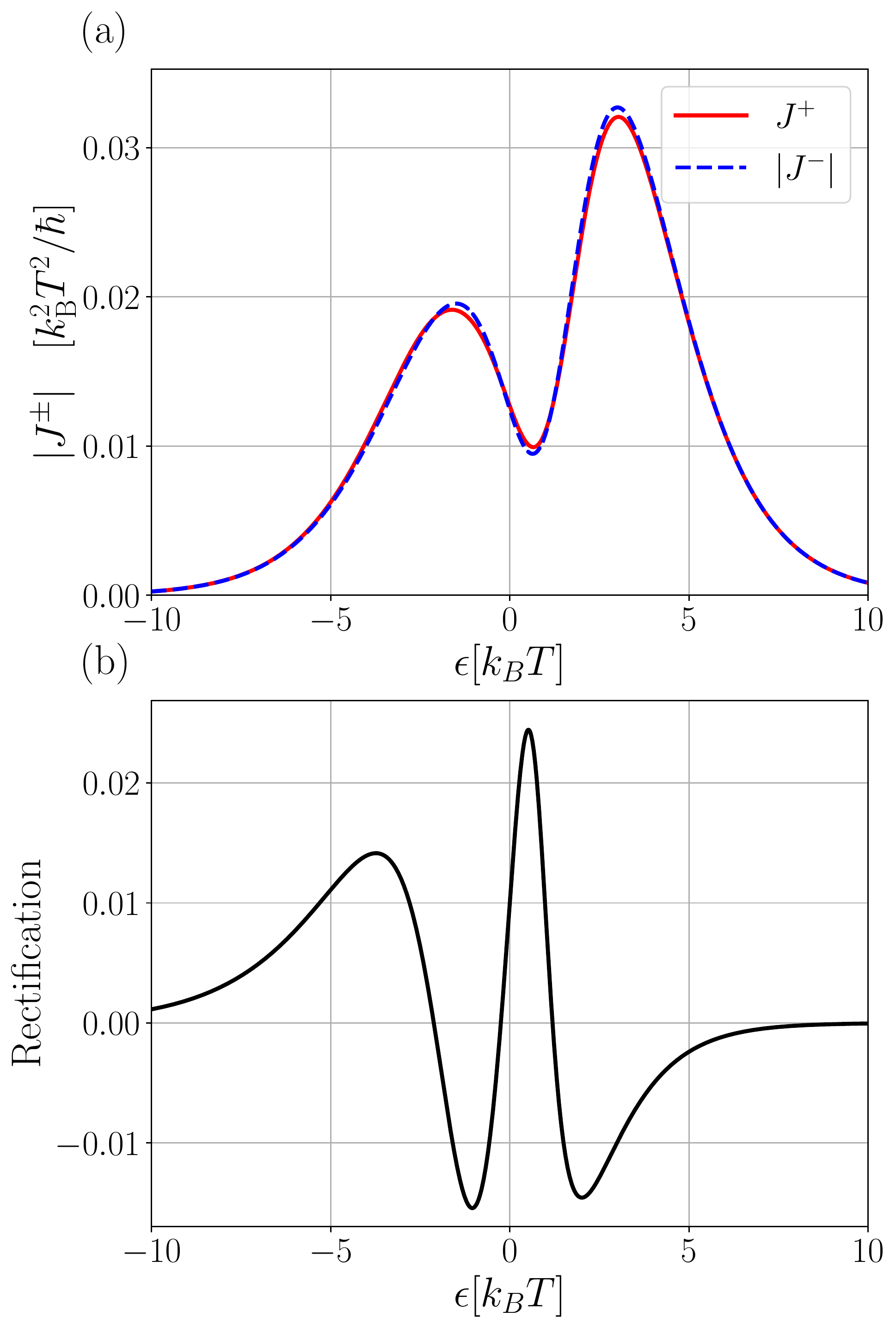}
	\caption{Closed-circuit setup for two non-degenerate levels of energies $\epsilon\pm\Delta\epsilon/2$.
	Absolute values of forward (red solid curve) and backward (blue dashed curve) heat currents [panel (a)], and the corresponding rectification (black solid curve) [panel (b)], are plotted as functions of $\epsilon$, measured with respect to $\mu_L=\mu_R=0$. We have used the same parameters as for Fig.~\ref{Sequential:1:grafici} and $\Delta\epsilon=2k_BT$.
Both panels are in agreement with the results of Ref.~\onlinecite{vicioso2018} (see Fig.~3).
}
\label{Sequential:2:grafici_CC}
\end{figure}
{\it Closed-circuit setup.--}
In this configuration we set $\mu_L=\mu_R=0$.
In Fig.~\ref{Sequential:2:grafici_CC} we plot the heat currents calculated as functions of the average levels' energies $\epsilon$.
We first note that the heat current [panel (a)] resembles the behavior found for the degenerate case, reported in Fig.~\ref{Sequential:1:grafici}, with important differences.
Like in the degenerate case, Sec.~\ref{sec2deg}, the currents have two asymmetric maxima and are suppressed exponentially at large $|\epsilon|$
(the currents are suppressed when $\epsilon\gg0$, since the QD is mostly empty, and when $\epsilon\ll 0$, the QD is mostly occupied by one electron).
However, at $\epsilon=0$, the sequential tunneling processes from the left lead do not cancel out with the ones from the right lead because now the levels have different energies.
As a consequence, when $\epsilon=0$ both heat currents are finite.
Moreover, they present a local minimum when $\epsilon\simeq \Delta\epsilon/2$, independently of the values of $\Delta T$ and tunneling constants.
This can be understood by noting that, at least when $\Delta\epsilon\gtrsim k_BT_L,k_BT_R$, electron transport is mainly due to the lower level ($\epsilon_\uparrow$), which is aligned with the common chemical potential of the leads, while the upper level is too high in energy, thus hardly populated ($P_\downarrow\simeq 0$).
In this situation, however, energy current is minimum since $\epsilon_\uparrow=0$.
This makes clear that the value of heat current at the minimum is finite if $\Delta\epsilon\simeq k_BT_L,k_BT_R$ (this is the case of Fig.~\ref{Sequential:2:grafici_CC}), while it vanishes when $\Delta\epsilon\gg k_BT_L,k_BT_R$, or when $\Delta\epsilon\ll k_BT_L,k_BT_R$, where the levels are nearly degenerate.
Notice that the distance between the two maxima is controlled by the average thermal energy $k_BT$.

The resulting rectification is plotted in panel (b) as a function of $\epsilon$.
For large values of $|\epsilon|$, $R$ behaves as in the degenerate case, while when $|\epsilon|<5k_BT$ the rectification oscillates between positive and negative values presenting an absolute maximum at $\epsilon\approx \Delta\epsilon/4$, close to the heat current minimum.
Moreover, the negative dip on the left occurs approximately at $-\Delta\epsilon/2$, while the negative dip on the right occurs approximately at $\Delta\epsilon$: both correspond to values of heat currents between the local minimum and the maxima.
The positions of such peaks and dips, however, depend also on the other parameters of the system.

The behavior of heat currents and rectification with the bias $\Delta T$ is essentially the same as the one found in Sec.~\ref{sec2deg}.
Notice that the upper bound found for the degenerate case, Eq.~\eqref{r15}, does not apply here.
For $\epsilon=k_BT/2$
and the parameters used in Figs.~\ref{Sequential:1:grafici} and \ref{Sequential:2:grafici_CC} we find $R\simeq 0.11$ for the largest value of $\Delta T$, which is larger than the bound $R_{\rm max}\simeq 0.067$.

\begin{figure}[h!]
\centering
	\includegraphics[width=\columnwidth]{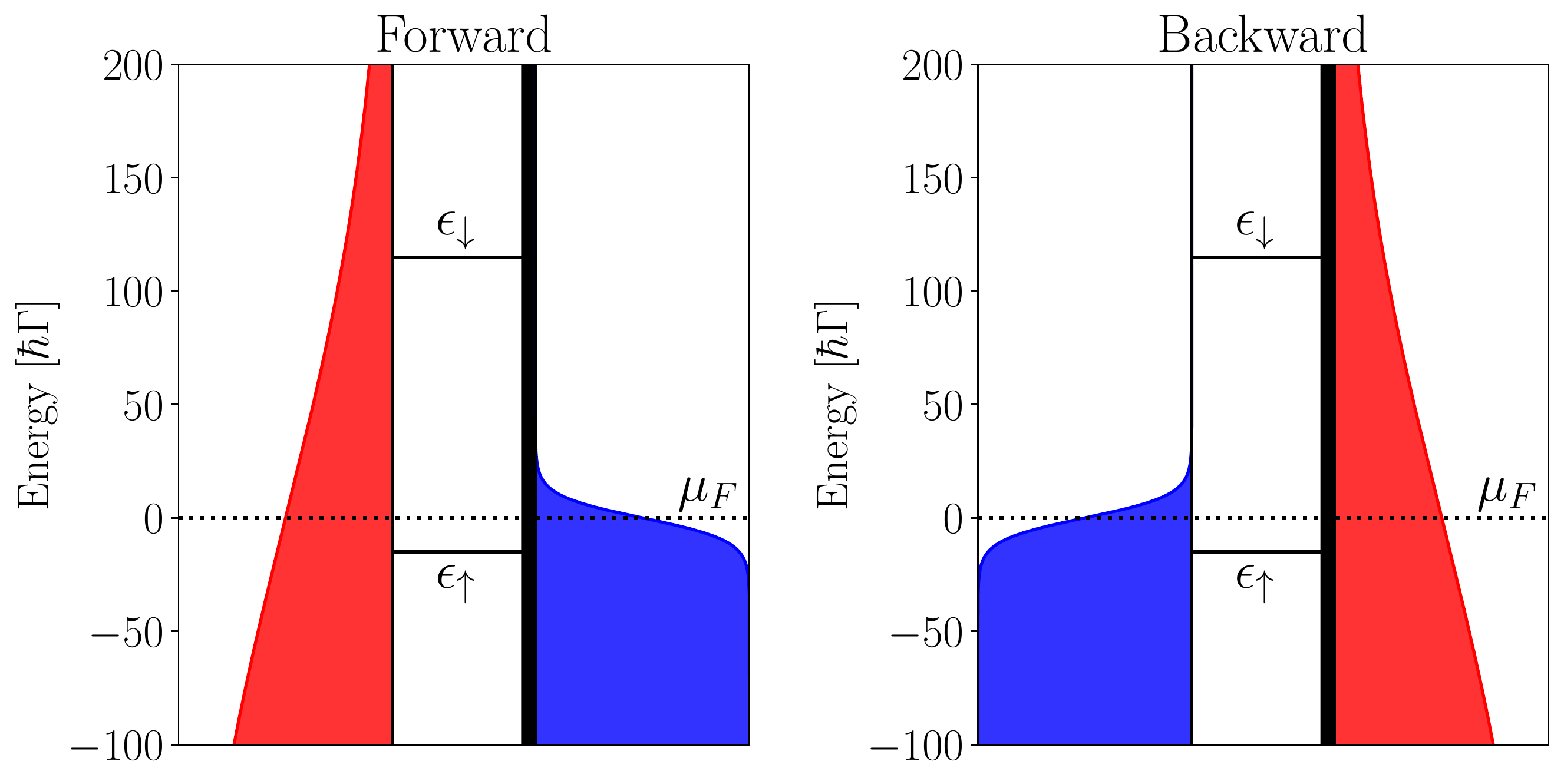}
	\caption{Energy diagrams in the closed-circuit setup relative to the following parameters: $\Gamma_L=0.1 k_BT/\hbar$, $\Gamma_R=0.02 k_BT/\hbar$, $\epsilon=k_BT$, $\Delta\epsilon=2.6k_BT$ and $\Delta T=1.8T$. 		The right barrier is thicker than the left one because the tunneling constants satisfy $\Gamma_R\ll\Gamma_L$.
		In the left panel the device is in the forward configuration, namely $T_L>T_R$, while, in the right panel the leads' temperatures are exchanged and the device is in the backward configuration.
		In both panels the chemical potentials of the leads ($\mu_F$) are set to zero (dotted line).
		The rectification coefficient turns out to be $R=0.31$.}
		\label{schemaCC}
\end{figure}

Intuitively one can expect the rectification to be optimized by maximizing the asymmetry between the tunneling constants $\Gamma_L$ and $\Gamma_R$, and for large temperature bias.
By using Eq.~(\ref{Sequential:J^u}) we can prove that in the closed-circuit setup it is possible to reach perfect rectification when the parameters satisfy the conditions: i) $\epsilon_\uparrow<0$, ii) $\Delta\epsilon\gg |\epsilon_\uparrow|$, iii) $T+|\Delta T|/2\gg \epsilon_\downarrow$, iv)  $T-|\Delta T|/2\ll |\epsilon_\uparrow|$.
Such conditions are represented, for the forward configuration, in the energy diagram in Fig.~\ref{schemaCC}(left panel), where the lead $L$ is hot and the lead $R$ is cold.
Because of iii) and iv), the distribution functions can be approximated as $f_{L\uparrow}\approx f_{L\downarrow}\approx1/2$ and $f_{R\uparrow}\approx1-f_{R\downarrow}\approx1$, so that the forward heat current takes the form
\begin{equation}
		J^+\approx\frac{\Gamma_L\Gamma_R}{2\Lambda}\left[\epsilon_\downarrow\frac{\Gamma_L}{2}-\epsilon_\uparrow\left(\frac{\Gamma_L}{2}+\Gamma_R\right)\right]
\end{equation}
(refer to App.~\ref{ME} for the notation).
By taking the limit $\Gamma_L\gg\Gamma_R$, one finds that $\Lambda\approx3\Gamma^2_L/4$ so that $J^+\approx\Gamma_R\Delta\epsilon/3$.
The physical origin of this expression for the heat current can be understood by looking at the energy diagram in Fig.~\ref{schemaCC}(left panel).
Since the right lead's temperature is much smaller than $|\epsilon_\uparrow|$, the tunneling rate of the process which transfer an electron from $\epsilon_\uparrow$ to lead $R$ and the one which transfer an electron from lead $R$ to $\epsilon_\downarrow$ are suppressed. Thus, the heat transport happens mainly through two following processes. The first one involves an electron tunneling from lead $L$ to $\epsilon_\downarrow$ and, from there, to lead $R$. In the second one, the electron starts in the right lead, tunnels into $\epsilon_\uparrow$, and then arrives in lead $L$. Such processes occur on the same typical time, of the order of $1/\Gamma_R$. Since they involve different QD levels, they transfer different amounts of heat.

For the backward configuration, where the lead $L$ is cold and the lead $R$ is hot, the energy diagram is represented in Fig.~\ref{schemaCC}(right panel).
Because of iii) and iv), the distribution functions can be approximated as $f_{L\uparrow}\approx1-f_{L\downarrow}\approx1$ and $f_{R\uparrow}\approx f_{R\downarrow}\approx1/2$, so that the backward heat current takes the form
\begin{equation}
\label{appHCb}
		J^-\approx\frac{\Gamma_L\Gamma_R}{2\Lambda}\left[\epsilon_\uparrow\left(\frac{\Gamma_R}{2}+\Gamma_L\right)-\epsilon_\downarrow\frac{\Gamma_R}{2}\right].
\end{equation}
By taking the limit $\Gamma_L\gg\Gamma_R$, one finds that $\Lambda\approx\Gamma^2_L$ so that $J^-\approx\Gamma_R\epsilon_\uparrow/2$. Note that $J^-$ is, correctly, a negative quantity.
Also in this case the physical origin of this expression for the heat current can be understood by looking at the energy diagram in Fig.~\ref{schemaCC}(right panel).
The low temperature of the lead $L$ suppresses the tunneling rate of the process in which an electron tunnels from  $\epsilon_\uparrow$ to the lead $L$ and of the one in which an electron tunnels from lead $L$ to  $\epsilon_\uparrow$. Therefore, heat transport is dominated by the process in which an electron in lead $L$ tunnels into  $\epsilon_\uparrow$ and, from there, tunnels into lead $R$, thus transferring an amount of heat equal to $\epsilon_\uparrow$ in a typical time $1/\Gamma_R$. This gives rise to the current in Eq.~(\ref{appHCb}).
By plugging in the expressions for $J^+$ and $J^-$ into the definition of $R$, Eq.~(\ref{erre}), we find $R=(\Delta\epsilon/3+\epsilon_\uparrow/2)/(\Delta\epsilon/3-\epsilon_\uparrow/2)$.
By imposing condition ii) we find $R\simeq 1$.
The drawback is that both heat currents are suppressed, since we have assumed a small value of $\Gamma_R$.

{\it Open-circuit setup.--}
Although the charge current is zero, the fact that the two levels have different energy allows heat transport, contrary to what happens in the degenerate case (Sec.~\ref{sec2deg}).
Indeed, to nullify the charge current [Eq.~(\ref{Sequential:J^c})] the rate of electrons tunneling into the lower level [first term in the square bracket of Eq.~(\ref{Sequential:J^c})] has to cancel out with the rate of electrons tunneling into the upper level (second term in the square bracket), namely
\begin{equation}\label{Sequential:2:OC_Jc=0}
	\Sigma_\downarrow^-(f_{L\uparrow}-f_{R\uparrow})=-\Sigma_\uparrow^-(f_{L\downarrow}-f_{R\downarrow}).
\end{equation}
In this condition the energy current is finite and reads
\begin{equation}\label{Sequential:2:OC_Ju_3}
\begin{split}
	I=&\frac{\Gamma_L\Gamma_R}{\Lambda}\Delta\epsilon\left[\Sigma_\uparrow^-(f_{L\downarrow}-f_{R\downarrow})\right]\\
	=&\Gamma_L\Gamma_R\Delta\epsilon P_\downarrow \frac{f_{L\downarrow}-f_{R\downarrow}}{\Gamma_Lf_{L\downarrow}+\Gamma_Rf_{R\downarrow}},
\end{split}
\end{equation}
where the last equality is obtained using Eq.~(\ref{solME}).

\begin{figure}[h!]
\centering
	\includegraphics[width=\columnwidth]{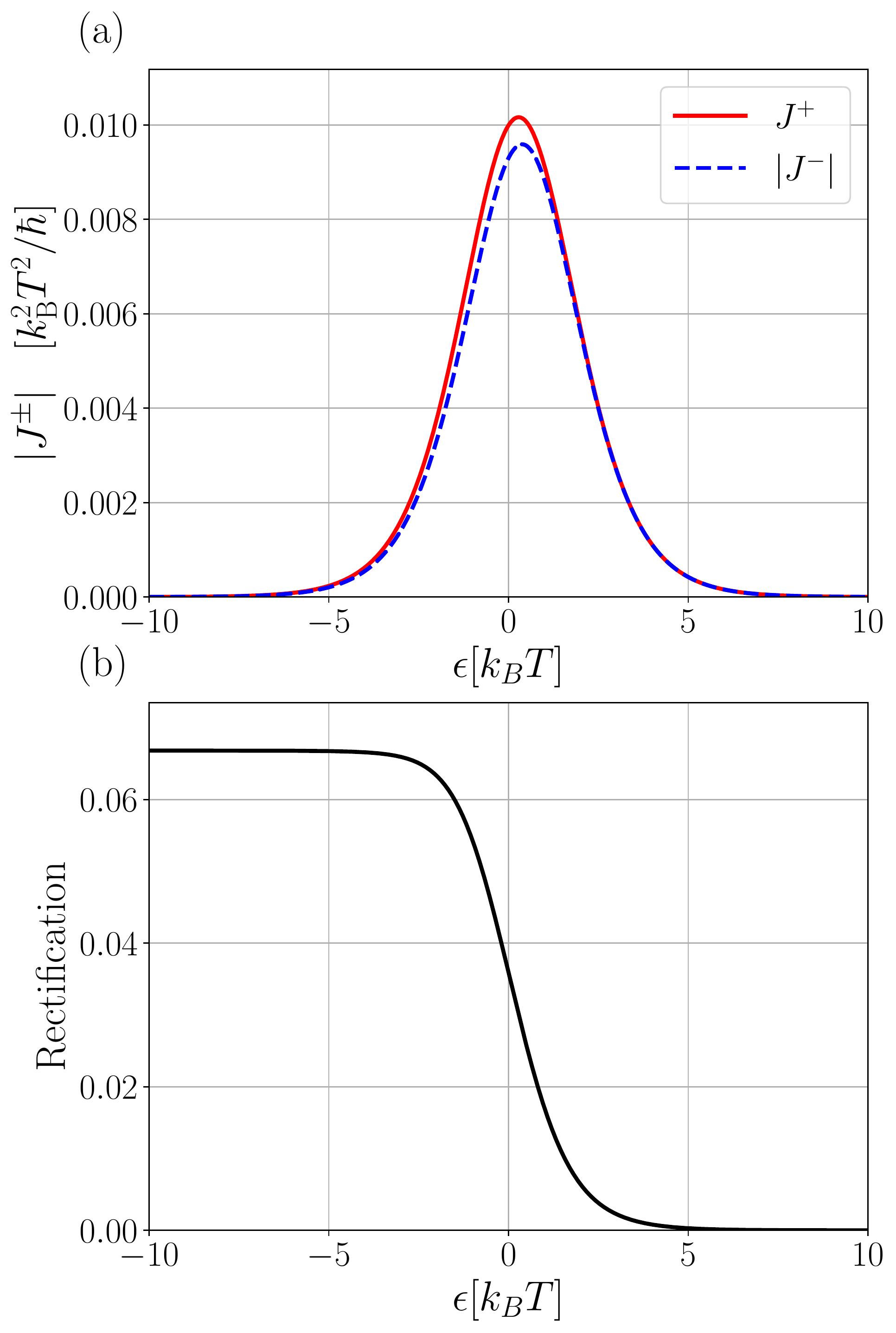}
	\caption{Open-circuit setup for two non-degenerate levels of energies $\epsilon\pm\Delta\epsilon/2$.
	Absolute values of forward (red solid curve) and backward (blue dashed curve) heat currents [panel (a)], and the corresponding rectification (black solid curve) [panel (b)], are plotted as functions of $\epsilon$, measured with respect to the average $(\mu_L+\mu_R)/2=0$. We have used the same parameters as for Fig.~\ref{Sequential:1:grafici} and $\Delta\epsilon=2k_BT$.
}
\label{Sequential:2:grafici_OC}
\end{figure}
In Fig.~\ref{Sequential:2:grafici_OC}, the heat currents and the rectification are plotted as functions of the levels' mean energy $\epsilon$.
Panel (a), where the absolute values of the heat currents $J^+$ and $J^-$ are displayed, shows that both are bell-shaped with the maximum occurring near $\epsilon=0$, while they are strongly suppressed when $|\epsilon|$ increases beyond $5k_BT$.
The bell-shape feature is a result of the compensation taking place in the open-circuit setup, whereby the charge current which would arise due to the temperature bias is counter-balanced by the appearance of a thermovoltage between the leads, effectively moving the weight of the curve towards the centre of the plot irrespective of the actual values of the temperature bias and of $\Delta\epsilon$.
Consistently, the same behavior was found~\cite{Zianni2007,Erdman2017} in the thermal conductance of a multilevel interacting QD.
Notice that the maximum occurs for a value of $\epsilon$ slightly away from zero.
The height of the maximum, however, does depend on $\Delta T$ (linearly up to $\Delta T\simeq 1.25 T$) and the energy separation between the levels.
$\Delta\epsilon\simeq 2.8k_BT$ is the value which yields the largest maximum for our choice of parameters.
The height of the peak goes rapidly to zero by moving $\Delta\epsilon$ away from this value, while the peak width remains virtually unaltered (in agreement with approximate analytical results for the thermal conductance in multilevel interacting QDs reported in Refs.~\onlinecite{Zianni2007,Erdman2017}).
Such width is slightly increasing with $\Delta T$, but only for $\Delta T>T$, while it is virtually independent of tunneling constants.
On the other hand, the rectification coefficient, plotted in panel (b) as a function of $\epsilon$, shows a peculiar behavior: when $\epsilon>5k_BT$ the rectification goes to zero, while when $\epsilon<5k_BT$ the rectification tends to a finite value.
The latter fact stems from the assumption that $E_C$ is the largest energy scale: indeed, as we will see in Fig.~\ref{Cotunneling_c&r:QDnondeg:OCcotcurr}(b), a different behavior is found for finite $E_C$.
The behavior of heat currents and rectification with the bias $\Delta T$ is essentially the same as the one found in Sec.~\ref{sec2deg}.

Using similar arguments as for the closed-circuit setup, we could find that one can reach ideal rectification under the following conditions: $\epsilon_\downarrow\simeq 0$, $\Delta\epsilon\gg k_BT$, $\Gamma_L\gg\Gamma_R$ and $T_L\gg T_R$.

\subsection{Cotunneling contributions}
\label{secCot}
In the previous section we studied the rectification of a QD in the sequential tunneling regime and under the assumption that the charging energy $E_C$ is much larger than any other energy in the system.
Such a condition allowed us to neglect the double occupation state of the QD and to find analytic expressions for the occupation probabilities ($P_0$, $P_\uparrow$ and $P_\downarrow$) and currents [see Eqs.~(\ref{solME}), (\ref{Sequential:J^c}) and (\ref{Sequential:J^u})].
In this section we include contributions from cotunnelling processes in the calculation of the current, allowing for a finite charging energy (i.e.~for the double occupation of the QD).
The latter account for coherent, second-order processes in the coupling Hamiltonian, that transfer an electron from one lead to the other via a virtual state either changing (inelastic) or not changing (elastic) the state of the QD.
The cotunneling transition rates (for charge and energy) are calculated taking into account that the QD can be initially empty [(0,0)], fully occupied [(1,1)], or occupied by one electron [either (0,1) or (1,0)], where in the notation $(i,j)$, $i=0,1$ refers to the level $\epsilon_\uparrow$ and $j=0,1$ refers to the level $\epsilon_\downarrow$.
Such cotunneling transition rates are calculated in details in App.~\ref{appQD}.
We stress that the inelastic cotunneling processes modify the state of the QD, thus modifying the MEs and their stationary solutions.
Such modified MEs are reported in App.~\ref{AppQDME} , see Eqs.~(\ref{meIN}).
Notice that the first square brackets on the right-hand-side of Eqs.~(\ref{meIN}) account for the sequential tunneling contribution only, where $P_2$ represents the probability for the QD to be doubly occupied.

The total currents are obtained by summing the cotunneling contributions to the sequential contribution.
Since charge and energy currents are conserved, in the following we will focus only on the currents flowing out of the left lead and express them as
\begin{equation}
	I^{c} = I^{c}_\text{seq} + I^{c}_\text{cot},
	\label{eq:j_cot}
\end{equation}
and
\begin{equation}
	I = I_\text{seq} + I_\text{cot},
	\label{eq:j_cot}
\end{equation}
respectively.
For consistency, in this section we account for the double occupancy of the QD even for the sequential currents.
Therefore, $I^{c}_\text{seq}(\Delta T)$ and $I_\text{seq}(\Delta T)$ are the currents calculated in the weak coupling regime, which, unlike Eqs.~(\ref{Sequential:J^c}) and (\ref{Sequential:J^u}), also accounts for the probability $P_2$ of finding two electrons in the QD and the related sequential processes.
The expressions for the currents $I^{c}_\text{seq}$ and $I_\text{seq}$ are reported in Eqs.~(\ref{Cotunneling_c&r:Jcseq}) and (\ref{Cotunneling_c&r:Juseq}).
Also the expressions for the cotunneling currents $I^{c}_\text{cot}$ and $I_\text{cot}$ are collected in App.~\ref{cot-cur}.

Before discussing the results on the specific situations and setups, in the following we show that the heat current is symmetric with respect to $\epsilon = -E_C/2$.
This can be understood by considering the symmetry properties of the Hamiltonian $H_{\rm QD}$.
Indeed, Eq.~(\ref{Hqd}) can be cast in the form
\begin{align}
H_\text{QD} = \sum_\sigma\left[-\epsilon_\sigma-E_C\right]\hat{h}_{\sigma}+ E_C\hat{h}_{\uparrow}\hat{h}_{h\downarrow} + \mathrm{const,}
\end{align}
where we have defined the the operator $\hat{h}_{\sigma} = 1-\hat{n}_{\sigma}$.
This proves that the Hamiltonian does not change by substituting $\epsilon_\sigma$ with $-\epsilon_\sigma-E_C$ and replacing the operator $\hat{n}_{\sigma}$ with the operator $\hat{h}_{\sigma}$.
This means that the Hamiltonian $H_{\rm QD}$ is particle-hole symmetric around $-E_C/2$, implying that $$I^c(\epsilon)=-I^c[-\epsilon-E_C]$$ and $$J(\epsilon)=J[-\epsilon-E_C]$$
as long as $E_C$ is finite and provided that the average of the chemical potentials is zero ($\mu_L+\mu_R=0$).
Note that $\mu_L=\mu_R=0$ in the closed-circuit setup and $\mu_L=-\mu_R$ in the open-circuit setup.

\subsubsection{Degenerate level}
\label{degCot}
Let us first consider the case of a QD with a doubly-degenerate level, namely $\Delta\epsilon=0$.
A few observations are in order.
The cotunnelling processes (either elastic or inelastic) in which both the initial and the final states have one electron in the same lead do not transfer energy, since $\Delta\epsilon=0$, and can be ignored.
Furthermore, the inelastic cotunneling processes that change the QD state from empty to doubly occupied occur rarely.
Indeed, for such processes to happen, the proper initial conditions must be fulfilled, namely the leads must provide available electrons at high energy (order of $E_C$) and the QD must be empty [see the first line of Eq.~(\ref{Cotunneling_c&r:Juine})].
However, when the electrons in the leads have large enough energy, the QD is rarely empty because of the occurrence of sequential tunneling processes, while when the QD is empty, the electrons in the leads do not have enough energy to overcome $E_C$.
The same happens for the processes that empty the initially doubly occupied QD.
Instead, for inelastic cotunneling processes that change the QD state from (1,0) to (0,1), and vice versa, the energy of the electrons involved in the process is enough to overcome the charging energy $E_C$, making the process more likely to happen, as shown in Fig.~\ref{fig:ineCurr}.
Therefore, the main contribution to the heat current comes from either elastic cotunneling processes or inelastic cotunneling processes that change the QD state $(1,0)\leftrightarrow(0,1)$.
However, at the end of this subsection we will find one scenario where cotunneling processes which move two electrons from/to the QD are responsible for the finiteness of the rectification.

\begin{figure}[h!]
	\centering
		\includegraphics[width=0.99\columnwidth]{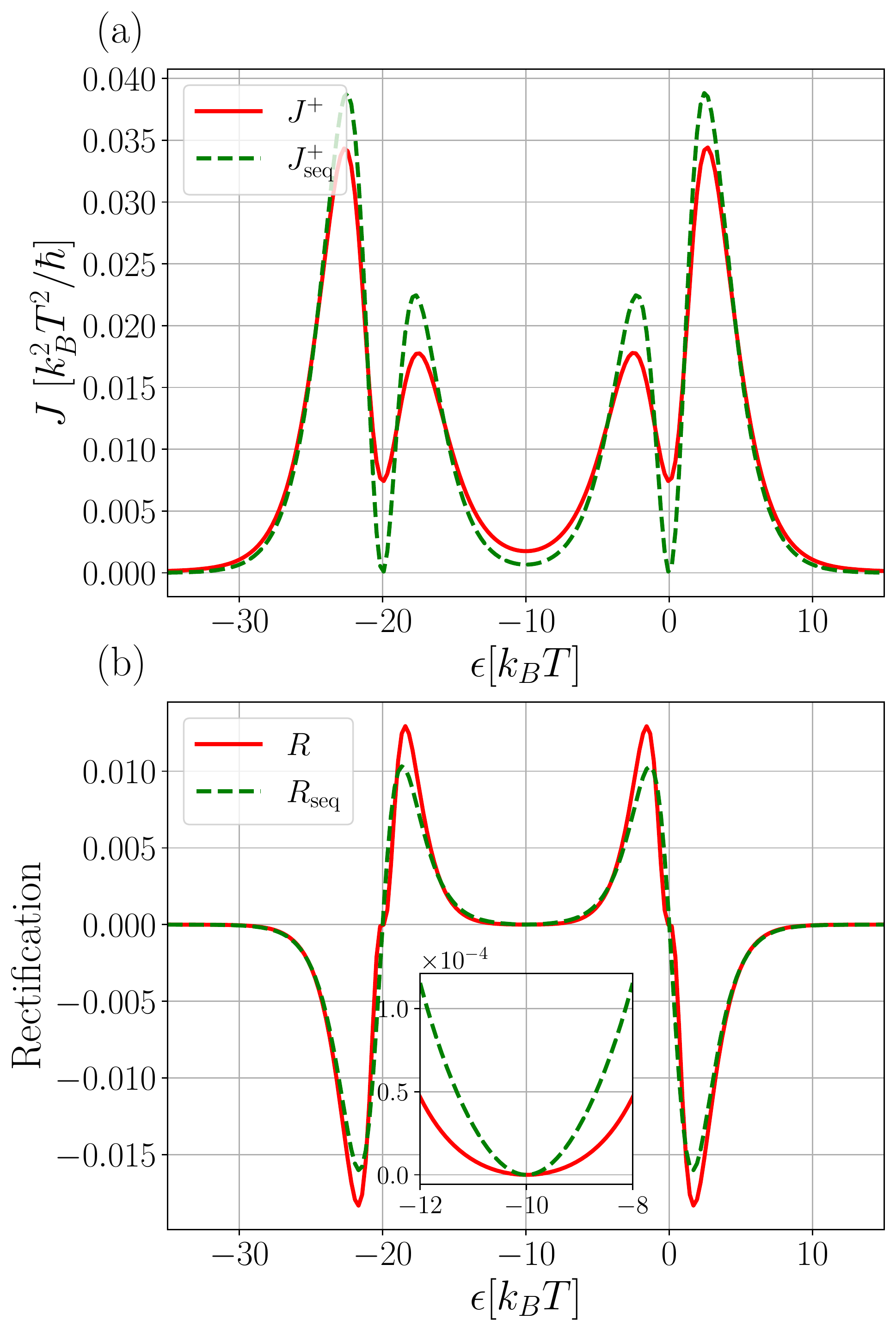}
		\caption{Closed-circuit setup. Heat currents (in units of $k_B^2T^2/\hbar$) and rectification coefficient for the case of a doubly-degenerate level as functions of the energy $\epsilon$ of the  level. Panel (a) shows the forward heat current which includes cotunnelings contribution $J^+$ (solid red curve) and in the presence of sequential processes only  $J^+_{\rm seq}$ (green dashed curve). Panel (b) shows the rectification $R$ which includes cotunneling contributions (solid red curve) and the rectification accounting for sequential tunneling processes only $R_{\rm seq}$ (green dashed curve). The inset in panel (b) contains a zoom of the main plot around the value $\epsilon=-10k_BT$, which shows that cotunneling contributions suppress rectification.
	All parameters are the same used for Fig.~\ref{Sequential:1:grafici} and $E_C=20k_BT$.}
		\label{Cotunneling_c&r:deg:CCcurrents}
	\end{figure}

{\it Closed-circuit setup.--}
In Fig.~\ref{Cotunneling_c&r:deg:CCcurrents}, forward heat current [panel (a)] and rectification coefficient [panel (b)] are plotted as functions of the energy $\epsilon$ of the level.
In panel (a), $J^+$ (solid red line) is obtained including the cotunneling contributions, while $J^+_{\rm seq}$ (dashed green line) accounts for the sequential tunnelling processes only.
Note that the latter curve resembles very much the corresponding curve, in the range of energies $\epsilon$ considered, in Fig.~\ref{Sequential:1:grafici}(a), which accounts for single occupation only.
The main message of Fig.~\ref{Cotunneling_c&r:deg:CCcurrents}(a) is that cotunneling contributions increase the heat current near $\epsilon=0$ and $\epsilon=-E_C=-20k_BT$, which correspond to the minima of the heat current in the sequential regime (dashed green line), while decrease the heat current for the values of $\epsilon$ which corresponds to the peaks.

Let us start discussing the sequential tunneling regime.
The heat current is nearly zero at $\epsilon=0$ and at $\epsilon=-E_C=-20k_BT$ because the energy carried by the electrons which tunnel through the QD in these two cases is zero, as one can understand from Eq.~(\ref{Cotunneling_c&r:Juseq}).
Indeed, when $\epsilon=0$, the QD has vanishing probability to be doubly occupied [there is not enough thermal energy for the QD to be in the state $(1,1)$], i.e.~$P_2\approx0$, while the functions $F_{L\uparrow}$ and $F_{L\downarrow}$ are also vanishing, since $E_C\gg k_BT$.
The remaining terms in Eq.~(\ref{Cotunneling_c&r:Juseq}), however, account for tunneling of electrons which carry no heat because $\epsilon=0$.
Similar arguments apply to the case $\epsilon=-E_C$.
In this situation the QD is very likely occupied (namely $P_0\approx0$), while $f_{L\uparrow}\approx f_{L\downarrow}\approx 1$, implying that $f^-_{L\uparrow}$ and $f^-_{L\downarrow}$ are vanishing.
The remaining terms account for tunneling of electrons which do not carry heat since $\epsilon+E_C=0$.

Cotunneling processes, however, allow electrons with energy different from $\epsilon$ to tunnel through the QD (through the virtual states), thus allowing a finite heat current to flow at $\epsilon=0$ and $\epsilon=-E_C=-20k_BT$ and giving rise to a reduction of the heat transfer.
Far from resonance, when the sequential forward heat current $J^+_{\rm seq}$ decreases exponentially, the cotunneling processes become the dominant transport processes, increasing the heat current.
Fig.~\ref{Cotunneling_c&r:deg:CCcurrents}(a) also confirms that $J^+$ is symmetric with respect to $\epsilon=-E_C/2=-10k_BT$.

The rectification coefficient is plotted in Fig.~\ref{Cotunneling_c&r:deg:CCcurrents}(b) as a function of $\epsilon$, with a solid red curve when cotunneling contributions are included, and with a dashed green curve when sequential tunneling processes only are accounted for.
First notice that the latter curve basically coincides with the curve in Fig.~\ref{Sequential:1:grafici}(c) in the range of energies $\epsilon$ considered.
We note that cotunneling increases the rectification in the ranges of values of $\epsilon$ where it lowers the magnitude of the currents and decreases the rectification where it increases them.
This behaviour can be understood as follows.
Let us define the cotunneling corrections to the forward and backward heat currents as $\Delta J^\pm=J^\pm-J^\pm_{\rm seq}$, respectively.
In turns out that the absolute values of $\Delta J^+$ and $\Delta J^-$ are nearly equal, but their sign is opposite.
This happens because the main contribution to the heat currents, as notice above, comes from cotunneling processes that are elastic and from inelastic processes occurring when the QD is occupied by one electron.
The cotunneling rates associated to such processes contain, under the integration symbol, the difference between the Fermi distributions of the leads [see for example Eqs.~(\ref{Cotunneling_ela:0curr}), (\ref{Cotunneling_ine:1currLR}) and (\ref{Cotunneling_ine:1currRL})] and, therefore, change sign under the inversion of the temperature bias.
Now, since
$\Delta J^+\approx -\Delta J^-$, we can express the effect of the cotunneling on the currents as $J^\pm=J^\pm_{\rm seq}\pm\Delta J^+$.
This implies that the absolute value of both currents are either increased or decreased, depending on whether $\Delta J^+$ is positive or negative, respectively.
Now, we can write the rectification with the cotunneling contributions as
\begin{equation}\label{Cotunneling_c&r:QDdeg:rect}
	R=\frac{J^{+}_{\rm seq}+J^{-}_{\rm seq}}{J^{+}_{\rm seq}-J^{-}_{\rm seq}+2\Delta J^{+}}.
\end{equation}
Therefore the rectification $R$ coincides with the sequential tunneling regime's rectification $R_{\rm seq}$ when the cotunneling correction is zero, namely $\Delta J^+=0$, is greater than $R_{\rm seq}$ when $\Delta J^+<0$, and is smaller than $R_{\rm seq}$ when $\Delta J^+>0$.
	\begin{figure}[h!]
		\centering
		\includegraphics[width=0.99\columnwidth]{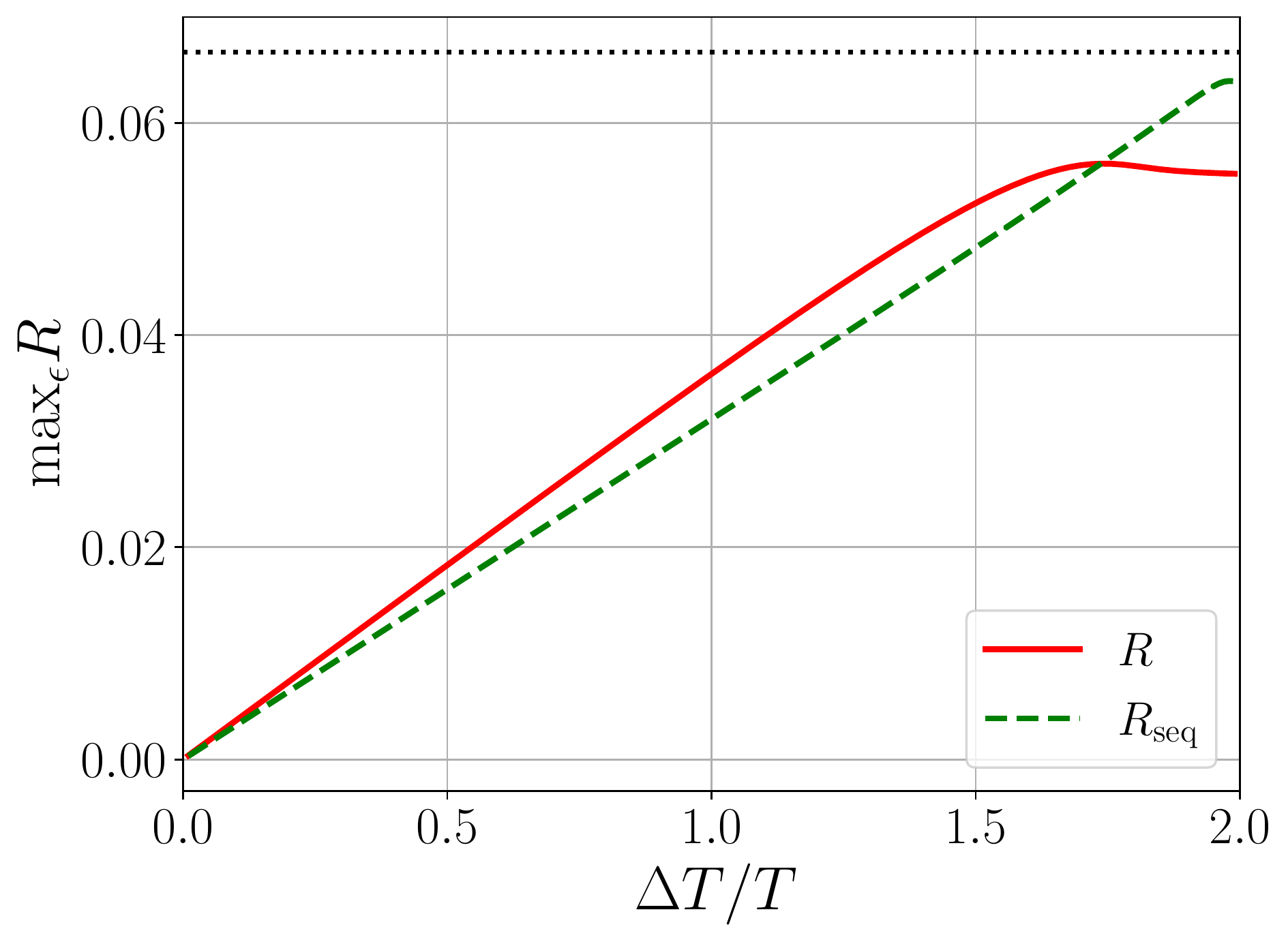}
		\caption{Maximum value (over the energy $\epsilon$ of the level) of rectification as a function of the temperature bias $\Delta T$.
		The solid red line represents the results obtained including the cotunneling contributions, while the dashed green line represents the results obtained with sequential processes only. The dotted black line is the upper-bound of the rectification found in the sequential tunneling regime, see Eq.~\eqref{r15}.
		All parameters are the same used for Fig.~\ref{Sequential:1:grafici} and $E_C=20k_BT$.
		}
		\label{Cotunneling_c&r:QDdeg:maxR}
	\end{figure}
	
In Fig.~\ref{Cotunneling_c&r:QDdeg:maxR}, the maxima over the QD level's energy $\epsilon$ of the rectifications with the cotunneling contributions $R$ (solid red line) and without the cotunneling contributions $R_{\rm seq}$ (dashed green line) are plotted as functions of the bias temperature $\Delta T$.
The dotted black line is the upper-bound of the rectification found in the sequential tunneling regime, see Eq.~\eqref{r15}, which, in the case of $\Gamma_L=2\Gamma_R$, is equal to $1/15\approx 0.067$.
We note that the cotunneling contributions increase the maximal rectification when the bias temperature is smaller than about $1.75 T$.
The fact that for larger values of $\Delta T$ cotunneling contributions decrease $R$ is a consequence of the fact that by increasing $\Delta T$, the local maxima of both $R$ and $R_{\rm seq}$, see Fig.~\ref{Cotunneling_c&r:deg:CCcurrents}(b), move towards $\epsilon=0$ and $\epsilon=-E_C$.
For such values of $\epsilon$, however, the cotunneling corrections $\Delta J^+$ to the heat currents are positive, see Fig.~\ref{Cotunneling_c&r:deg:CCcurrents}(a), thus producing,  according to Eq.~(\ref{Cotunneling_c&r:QDdeg:rect}), a decrease of $R$ with respect to $R_{\rm seq}$.

{\it Open-circuit setup.--}
\begin{figure}[h!]
	\centering
		\includegraphics[width=0.99\columnwidth]{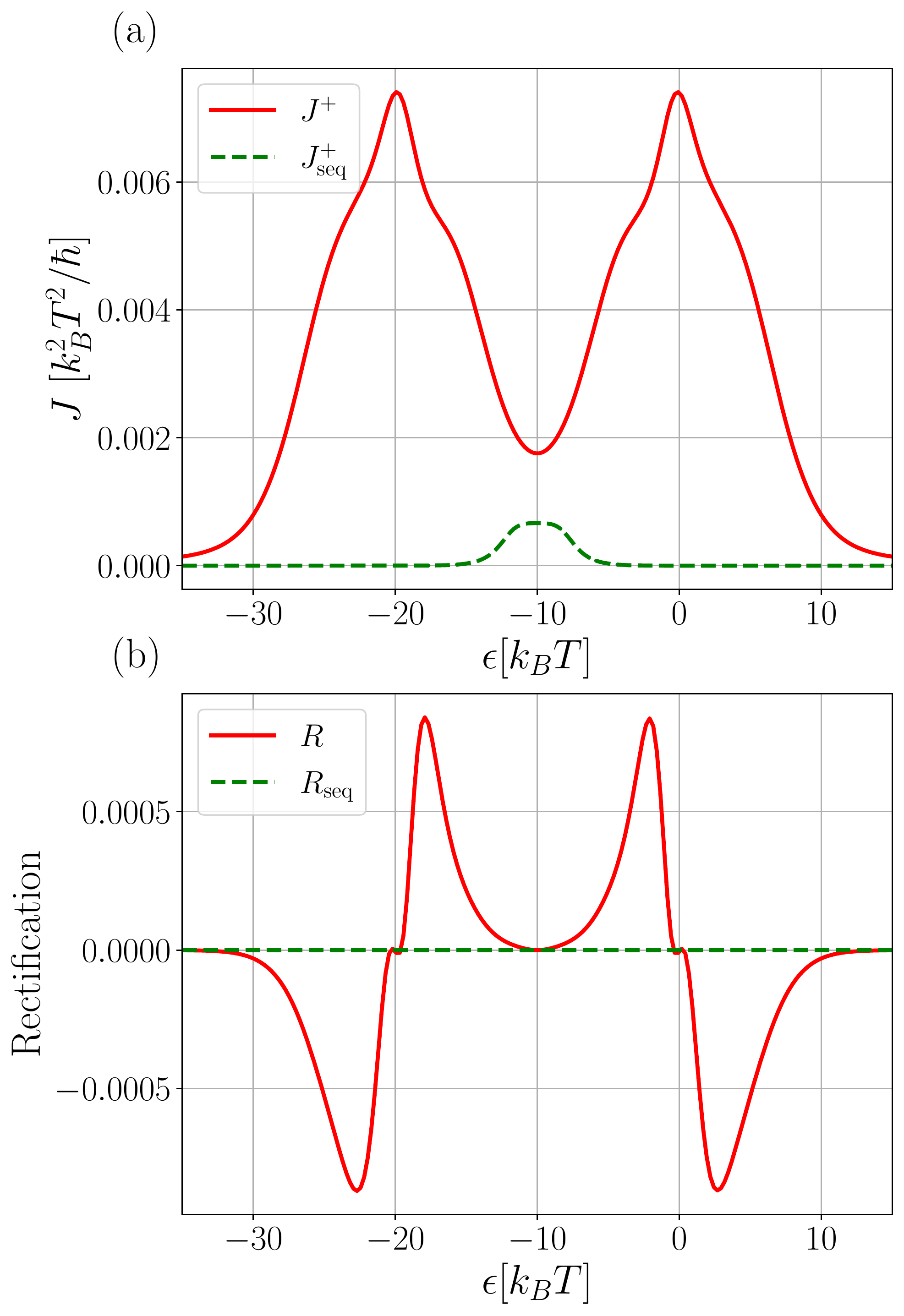}
		\caption{Open-circuit setup. Heat currents (in units of $k_B^2T^2/\hbar$) and rectification coefficient for the case of a doubly-degenerate level as functions of the energy $\epsilon$ of the  level. Panel (a) shows the forward heat current and panel 
		(b) shows the rectification $R$, which in this case is entirely due to the cotunneling contributions. Same color code as in Fig.~\ref{Cotunneling_c&r:deg:CCcurrents}.
    		All parameters are the same used for Fig.~\ref{Sequential:1:grafici}, while $E_C=20k_BT$.}
		\label{Cotunneling_c&r:deg:OCcurrrect}
\end{figure}
When $E_C$ is finite, the heat current (in the sequential tunneling regime) can be finite in the open-circuit setup since charge current and energy currents are not proportional to each other [see Eqs.~(\ref{Cotunneling_c&r:Jcseq}) and (\ref{Cotunneling_c&r:Juseq})], contrary to what we found in Sec.~\ref{sec2deg}.
This is due to the fact that, when $E_C$ is finite, the processes which involve the charging energy $E_C$ [the ones proportional to $F_{L\sigma}^\pm$ in Eq.~(\ref{Cotunneling_c&r:Juseq})] transfer a different amount of energy, namely $(E_C+\epsilon_\sigma)$, with respect to those which do not involve $E_C$, while transferring the same charge.
It is possible, however, to prove that the rectification still vanishes.
Indeed, by substituting the solution of the master equations~(\ref{meIN}), accounting for sequential tunneling processes only, in the expression of the charge current [Eq.~(\ref{Cotunneling_c&r:Jcseq})], one finds that the condition which nullifies such current is $f_L F_R^-=f_R F_L^-$, independently of the tunneling constants $\Gamma_L$ and $\Gamma_R$.
By plugging in this condition into the expression of the energy current one finds that $J^+=-J^-$, i.e.~there is no rectification.
Cotunneling processes, however, can generate rectification.

In Fig.~\ref{Cotunneling_c&r:deg:OCcurrrect}, forward heat currents [panel (a)] and rectification coefficient [panel (b)] are plotted as functions of the energy $\epsilon$ of the levels.
In panel (a), the forward heat current ($J^+$), obtained including the cotunneling contributions, is plotted together with the heat current $J^+_{\rm seq}$ ($=-J^-_{\rm seq}$) relative to the sequential processes only.
We first notice that $J^+_{\rm seq}$ is small but finite, as expected, although only around $\epsilon=-E_C/2=-10k_BT$, where the sequential processes involving the empty and doubly occupied states coexist.
Indeed, around $\epsilon=-E_C/2$, $P_0$ and $P_2$ turn out to be both small but finite.
Notice that for values of $\epsilon$ for which only one of the two ($P_0$ or $P_2$) is non-zero one finds that Eqs.~(\ref{Cotunneling_c&r:Jcseq}) and (\ref{Cotunneling_c&r:Juseq}) are proportional to each other, implying that the heat current is zero.
This is the case for $\epsilon<-E_C=-20k_BT$, where one finds that $P_0\approx 0$ (the QD is at least singly occupied) and $f_{L\sigma}^-\approx 0$, or for $\epsilon>0$, where $P_2\approx 0$ (the QD cannot be doubly occupied) and $F_{L\sigma}\approx 0$.

Remarkably, Fig.~\ref{Cotunneling_c&r:deg:OCcurrrect}(a) shows that the cotunneling processes contribute substantially to the overall heat current, which then grossly deviates from the sequential result.
On the one hand, this is due to the fact that cotunneling contributions to the heat current are essentially unrelated to the (overall) charge current, which is zero~\cite{nota2}.
On the other hand, the cotunneling contributions to the charge current modify the thermovoltage (with respect to the sequential situation) that establishes between the leads in the open-circuit condition. In turn, such a thermovoltage influences  the heat current (through the distribution functions entering the expressions of the sequential and cotunneling contributions).
This leads to an additional indirect modification of the overall heat current, with respect to the sequential result.
According to Fig.~\ref{Cotunneling_c&r:deg:OCcurrrect}(a), $J^+$ reaches its maximum at $\epsilon=0$ and at $\epsilon=-E_C=-20k_BT$, while being symmetric with respect to $\epsilon=-E_C/2$.

In Fig.~\ref{Cotunneling_c&r:deg:OCcurrrect}(b), the rectification $R$ which includes the cotunneling contributions is plotted as a function of the energy of the levels $\epsilon$.
$R$ presents two maxima at $|\epsilon|\approx 2.5k_BT$ and at $|\epsilon-E_C|\approx 2.5k_BT$.
However, the rectification is very small, at least one order of magnitude smaller than in the closed-circuit setup.
The reason is that the main cotunneling contributions to the heat current, mentioned in the beginning of Sec.~\ref{degCot}, for degenerate levels change sign under the inversion of the temperature bias.
More precisely, this is the case for the differences ${\cal J}^u_{L\sigma\rightarrow R\bar{\sigma}}-{\cal J}^u_{R\sigma\rightarrow L\bar{\sigma}}$ in the second and third term of Eq.~(\ref{Cotunneling_c&r:Juine}), which are proportional to the energy integral of the Fermi functions, calculated at equal energy, of the two leads [see Eqs.~(\ref{Cotunneling_ine:1currLR}) and  (\ref{Cotunneling_ine:1currRL})]. Moreover, it is important to mention that, for degenerate levels, $P_\uparrow$ and $P_\downarrow$ are virtually independent of the sign of the temperature bias.
As already mentioned above, also the quantities ${\cal J}^u_{ij,\sigma}$ in Eq.~(\ref{Cotunneling_c&r:Juela}) change sign under the inversion of the temperature bias.
In conclusion, rectification is generated by the rare cotunneling events, represented by the first and forth term in Eq.~(\ref{Cotunneling_c&r:Juine}), that move two electrons from the lead to the QD or vice versa.

\subsubsection{Non-degenerate levels}
\label{ndegCot}
Let us now consider the case of a QD with two non-degenerate levels, namely $\Delta\epsilon\neq 0$.
In this situation both elastic and inelastic co-tunnelling processes contribute significantly to the heat current.

{\it Closed-circuit setup.--}
\begin{figure}[h!]
\centering
	\includegraphics[width=0.99\columnwidth]{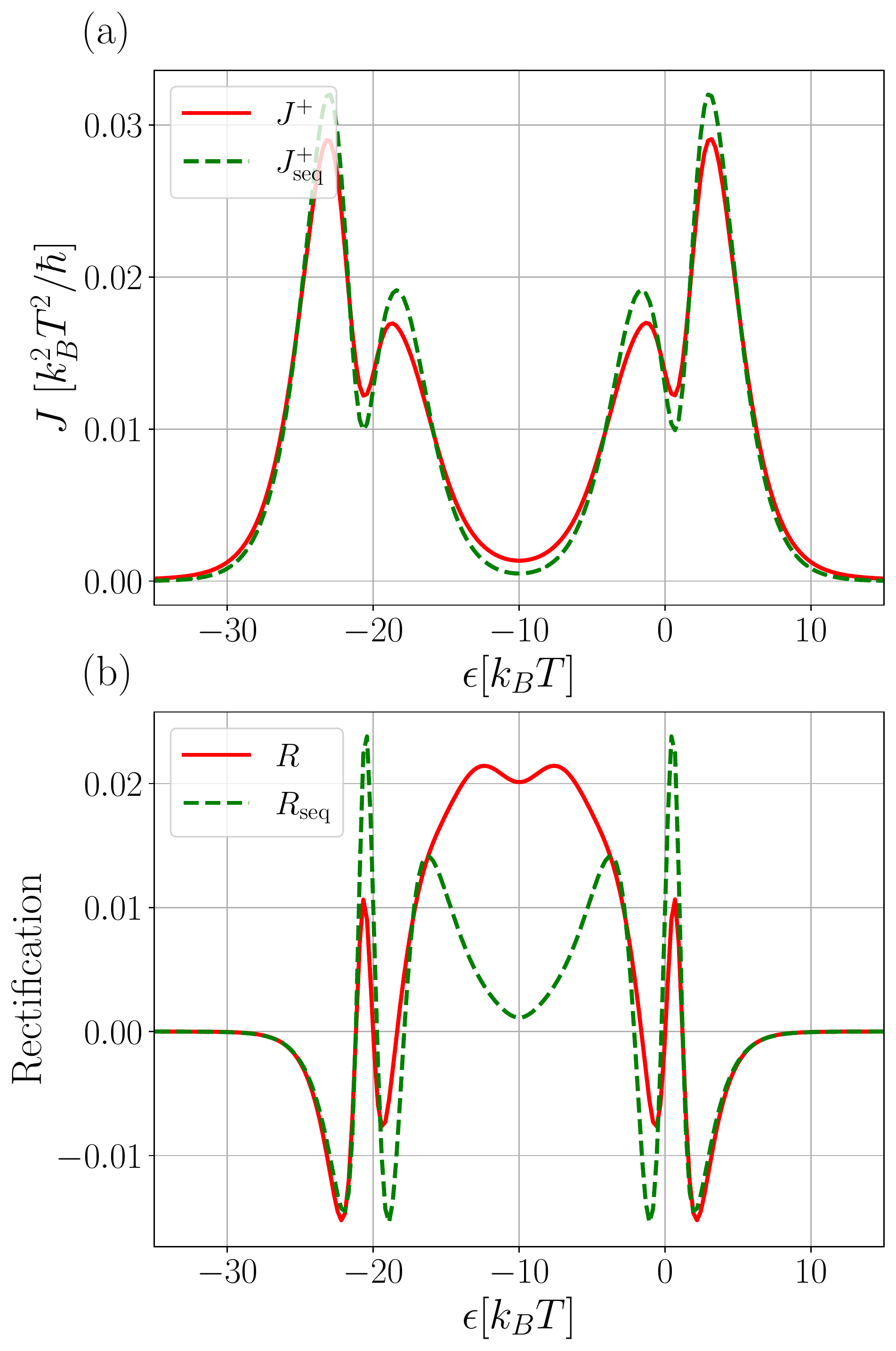}
	\caption{Closed-circuit setup. Heat currents (in units of $k_B^2T^2/\hbar$) and rectification coefficient for the case of two non-degenerate levels as functions of the average energy of the  levels $\epsilon$. Panel (a) shows the forward heat current and panel
	(b) shows the rectification $R$. Same color code as in Fig.~\ref{Cotunneling_c&r:deg:CCcurrents}.
	All parameters are the same used for Fig.~\ref{Sequential:1:grafici}, while $\Delta\epsilon=2k_BT$ and $E_C=20k_BT$.}
	\label{Cotunneling_c&r:QDnondeg:CCcotcurr}
\end{figure}
In Fig.~\ref{Cotunneling_c&r:QDnondeg:CCcotcurr}, the heat currents and rectification coefficient are plotted as functions of the average QD levels' energy $\epsilon$.
In panel (a), we plot the forward heat current which includes cotunneling contributions $J^+$ as a solid red curve and in the presence of sequential tunneling processes only $J^+_{\rm seq}$ as a dashed green curve.
The latter curve closely resembles the one plotted in Fig.~\ref{Sequential:2:grafici_CC}(a) for $\epsilon>-10k_BT$, meaning that double occupancy, at least for the value of $E_C$ considered, does not modify the results substantially.
The behavior of the heat current is similar to the degenerate case, see Fig.~\ref{Cotunneling_c&r:deg:CCcurrents}, with a global minimum at $\epsilon=-E_C/2=-10k_BT$, and two symmetric local minima at $\epsilon\approx0$ and $\epsilon\approx- E_C=-20k_BT$, which however do not touch zero, as in the degenerate case.
We emphasize that the cotunneling contributions $\Delta J^+$ and $-\Delta J^-$, unlike in the degenerate case, do not coincide (i.e.~ $\Delta J^+\ne -\Delta J^-$). The reason for this is that the inelastic cotunneling processes that occur when the QD is occupied by one electron transfer a finite amount of heat and are not antisymmetric under the exchange of the leads' temperatures.

In Fig.~\ref{Cotunneling_c&r:QDnondeg:CCcotcurr}(b) we plot the rectifications with (solid red curve, $R$) and without (dashed green curve, $R_{\rm seq}$) cotunneling contributions as functions of the average QD levels' energy $\epsilon$.
The curve $R_{\rm seq}$, for $\epsilon>-10k_BT$, closely resembles the rectification reported in Fig.~\ref{Sequential:2:grafici_CC}(b), calculated in the limit of infinite $E_C$.
Remarkably, the cotunneling contributions increase the rectification with respect to both $R_{\rm seq}$ and the rectification obtained in the degenerate case for values of $\epsilon$ between $-E_C=-20k_BT$ and 0.
Outside this range the rectification is mainly suppressed with respect to the sequential only result.
As already noticed in Sec.~\ref{degCot}, heat currents are symmetric around the axis $\epsilon=-E_C/2=-10k_BT$.

\begin{figure}[h!]
\centering
	\includegraphics[width=0.99\columnwidth]{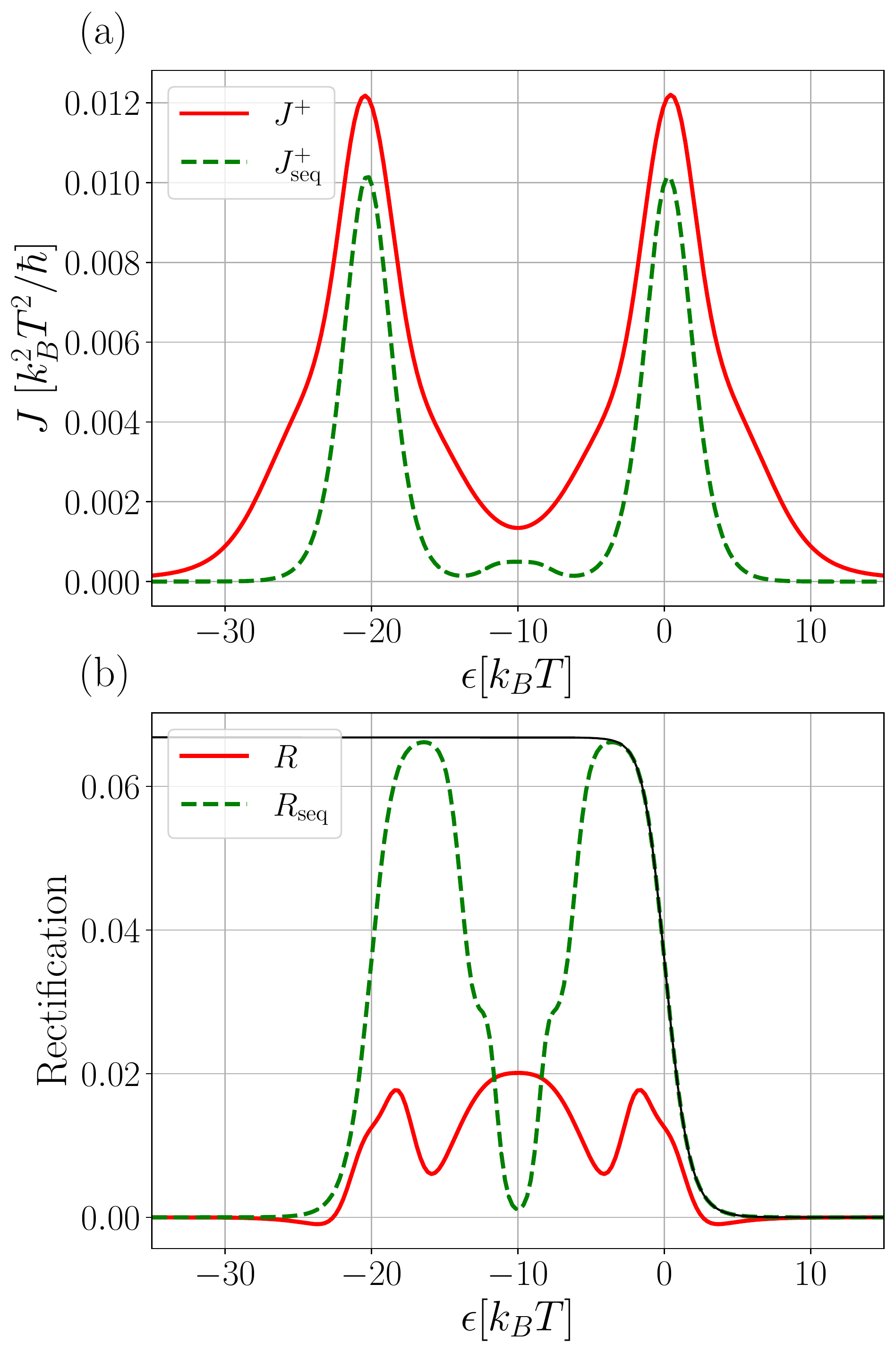}
	\caption{Open-circuit setup. Heat currents (in units of $k_B^2T^2/\hbar$) and rectification coefficient for the case of two non-degenerate levels as functions of the average energy of the  levels $\epsilon$. Panel (a) shows the forward heat current and panel
	(b) shows the rectification $R$.
	Same color code as in Fig.~\ref{Cotunneling_c&r:deg:CCcurrents}, while the thin black curve is the one relative to single occupation, taken from Fig.~\ref{Sequential:2:grafici_OC}(b).
	All parameters are the same used for Fig.~\ref{Sequential:1:grafici}, while $\Delta\epsilon=2k_BT$ and $E_C=20k_BT$.
	}
	\label{Cotunneling_c&r:QDnondeg:OCcotcurr}
\end{figure}

{\it Open-circuit setup.--}
In Fig.~\ref{Cotunneling_c&r:QDnondeg:OCcotcurr} we plot the heat currents and rectification as functions of the average QD levels' energy $\epsilon$.
In panel (a), we plot the forward heat current $J^+$ (solid red line) along with the one accounting only for sequential tunneling processes $J^+_{\rm seq}$ (dashed green line).
The latter presents a local maximum at $\epsilon=-E_C/2=-10k_BT$ of similar shape and height as in the degenerate case [see Fig.~\ref{Cotunneling_c&r:deg:OCcurrrect}(a)].
Notice that such a maximum does not appear in Fig.~\ref{Sequential:2:grafici_OC}(a), where double occupancy of the QD was not allowed.
$J^+_{\rm seq}$ peaks also at $\epsilon=0$, resembling the curve in Fig.~\ref{Sequential:2:grafici_OC}(a), and $\epsilon=-E_C=-20k_BT$: here heat transport is made possible by the energy difference $\Delta\epsilon$ between the levels (i.e.~charge and heat currents are not proportional to each other).
Interestingly, the red curve is always above the green curve, meaning that cotunneling contributions increase the heat current for all values of $\epsilon$.
In addition, the main peaks are widened, while at $\epsilon=-E_C/2=-10k_BT$ we have now a minimum.
Also in this case the heat currents are symmetric around the axis $\epsilon=-E_C/2=-10k_BT$, as noticed in Sec.~\ref{degCot}

In panel (b) of Fig.~\ref{Cotunneling_c&r:QDnondeg:OCcotcurr} we plot the rectification coefficients when cotunneling contributions are included ($R$, solid red curve) and when only sequential tunneling processes are allowed ($R_{\rm seq}$, dashed green curve).
Notice that $R_{\rm seq}$ virtually coincides with the rectification plotted in Fig.~\ref{Sequential:2:grafici_OC}(b) (here plotted as a thin black curve) only for $\epsilon>-3k_BT$, whereas the two curves largely depart for other values of $\epsilon$.
This means that the rectification is much more sensitive to the finiteness of $E_C$ than the current, making clear that in the range $-17k_BT<\epsilon<-3k_BT$ the limit of infinite $E_C$ does not apply (electrons have the energy to overcome the Coulomb repulsion, represented by the value of $E_C$, and the processes that involve $E_C$ can occur).
As a result, $R_{\rm seq}$ drops rapidly by decreasing $\epsilon$ below $-3k_BT$, presenting a minimum at $\epsilon=-E_C/2=-10k_BT$, thus leaving a maximum at $\epsilon\approx-4k_BT$.

As shown by the solid red curve, representing $R$, the cotunneling contributions affects very much the rectification by lowering it around the maxima of $R_{\rm seq}$ and increasing it around $\epsilon=-E_C/2=-10k_BT$.
In particular, $R$ reaches its maximum at $\epsilon=-E_C/2=-10k_BT$, presents two symmetric local maxima at $\epsilon\approx-2k_BT$ and at $\epsilon\approx-18k_BT$, and has two symmetric local minima at $\epsilon\approx-4k_BT$ and $\epsilon\approx-16k_BT$.

\section{Four-terminal device}
\label{secNL}
The set up   is shown in Fig.~\ref{setup}(c-d) and consists of a pair of Coulomb-coupled QDs, each attached to two leads.
On the left-hand-side, QD$_\uparrow$ is attached to $L1$, at temperature $T+\Delta T/2$, and to $L2$, at temperature $T-\Delta T/2$.
On the right-hand-side, QD$_\downarrow$ is attached to R1 and R2, both at temperature $T$.
All reservoirs are kept at the same chemical potential, which is set to zero without loss of generality.
There is natural flow of heat through QD$_\uparrow$ from the hot to the cold reservoir, depending on the sign of $\Delta T$.
QD$_\uparrow$ along with the two reservoirs attached to it constitute the drive circuit, while QD$_\downarrow$ and reservoirs $R1$ and $R2$ constitute the drag circuit~\cite{bhandari2018}.
The drag circuit is coupled to the drive circuit via the Coulomb interaction: there is no particle exchange between the two circuits.
The exchange of energy between the two circuits leads to a finite heat flow in the drag circuit~\cite{bhandari2018}. 
Non-local heat rectification, i.e.~rectification in the drag currents, occurs when the absolute value of the heat flowing between the drive and drag circuit depends on the sign of the temperature bias $\Delta T$ applied to the drive circuit.
We must stress here that the currents $J_{R1}$ and  $J_{R2}$ need not be equal, since there is an energy flow between drive and drag circuits.
The sketch in Fig.~\ref{setup}(c) represents the forward bias configuration, with $\Delta T>0$ and drag currents indicated by $J_{R1}^+$ and $J_{R2}^+$, while Fig.~\ref{setup}(d) represents the backward bias configuration, with $\Delta T<0$ and drag currents indicated by $J_{R1}^-$ and $J_{R2}^-$.
We fix the convention where heat currents are positive when they enter the QDs.

The state of the system is described by the following set of occupancy (see Sec.~\ref{sec:model}): $(n_{1\uparrow},n_{1\downarrow})=\{00,10,01,11\}$, where $n_{1\sigma}$ represents the number of electrons in QD$_\sigma$. 
Note that here, as in Sec.~\ref{secCot}, we allow for double occupation but we consider sequential tunneling only.
As in App.~\ref{ME} and Sec.~\ref{secCot}, we describe the state of the system by the probabilities $P_0$, $P_\uparrow$, $P_\downarrow$ and $P_2$, the latter referring to double occupancy.
The MEs which allow to determine such probabilities are formally equal to Eqs.~(\ref{meIN}), reported in App.~\ref{AppQDME}.
The heat currents (which coincides with the energy currents) relative to the drag circuit are given by
\begin{multline}
J_\beta=\epsilon_\downarrow\Gamma_\beta[f_\beta(\epsilon_\downarrow)P_0-f_\beta^-(\epsilon_\downarrow)P_\downarrow]
+(\epsilon_\downarrow+E_C)\times\\
\Gamma_\beta[f_\beta(\epsilon_\downarrow+E_C)P_\uparrow-f_\beta^-(\epsilon_\downarrow+E_C)P_2],
\end{multline}
where $\beta={R1, R2}$.

Before discussing the results, some general observations are in order.
When, in the drive circuit, the coupling to the hot reservoir is stronger with respect to the coupling to the cold reservoir, and setting $\epsilon_\uparrow=\epsilon_\downarrow$, we notice that both currents in the drag circuit ($J_{R1}$ and $J_{R2}$) are negative (entering the leads), irrespective of all other parameters.
In the opposite situation, where the coupling to the cold reservoir is stronger with respect to the coupling to the hot reservoir, the currents $J_{R1}$ and $J_{R2}$ are both positive (exiting the leads).
This is not the case, however, when $\epsilon_\uparrow\neq\epsilon_\downarrow$, where the sign of $J_{R1}$ and $J_{R2}$ can be different and depend on all the parameters~\cite{cooling}. 
In particular, the sign of $J_{R1}$ and $J_{R2}$ does not depend on the direction of the temperature bias (forward or backward).
However, when $\Gamma_{R1}= \Gamma_{R2}$, $J_{R1}$ and $J_{R2}$ are equal even when $\epsilon_\uparrow\neq\epsilon_\downarrow$ and regardless of the values of $\Gamma_{L1}$ and $\Gamma_{L2}$.

Remarkably, non-local rectification takes place only when the couplings in the drive circuit are asymmetric, i.e.~when $\Gamma_{L1}\ne \Gamma_{L2}$.
In what follows, for simplicity, we fix $\Gamma_{R1}= \Gamma_{R2}=0.05k_BT$, and we identify $J^\pm\equiv J^\pm_{R1}=J^\pm_{R2}$, with the rectification coefficient defined as in Eq.~(\ref{erre}).

\begin{figure}[!thb]
	\centering	\includegraphics[width=0.98 \columnwidth]{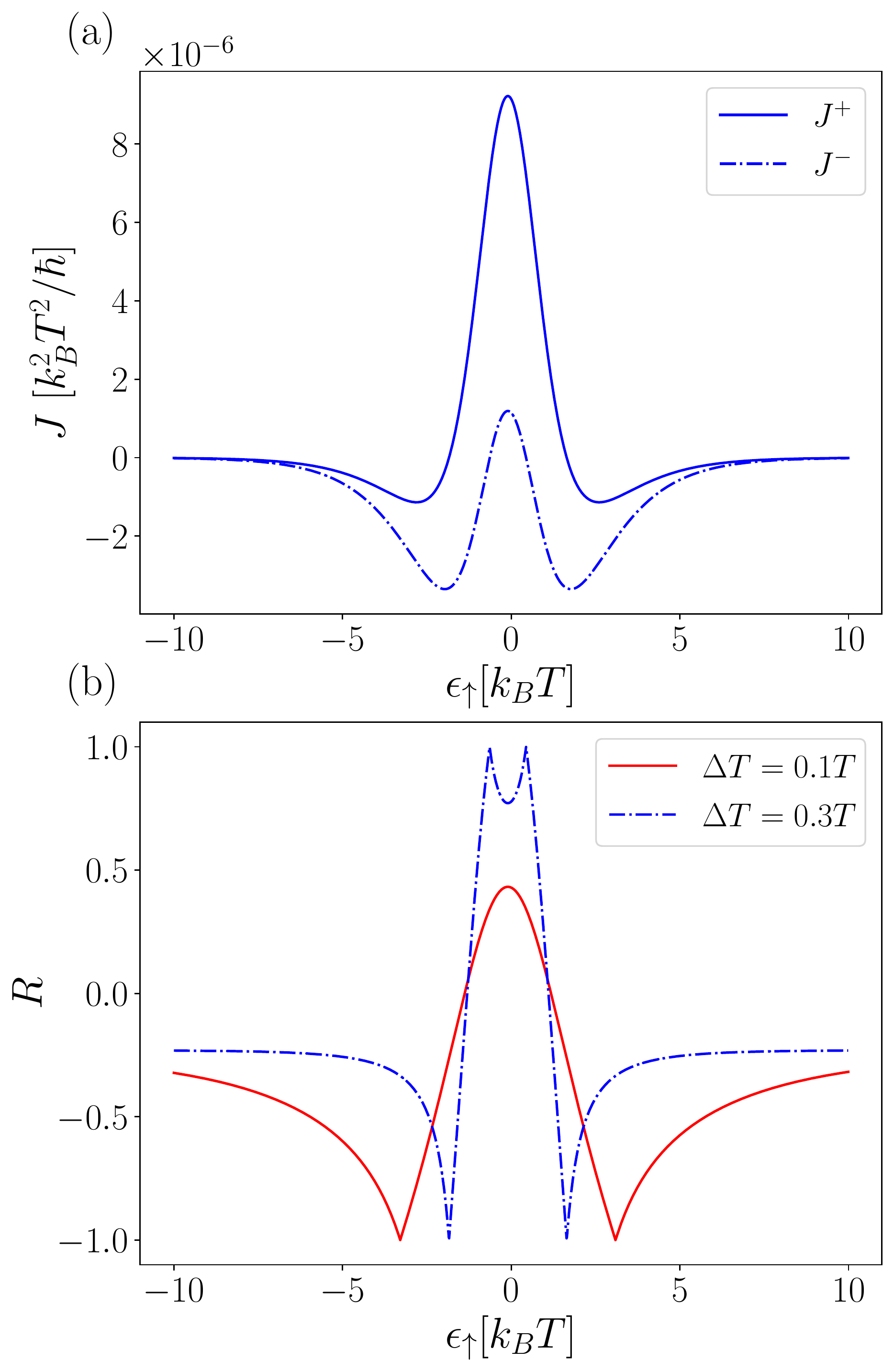}
	\caption{Non-local heat current (a) and non-local rectification coefficient (b) as a function of $\epsilon_\uparrow$. The parameters are: $\epsilon_\downarrow = k_BT$, $\Gamma_{R1}=\Gamma_{R2}=\Gamma_{L1}=0.05 k_BT/\hbar$, $\Gamma_{L2}=0.08 k_BT/\hbar$ and $E_C = 0.2 k_BT$. In panel (a) the two curves are relative to $\Delta T=0.3 T$, while in panel (b) the red solid curve is relative to $\Delta T=0.1 T$ and the blue dashed-dotted curve is relative to $\Delta T=0.3 T$.}
	\label{figrecEnIn}
\end{figure}

In Fig.~\ref{figrecEnIn} we plot the heat current $J$ [panel (a)] and the rectification coefficient [panel (b)] as a function of the energy level of the QD in the drive circuit $\epsilon_\uparrow$ in the presence of an asymmetry in the couplings in the drive circuit. For panel (a) the blue solid curve refers to the forward bias and the blue dashed-dotted curve refers to the backward bias, calculated at $\Delta T = \pm 0.3T$, respectively.
Both curves present a (positive) maximum around $\epsilon_\uparrow=0$ (the energy level is aligned with the chemical potential of the leads), in agreement with the results of Ref.~\onlinecite{bhandari2018}.
Both currents vanish for large values of $\epsilon_\uparrow$, since in this situation transport cannot occur even in the drive circuit, but for intermediate values of $\epsilon_\uparrow$ they show negative minima.
Despite the relatively small difference between the coupling strengths in the drive circuit ($\Gamma_{L1}=0.05 k_BT$ and $\Gamma_{L2}=0.08 k_BT$), the two curves depart significantly.
This is quantified by the non-local rectification coefficient plotted in Fig.~\ref{figrecEnIn}(b), where we consider two values of temperature bias: $\Delta T=0.1 T$ (red solid curve) and $\Delta T=0.3 T$ (blue dashed-dotted curve).
We first notice that $R$, as a function of $\epsilon_\uparrow$, spans the whole range of values ($[-1,1]$) for $\Delta T=0.3 T$, while taking values in the range $[-1,0.4]$ for $\Delta T=0.1 T$.
In particular, $R=1$ is reached when $J^-$ crosses zero [see panel (a)]), while $R=-1$ is attained for the value of $\epsilon_\uparrow$ for which  $J^+=0$.
Overall, rectification is large in quite large ranges of values of $\epsilon_\uparrow$.
$J_{R1}$ and $J_{R2}$, however, are rather suppressed if compared with the heat currents relative to the setup with a single QD [Fig.~\ref{setup}(a-b)].
Indeed, the heat current flowing in the drive circuit, in the case where drive and drag circuits are decoupled ($E_C=0$), turns out to be 3 orders of magnitude larger than in Fig.~\ref{figrecEnIn}(a).

A richer behavior occurs if one now assumes that the effective tunneling amplitudes $t_\alpha$ [see Eq.~(\ref{ht2})] are energy-dependent, thus depending on the charge state of the QDs.
This situation was actually experimentally observed in Ref.~\onlinecite{keller2016}, where the tunneling probabilities between the QD and the electron reservoirs in the drag circuit were found to depend on the charge state of the QD in the drive circuit. 
We can account for this situation by replacing the definition of the tunneling constants in the drag circuit [see Eq.~(\ref{tunc})] with
\begin{align}
&\Gamma_{R1/R2}^{(0)}= \frac{2\pi}{\hbar} D_{R1/R2}\; |t_{R1/R2}^{(0)}|^2=\kappa_{R1/R2}^{(0)}\Gamma_{R1/R2}\nonumber \\
&\Gamma_{R1/R2}^{(1)}= \frac{2\pi}{\hbar} D_{R1/R2}\; |t_{R1/R2}^{(1)}|^2=\kappa_{R1/R2}^{(1)}\Gamma_{R1/R2},
\end{align}
where the superscript (0) and (1) refer to the charge state (empty and occupied) of the QD in the drive circuit QD$_\uparrow$.
The second equalities define the charge state-dependent coefficients $\kappa_{R1/R2}^{(0/1)}$.

\begin{figure}[!htb]
	\centering	\includegraphics[width=0.98 \columnwidth]{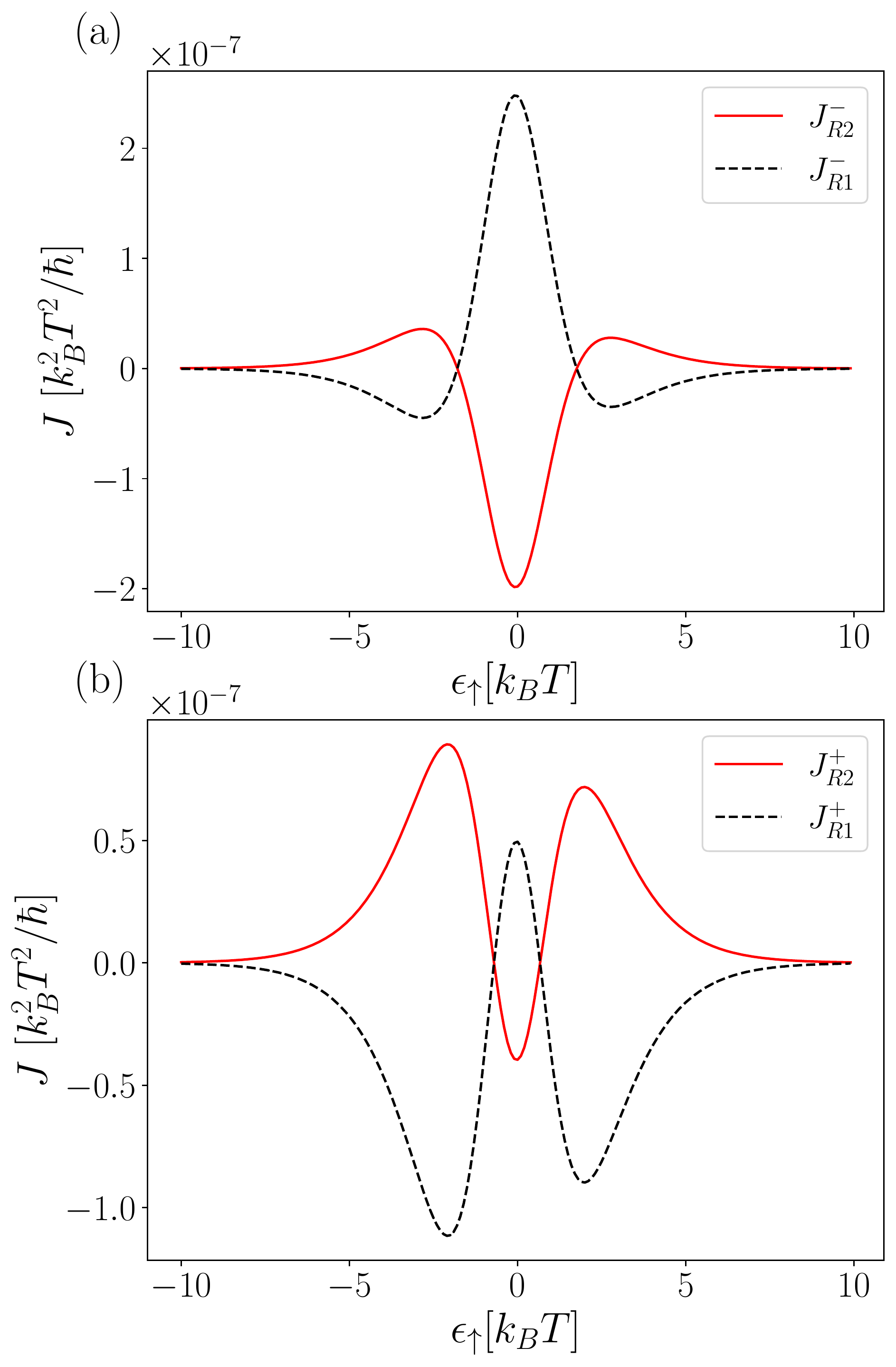}
	\caption{Energy-dependent couplings: non-local heat currents relative to the drag circuit as a function of $\epsilon_\uparrow$ for $\Delta T=0.1 T$. Panel (a) is relative to backward bias, while panel (b) to forward bias. Black dashed curves is for $J_{R1}$ and red curve is for $J_{R2}$. The other parameters are: $\epsilon_\downarrow=0.4 k_{B}T$, $\Gamma_{L1}=0.08k_BT/\hbar$, $\Gamma_{L2}=0.07k_{B}T/\hbar$, $\Gamma_{R1}=0.06k_{B}T/\hbar$, $\Gamma_{R2}=0.05k_BT/\hbar$, $E_C=0.0075 k_{B}T$, $\kappa_{L2}^{(0/1)}=\kappa_{R2}^{(0)}=1$, and $\kappa_{R2}^{(1)}=1/5$.}
	\label{fig:rectcs}
\end{figure}

In Fig.~\ref{fig:rectcs}(a) and (b) we plot the heat currents in the drag circuit as functions of the energy level of the QD in the drive circuit $\epsilon_\uparrow$ [backward  bias in panel (a) and forward bias in panel (b)].
Notice that in this case $J_{R1}$ and $J_{R2}$ are different since $\Gamma_{R1}\neq \Gamma_{R2}$.
Perfect heat rectification ($R=1$) occurs also in this case for both currents $J_{R1}$ and $J_{R2}$, and for the same values of $\epsilon_\uparrow$ (corresponding to points where the currents vanish).
What is remarkable in Fig.~\ref{fig:rectcs} is that
$J_{R1}$ and $J_{R2}$ have similar amplitude but opposite signs, both in the forward and backward temperature bias.
This means that if heat is extracted from lead $R1$, a similar amount of heat is deposited into lead $R2$ (or the other way around).
In particular, the extraction of heat from reservoir $R1$ can be used to lower its temperature, thus realizing an absorption refrigerator of the kind studied in Ref.~\onlinecite{Erdman2018}, where cooling is driven by a non-local temperature difference, with no work provided to the system.

We conclude the section by noting that we have checked that perfect rectification is obtained even when cotunneling processes are taken into account.

\section{Conclusions}
\label{secConc}
In this paper we have studied heat rectification in two different quantum dot (QD) systems: i) a single QD in a two-terminal device and, ii) a pair of coupled QDs in a four-terminal device.
Heat rectification occurs when a QD is coupled asymmetrically to two terminals.
Heat currents have been calculated using the master equation approach, up to the second order (cotunneling corrections) for a single QD.
In case i) we have considered a QD with either a doubly-degenerate level or two non-degenerate levels, each attached to two reservoirs.
Furthermore, we have assumed that the device is either in the open-circuit setup, where the charge current vanishes, or in the closed-circuit setup, where the two reservoirs are kept at equal chemical potentials. In both cases energy current and heat current coincide.
Within the sequential tunneling regime, we have first considered the case where the charging energy is very large such that the QD can only be occupied by a single electron.
In this situation charge and heat current are proportional, so that no heat current can flow in the open-circuit setup in the case of a doubly-degenerate level.
In the closed-circuit setup, however, the heat current is finite and, remarkably, we could derive an upper bound to the rectification coefficient $R$ which only depends on the tunneling constants to the left and to the right as
\begin{equation*}
|R| \leq \frac{1}{5}\left| \frac{\Gamma_L-\Gamma_R}{\Gamma_L+\Gamma_R} \right| .
\end{equation*}
On the contrary, we have found that no bound exists in the case of two non-degenerate levels and we have identified the parameters' set which allow $R$ to reach 1, its maximum value.

Very interesting results are related to the effect of cotunneling contributions, including both elastic and inelastic processes, in the two-terminal configuration.
In general, we have found that there exists ranges of values of the levels' position where rectification is enhanced by cotunneling.
Moreover, we have found that
\begin{itemize}
\item in the open-circuit setup of the degenerate level case, cotunneling processes, while permitting a finite heat flow, yield a finite (though small) rectification;
\item in the closed-circuit setup of the degenerate case, cotunneling corrections to the forward heat current are opposite to the corrections to the backward heat current, so that the magnitude of both currents are either increased or decreased by cotunneling. Rectification enhancement occurs when cotunneling lowers such magnitude (see also Ref.~\onlinecite{bhandari2021});
\item in the open-circuit setup of the non-degenerate case, cotunneling always increases the currents with respect to the sequential regime.
\end{itemize}

In the case ii), we have considered the case of two Coulomb-coupled QDs, each connected to two terminals.
This is a non-local configuration, where the temperature bias is applied to two terminals (the drive circuit), while the heat currents of interest are calculated in the other two terminals (the drag circuit).
In this situation we have found, remarkably, that the rectification coefficient can reach the ideal value  (i.e.~$R=1$), although with a rather small absolute value of the the currents.
Moreover, we have considered the experimentally relevant case~\cite{keller2016} where the tunnel couplings between QDs and leads depend on energy.
We have found that the heat currents in the drag circuit have similar amplitude but opposite signs, meaning that if heat is extracted from one reservoir, a similar amount of heat is deposited into the other.

\section{Acknowledgements}
We acknowledge support from the SNS-WIS joint lab QUANTRA, and E.P. acknowledges support by the University of Catania, Piano di Incentivi per la Ricerca di Ateneo 2020/2022, proposal Q-ICT and by the CNR-QuantERA grant SiUCs.
B.B. was supported by the U.S. Department of Energy (DOE), Office of Science, Basic Energy Sciences (BES), under Award No. DESC0017890, and L.T. was supported by the Knut and Alice Wallenberg Foundation.

\begin{appendix}

\section{Master equation in the sequential tunneling regime}
\label{ME}
Here we assume that the charging energy $E_C$ is the largest energy scale, so that we can neglect all electronic configurations in which the total number of electrons in the QD exceeds one.
We can describe the state of the QD by specifying the probability of finding the QD in the state with zero electrons $P_0$, with one electron in a level with spin up ($P_\uparrow$), and with one electron in a level with spin down ($P_\downarrow$).
The master equations can be written in matrix form as
\begin{equation}\label{Sequential:mastereq3}
		\frac{d}{dt}\left(\begin{array}{c} P_0	\\	P_\uparrow	\\	P_\downarrow\end{array}\right)
		=
		\left(
		\begin{array}{ccc}
		-\Sigma^+_{\uparrow}-\Sigma^+_{\downarrow}	&	\Sigma^-_{\uparrow}	&	\Sigma^-_{\downarrow}	\\
		\Sigma^+_{\uparrow}	&	-\Sigma^-_{\uparrow}	&	0	\\
		\Sigma^+_{\downarrow}	&	0	&	-\Sigma^-_{\downarrow}	\\
		\end{array}
		\right)
		\left(\begin{array}{c} P_0	\\	P_\uparrow	\\	P_\downarrow\end{array}\right)
\end{equation}
where
\begin{equation}\label{Sequential:Sigma}
		\Sigma^+_\sigma=\Gamma_Lf_{L\sigma}+\Gamma_Rf_{R\sigma},\quad \Sigma^-_\sigma=\Gamma_Lf^-_{L\sigma}+\Gamma_Rf^-_{R\sigma}.
\end{equation}
Here we have defined
\begin{equation}
	f_{\alpha\sigma}=f_\alpha(\epsilon_\sigma)
\end{equation}
as the Fermi distribution function [$f_\alpha(E)$] of lead $\alpha=L,R$ evaluated at the QD levels' energies $\epsilon_\sigma$, while
\begin{equation}
	f_{\alpha\sigma}^-=1-f_\alpha(\epsilon_\sigma),
\end{equation}
where
\begin{equation*}
	f_\alpha(E)=\left[ 1+\exp(\frac{E-\mu_\alpha}{k_BT_\alpha}) \right]^{-1} .
\end{equation*}

The stationary master equation is obtained by equating the time derivative of $P$ to zero and imposing the normalization of the probabilities $P_0 + P_\uparrow + P_\downarrow = 1$.
We obtain the following stationary solutions
\begin{equation}\label{solME}
\begin{split}
P_0&=\frac{\Sigma_\uparrow^- \Sigma_\downarrow^-}{\Lambda}\\
P_\uparrow&=\frac{\Sigma_\uparrow^+ \Sigma_\downarrow^-}{\Lambda}\\
P_\downarrow&=\frac{\Sigma_\uparrow^- \Sigma_\downarrow^+}{\Lambda}
\end{split}
\end{equation}
with
\begin{equation}
\Lambda=\Sigma_\uparrow^- \Sigma_\downarrow^- + \Sigma_\uparrow^+ \Sigma_\downarrow^- + \Sigma_\uparrow^- \Sigma_\downarrow^+ .
\end{equation}
Note that $\Lambda$ changes non trivially under the exchange of the leads' temperatures. Such a behaviour results in a change in the QD state distribution which leads to rectification.

We have also calculated the master equation accounting for up to two electrons in the QD.
When $E_C$ is two order of magnitude larger than $k_BT$ we have proven that the probability of occupation of the states with two electrons is negligible.

The charge current entering the QD from the left lead can be written as ($-e$ is the electronic charge)
\begin{equation}\label{Sequential:J^cP}
		\begin{split}
		I^c=-e P_0&\left[ \Gamma_Lf_{L\uparrow}+\Gamma_Lf_{L\downarrow}\right]+
		\\
		&+eP_\uparrow\Gamma_Lf_{L\uparrow}^-+eP_\downarrow\Gamma_Lf_{L\downarrow}^-,
		\end{split}
\end{equation}

and becomes
\begin{equation}\label{Sequential:J^c}
	I^c=-e \frac{\Gamma_L\Gamma_R}{\Lambda} \left[\Sigma^-_\downarrow
	\left(f_{L\uparrow}-f_{R\uparrow}\right)+\Sigma^-_\uparrow\left(f_{L\downarrow}-f_{R\downarrow}\right) \right] 
\end{equation}
after substituting the solutions of the master equation, Eq.~\eqref{solME}.
Similarly, the energy current takes the form
\begin{equation}\label{Sequential:J^uP}
		\begin{split}
		I= P_0&\left[\epsilon_\uparrow\Gamma_Lf_{L\uparrow}+\epsilon_\downarrow\Gamma_Lf_{L\downarrow}\right]
		\\
		&-P_\uparrow\Gamma_L\epsilon_\uparrow f_{L\uparrow}^--P_\downarrow\Gamma_L\epsilon_\downarrow f_{L\downarrow}^-, 		
		\end{split}
\end{equation}
and becomes
\begin{equation}\label{Sequential:J^u}
I=\frac{\Gamma_L\Gamma_R}{\Lambda} \left[\Sigma^-_\downarrow\epsilon_\uparrow\left(f_{L\uparrow}-f_{R\uparrow}\right)+\Sigma^-_\uparrow\epsilon_\downarrow\left(f_{L\downarrow}-f_{R\downarrow}\right) \right].
\end{equation}
Finally, from the definition of heat current
\begin{equation}
J=I-\frac{\mu_L}{(-e)} I^c ,
\end{equation}
we have that heat and energy current coincide ($J=I$) in both open- and closed-circuit setups, since $I^c=0$ in the first case, and $\mu_L=0$ in the second~\cite{nota1}.

\section{Cotunneling contributions for a single QD with two states}
\label{appQD}
We consider the QD in the weak coupling condition, so that we can treat the tunnel Hamiltonian $H_{\rm T}$ in Eq.~(\ref{ht}) as a perturbation to the system.
Therefore, we describe the system through the eigenstates of the free Hamiltonian $H_0=H_{\rm QD}+H_{\rm L}+H_{\rm R}$ and calculate the transition rates between two of such states using the generalized Fermi golden rule
\begin{equation}\label{ThMod:FermiGRule}
	\gamma_{if}=\frac{2\pi}{\hbar}|\mathcal{A}_{if}|^2\delta (E_f-E_i),
\end{equation}
where $\gamma_{if}$ is the rate associated with the process that starts from the initial state $| i\rangle$ with energy $E_i$, and arrives in the final state $|f\rangle$ with energy $E_f$.
Since the perturbation $H_{\rm T}$ is time-independent, the delta function in equation \eqref{ThMod:FermiGRule} imposes the energy conservation between the initial and final states $E_i=E_f$.
Moreover, the amplitude $\mathcal{A}_{if}$ contains the perturbation term $H_{\rm T}$
\begin{equation}\label{ThMod:FermiAif}
	\mathcal{A}_{if}=\langle i|H_{\rm T}+H_{\rm T}\frac{1}{E_i-H_0}H_{\rm T}+\dots |f\rangle,
\end{equation}
which gives rise to a natural expansion in the powers of $H_{\rm T}$.
The first order describes the sequential tunneling processes, while the second order describes the cotunneling processes~\cite{nazarov2009,schon1997}.
In this appendix we calculate the transition amplitudes of the latter processes.

In a cotunneling process, the system evolves from an initial state to a final one passing through a virtual state.
Since there can be more than one virtual state, the cotunneling processes can exhibit quantum interference.
The transition amplitude $\mathcal{A}_{if}$ that enters the generalized Fermi golden rule can be written as
\begin{equation}\label{Cotunneling_intro:Aif_0}
	\mathcal{A}_{if}=\sum_{\nu}\frac{\langle f|H_{\rm T}|\nu\rangle\langle\nu|H_{\rm T}|i\rangle}{E_i-E_\nu+i\eta},
\end{equation}
where the sum is made over all the virtual states of the system, and the parameter $\eta$ goes to zero and is needed to eliminate the divergences in the calculation of the rates~\cite{turek2002}, see Appendix \ref{AppIntegrals}.
Such divergences are due to the sequential tunneling processes.
Indeed, the system can evolve from the initial state to the final state also through two consecutive sequential tunneling events.
For example, an electron of the left lead can tunnel sequentially into the QD and, from there, it can tunnel sequentially into the right lead.
Such a process transfers an electron from the left to the right one, but it is made of two sequential tunneling events.
When integrating over all possible initial and final states, both the transition rates that are given by the second-order Fermi golden rule and the processes made of two consecutive sequential tunneling events contribute to the integral.
However, since we have already accounted for the sequential tunneling, we have to remove the contributions of the sequential tunneling events from the transition rates and the currents of the cotunneling removing the divergences of the integrals.

For every pair of initial and final states, we have to calculate the transition rate of each given process and the currents associated with it by multiplying the Fermi golden rule rate by the Fermi distributions relevant to the tunneling process. The Fermi distributions are necessary to describe the probability of having the starting electrons in the initial state and the final electron levels empty so that they can be occupied by the incoming electrons. Then, we obtain the total rates by summing over all possible initial and final states.

We separate the cotunneling processes in two kinds: the elastic processes, in which the energy of the QD does not change between the initial and final state, and the inelastic processes, which modify the QD energy.
Thus, the inelastic processes modify the state of the QD and enter the master equation.
Whereas, the elastic processes do not.
However, both elastic and inelastic processes contribute to the transport of charge and heat.

\subsection{Cotunnelling rates: elastic processes}
\label{AppCotEl}
In this section we derive the cotunneling rates and the currents for the various elastic processes.
The elastic processes do not change the energy of the QD, therefore the initial and final states of the QD must be the same.
Since the QD can be occupied by either zero, one, or two electrons, we separate the cotunneling contributions according to the QD state.
For each QD state, we find the possible initial and final states and the corresponding transition rate.
Of course, we do not consider the processes in which the initial and final states are in the same lead because such processes do not contribute to the transport of heat nor charge.
In general, the cotunneling rate for an electron to go from lead $\alpha$ to lead $\beta$, while the QD is initially in the state ``in'', can be calculated as
\begin{equation}
	\label{cotunel}
	T^{\rm (in)}_{\alpha\rightarrow\beta}=\frac{2\pi}{\hbar}\int D_{\alpha}(E) D_{\beta}(E)f_\alpha(E)f_\beta^-(E) |\mathcal{A}_{if}|^2 dE,
\end{equation}
where $D_{\alpha}$ is the density of states of lead $\alpha$ and the Fermi distribution $f_\alpha$ describes the probability of finding an occupied electronic state in the lead $\alpha$, while $f_\beta^-=1-f_\beta$ is the probability of finding an unoccupied state in the lead $\beta$.
Of course, the cotunneling rate for the opposite process, i.e.~for an electron to go from lead $\beta$ to lead $\alpha$, is obtained by exchanging the leads indices in Eq.~(\ref{cotunel}).

On the other hand, to calculate a net current we have to account for both the $L\rightarrow R$ and the $R\rightarrow L$ processes.
The net {\it single-process} charge current, from left to right, can thus be written as
\begin{equation}
	\mathcal{J}^{{\rm in},c}_\sigma=e\frac{2\pi}{\hbar}\int D_{L}(E) D_{R}(E)[f_{L}(E)- f_{R}(E)] |\mathcal{A}_{if}|^2 dE,
\end{equation}
while the net {\it single-process} energy current, from left to right, can be written as
\begin{equation}
	\mathcal{J}^{{\rm in},u}_\sigma=\frac{2\pi}{\hbar}\int D_{L}(E) D_{R}(E)[f_{L}(E)- f_{R}(E)] |\mathcal{A}_{if}|^2 EdE,
\end{equation}
where $e$ is the electron charge and ``in'' refers to the state of the QD (with spin $\sigma$).

In the following we list the expressions for the transition rates and the currents depending on the initial state of the QD.
We will use the superscript $ij$ (with $i,j=0,1$) to indicate that the QD is initially in the configuration (in)=$(i,j)$, i.e.~there are $i$ electrons in the level $\uparrow$ and $j$ electrons in the level $\downarrow$.
\begin{figure}[!htb]
	\centering	\includegraphics[width=0.9 \columnwidth]{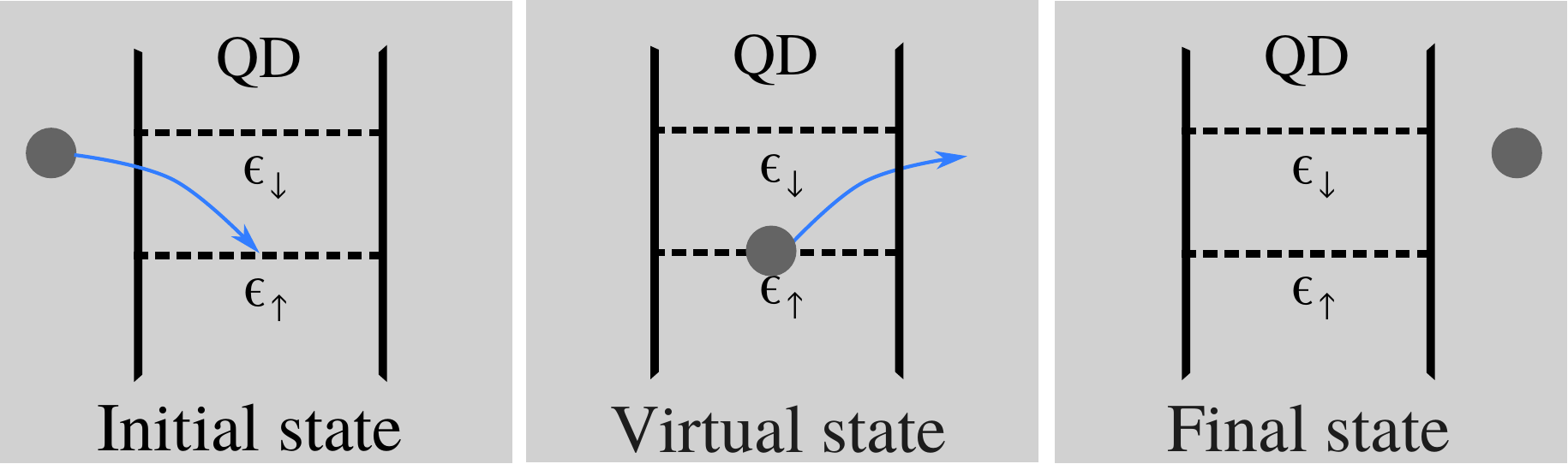}
	\caption{Diagram of the elastic cotunneling process $L\rightarrow R$ with incoming electron with spin $\uparrow$ and QD initially in the configuration $(0,0)$.}
	\label{fig:ela00}
\end{figure}
Namely, when the QD is initially empty and in its intermediate state the level $\epsilon_\sigma$ is occupied, see for example Fig.~\ref{fig:ela00}, we find
\begin{equation}\label{Cotunneling_ela:0T}
	T^{00}_{L\rightarrow R, \sigma}=\frac{\hbar}{2\pi}\Gamma_{L}\Gamma_{R}\int \frac{f_L(E)f_R^-(E)}{|E-\epsilon_\sigma+i\eta|^2} dE,
\end{equation}
and
\begin{equation}\label{Cotunneling_ela:0curr}
	\mathcal{J}^{00,c/u}_{\sigma}=\frac{\hbar}{2\pi}\Gamma_L\Gamma_R\int \frac{f_L(E)-f_R(E)}{|E-\epsilon_\sigma+i\eta|^2} KdE,
\end{equation}
where $K = -e$ for the charge current (superscript $c$) and $K=E$ for the energy current (superscript $u$).
When the QD is initially fully occupied and in the intermediate state the level $\epsilon_\sigma$ is empty, we find
\begin{equation}\label{Cotunneling_ela:2T}
	T^{11}_{L\rightarrow R, \sigma}=\frac{\hbar}{2\pi}\Gamma_L\Gamma_R\int  \frac{f_L(E)f_R^-(E)}{|\epsilon_\sigma+E_C-E+i\eta|^2} dE,
\end{equation}
and
\begin{equation}\label{Cotunneling_ela:2curr}
	\mathcal{J}^{11,c/u}_{\sigma}=\frac{\hbar}{2\pi}\Gamma_L\Gamma_R\int \frac{f_L(E)-f_R(E)}{|\epsilon_\sigma+E_C-E+i\eta|^2} KdE.
\end{equation}
When the QD is initially occupied by one electron with spin up and in its intermediate state the level $\epsilon_\sigma$ is empty we find
\begin{equation}\label{Cotunneling_ela:10T}
\begin{split}
&T^{10}_{L\rightarrow R, \sigma}=\frac{\hbar}{2\pi}\Gamma_L\Gamma_R\int f_L(E)f_R^-(E)
\\
&\times\left(\frac{\delta_{\sigma\uparrow}}{|\epsilon_\sigma-E+i\eta|^2}+\frac{\delta_{\sigma\downarrow}}{|E-\epsilon_\sigma-E_C+i\eta|^2}\right) dE,
\end{split}
\end{equation}
and
\begin{equation}\label{Cotunneling_ela:10curr}
\begin{split}
	&\mathcal{J}^{10,c/u}_{\sigma}=\frac{\hbar}{2\pi}\Gamma_L\Gamma_R\int [f_L(E)-f_R(E)]
	\\
	&\times\left(\frac{\delta_{\sigma\uparrow}}{|\epsilon_\sigma-E+i\eta|^2}+\frac{\delta_{\sigma\downarrow}}{|E-\epsilon_\sigma-E_C+i\eta|^2}\right) KdE.
\end{split}
\end{equation}
\begin{figure}[!htb]
	\centering	\includegraphics[width=0.9 \columnwidth]{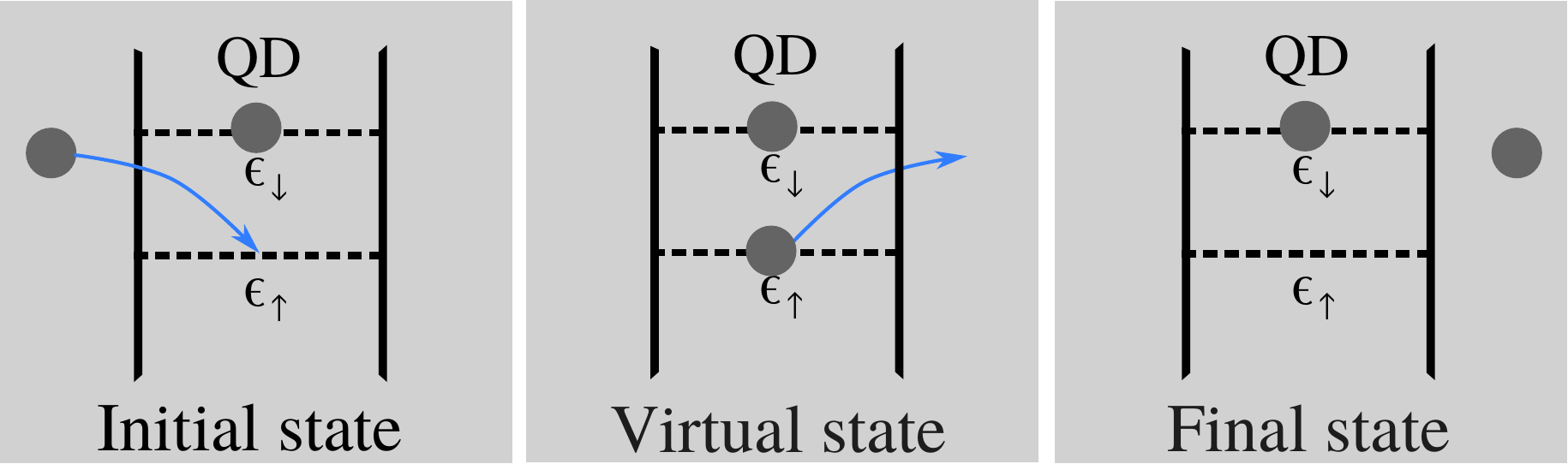}
	\caption{Diagram of the elastic cotunneling process $L\rightarrow R$ with incoming electron with spin $\uparrow$ and QD initially in the configuration $(0,1)$.}
	\label{fig:ela01}
\end{figure}
Finally, when the QD is initially occupied by one electron with spin down and in its intermediate state the level $\epsilon_\sigma$ is empty, see for example Fig~\ref{fig:ela01}, we find
\begin{equation}\label{Cotunneling_ela:01T}
\begin{split}
		&T^{01}_{L\rightarrow R, \sigma}=\frac{\hbar}{2\pi}\Gamma_L\Gamma_R\int f_L(E)f_R^-(E)
		\\
		&\times\left(\frac{\delta_{\sigma\downarrow}}{|\epsilon_\sigma-E+i\eta|^2}+\frac{\delta_{\sigma\uparrow}}{|E-\epsilon_\sigma-E_C+i\eta|^2}\right) dE,
\end{split}
\end{equation}
and
\begin{equation}\label{Cotunneling_ela:01curr}
\begin{split}
	&\mathcal{J}^{01,c/u}_{\sigma}=\frac{\hbar}{2\pi}\Gamma_L\Gamma_R\int [f_L(E)-f_R(E)]
	\\
	&\times\left(\frac{\delta_{\sigma\downarrow}}{|\epsilon_\sigma-E+i\eta|^2}+\frac{\delta_{\sigma\uparrow}}{|E-\epsilon_\sigma-E_C+i\eta|^2}\right) KdE.
\end{split}
\end{equation}

After removing the divergent part of the integrals of Eqs.~\eqref{Cotunneling_ela:0T}, \eqref{Cotunneling_ela:2T}, \eqref{Cotunneling_ela:10T} and \eqref{Cotunneling_ela:01T}, see Appendix \ref{AppIntegrals} for the details, the transition rates of the cotunneling processes can be negative.
Despite this, the total transition rate, which also accounts for two consecutive sequential tunneling events, is always positive as it must be.

\subsection{Cotunnelling rates: inelastic processes}
\label{AppCotInel}
In this section we derive the cotunneling rates and the currents for the various inelastic processes.
In an inelastic process the state of the QD gets modified, so that the energy of the QD changes.
This can take place in two different ways only, either by adding/removing two electrons to the QD, or by removing an electron from one level and adding one in the other level.
In particular, the change of states are
\begin{itemize}
	\item $(0, 0)\longrightarrow (1,1)$, the QD is initially empty and, through the inelastic process, becomes fully occupied;
	\item $(1, 1)\longrightarrow (0,0)$, the QD is initially fully occupied and, through the inelastic process, becomes empty;
	\item $(1, 0)\longleftrightarrow (0,1)$, in both initial and final state the QD has one electron inside, but the inelastic process changes the occupied level.
\end{itemize}
Analogously to App.~\ref{AppCotEl}, we can organize the different processes on the basis of the initial state of the QD.

\begin{figure}[!htb]
	\centering	\includegraphics[width=0.9 \columnwidth]{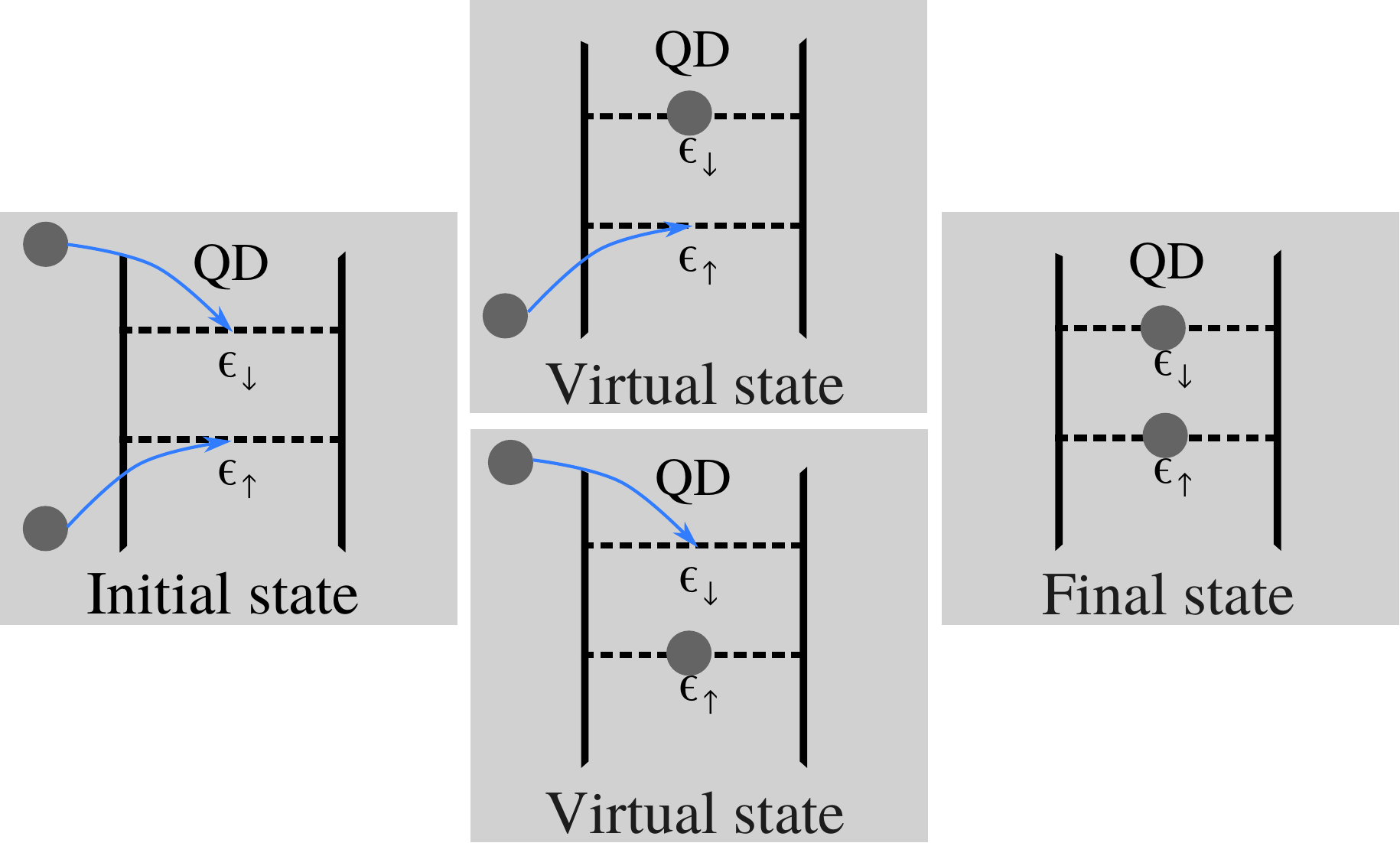}
	\caption{Diagram of the inelastic cotunneling process $(0,0)\rightarrow(1,1)$ with both electrons coming from the left lead.}
	\label{fig:ine00}
\end{figure}
When the QD is initially empty, the only possible final state is the one in which there are two electrons in the QD, since cotunneling comprises two tunneling processes.
The two electrons can both come from the left lead (i), or from the right lead (ii), or one from the left and one from the right lead (iii).
In case (i), there are two possible intermediate states, depending on which level is occupied first in the QD, as shown in Fig.~\ref{fig:ine00}.
Imposing the energy conservation between the initial and the final states, we obtain the following cotunneling rate
\begin{equation}\label{Cotunneling_ine:0LLT}
\begin{split}
	&T^{00\rightarrow11}_{LL}=\frac{\hbar}{2\pi}\Gamma_L^2\int f_L(E)f_L(\epsilon_\uparrow+\epsilon_\downarrow+E_C-E)
	\\
	&\times\left|\frac{1}{E-\epsilon_\uparrow+i\eta}+\frac{1}{\epsilon_\uparrow+E_C-E+i\eta}\right|^2 dE.
\end{split}
\end{equation}
Since the electrons involved in the process carry both charge and energy out of the left lead, the current takes the form
\begin{equation}\label{Cotunneling_ine:0LLcurr}
\begin{split}
	&\mathcal{J}^{00\rightarrow11,c/u}_{LL}=\frac{\hbar}{2\pi}\Gamma_L^2\int f_L(E)f_L(\epsilon_\uparrow+\epsilon_\downarrow+E_C-E)
	\\
	&\times\left|\frac{1}{E-\epsilon_\uparrow+i\eta}+\frac{1}{\epsilon_\uparrow+E_C-E+i\eta}\right|^2 KdE,
\end{split}	
\end{equation}
where $K=-2e$ for the charge current (superscript $c$), since both  electrons that tunnel into the QD come from the left lead, and $K=\epsilon_\uparrow+\epsilon_\downarrow+E_C$ for the energy current (superscript $u$), which is the energy removed from the left lead in the cotunneling process.
Notice that $K$ does not depend on the integration variable, therefore the charge and energy currents are proportional to each other.
Case (ii) is analogous to case (i) and we can calculate the cotunneling rate by exchanging the leads' incides $L\rightarrow R$ in Eq.~\eqref{Cotunneling_ine:0LLT} obtaining
\begin{equation}\label{Cotunneling_ine:0RRT}
\begin{split}
	&T^{00\rightarrow11}_{RR}=\frac{\hbar}{2\pi}\Gamma_R^2\int f_R(E)f_R(\epsilon_\uparrow+\epsilon_\downarrow+E_C-E)
	\\
	&\times\left|\frac{1}{E-\epsilon_\uparrow+i\eta}+\frac{1}{\epsilon_\uparrow+E_C-E+i\eta}\right|^2 dE.
\end{split}
\end{equation}
Note that there is no current associated to case (ii), since neither electron tunnels from the left lead.

In case (iii) there are two possible initial states, depending on whether the electron with spin up comes from the left or right lead.
For each initial state there are two possible intermediate states, depending on whether the electron occupying the QD comes from the left or the right lead. 
Imposing the energy conservation between the initial and the final states, we obtain the following cotunneling rate
\begin{equation}\label{Cotunneling_ine:0LRT}
\begin{split}
	&T^{00\rightarrow11}_{LR, \sigma}=\frac{\hbar}{2\pi}\Gamma_L\Gamma_R\int f_L(E)f_R(\epsilon_\uparrow+\epsilon_\downarrow+E_C-E)
	\\
	&\times\left|\frac{1}{E-\epsilon_\sigma+i\eta}+\frac{1}{\epsilon_\sigma+E_C-E+i\eta}\right|^2 dE,
\end{split}
\end{equation}
which depends on the spin variable $\sigma$.
The corresponding current leaving the left lead takes the form
\begin{equation}\label{Cotunneling_ine:0LRcurr}
\begin{split}
	&\mathcal{J}^{00\rightarrow11,c/u}_{LR, \sigma}=\frac{\hbar}{2\pi}\Gamma_L\Gamma_R\int f_L(E)f_R(\epsilon_\uparrow+\epsilon_\downarrow+E_C-E)
	\\
	&\times \left|\frac{1}{E-\epsilon_\sigma+i\eta}+\frac{1}{\epsilon_\sigma+E_C-E+i\eta}\right|^2 KdE,
\end{split}
\end{equation}
where $K=-e$ for the charge current (superscript $c$) and $K=E$ for the energy current (superscript $u$).
Notice that in this case charge and energy currents are not proportional to each other since $K=E$ cannot be taken out of the integration.

Let us now consider the case where the QD is initially occupied by two electrons.
The inelastic cotunneling processes, that move both electrons out of the QD, are the inverse processes with respect to the one discussed above (relative to the QD initially empty).
Therefore, the initial and the final state of the processes which empty the QD are, respectively, the final and the initial state of the processes which fill the QD.
Moreover, the intermediate states are the same.
Therefore, we obtain the following expressions for the cotunneling rates
\begin{equation}\label{Cotunneling_ine:2LLT}
\begin{split}
	&T^{11\rightarrow00}_{LL}=\frac{\hbar}{2\pi}\Gamma_L^2\int f^-_L(E)f^-_L(\epsilon_\uparrow+\epsilon_\downarrow+E_C-E)
	\\
	&\times\left|\frac{1}{E-\epsilon_\uparrow+i\eta}+\frac{1}{\epsilon_\uparrow+E_C-E+i\eta}\right|^2 dE,
\end{split}
\end{equation}
\begin{equation}\label{Cotunneling_ine:2LRT}
\begin{split}
	&T^{11\rightarrow00}_{LR,\sigma}=\frac{\hbar}{2\pi}\Gamma_L\Gamma_R\int f^-_L(E)f^-_R(\epsilon_\uparrow+\epsilon_\downarrow+E_C-E)
	\\
	&\times\left|\frac{1}{E-\epsilon_\sigma+i\eta}+\frac{1}{\epsilon_\sigma+E_C-E+i\eta}\right|^2 dE,
\end{split}
\end{equation}
which is analogous to the transition rate of Eq.~\eqref{Cotunneling_ine:0LRT}, and
\begin{equation}\label{Cotunneling_ine:2RRT}
\begin{split}
	&T^{11\rightarrow00}_{RR}=\frac{\hbar}{2\pi}\Gamma_L^2\int f^-_R(E)f^-_R(\epsilon_\uparrow+\epsilon_\downarrow+E_C-E)
	\\
	&\times\left|\frac{1}{E-\epsilon_\uparrow+i\eta}+\frac{1}{\epsilon_\uparrow+E_C-E+i\eta}\right|^2 dE.
\end{split}
\end{equation}
For the currents flowing out of the left lead we obtain
\begin{equation}\label{Cotunneling_ine:2LLcurr}
\begin{split}
	&\mathcal{J}^{11\rightarrow00,c/u}_{LL}=\frac{\hbar}{2\pi}\Gamma_L^2\int f^-_L(E)f^-_L(\epsilon_\uparrow+\epsilon_\downarrow+E_C-E)
	\\
	&\times\left|\frac{1}{E-\epsilon_\uparrow+i\eta}+\frac{1}{\epsilon_\uparrow+E_C-E+i\eta}\right|^2 KdE,
\end{split}
\end{equation}
where $K=-2e$ for the charge current (superscript $c$) and and $K=\epsilon_\uparrow+\epsilon_\downarrow+E_C$ for the energy current (superscript $u$),
\begin{equation}\label{Cotunneling_ine:2LRcurr}
\begin{split}
	&\mathcal{J}^{11\rightarrow00,c/u}_{LR, \sigma}=\frac{\hbar}{2\pi}\Gamma_L\Gamma_R\int f^-_L(E)f^-_R(\epsilon_\uparrow+\epsilon_\downarrow+E_C-E)
	\\
	&\times\left|\frac{1}{E-\epsilon_\sigma+i\eta}+\frac{1}{\epsilon_\sigma+E_C-E+i\eta}\right|^2 KdE,
\end{split}
\end{equation}
where $K=-e$ for the charge current (superscript $c$) and $K=E$ for the energy current (superscript $u$).

Finally, let us now consider the case where initially a certain level of the QD is occupied by one electron.
After the inelastic processes, in the final state the QD will still contain one electron, but in other level.
When one of the electron is initially in the left lead, there are two possible initial states, one for each value of the spin.
For each initial state there are two final states: (i) the electron in the QD tunnels into the right lead, so that both charge and energy are transferred in the cotunneling process; (ii) the electron in the QD tunnels into the left lead, in which case only energy is transferred (the QD levels have different energies).
\begin{figure}[!htb]
	\centering	\includegraphics[width=0.9 \columnwidth]{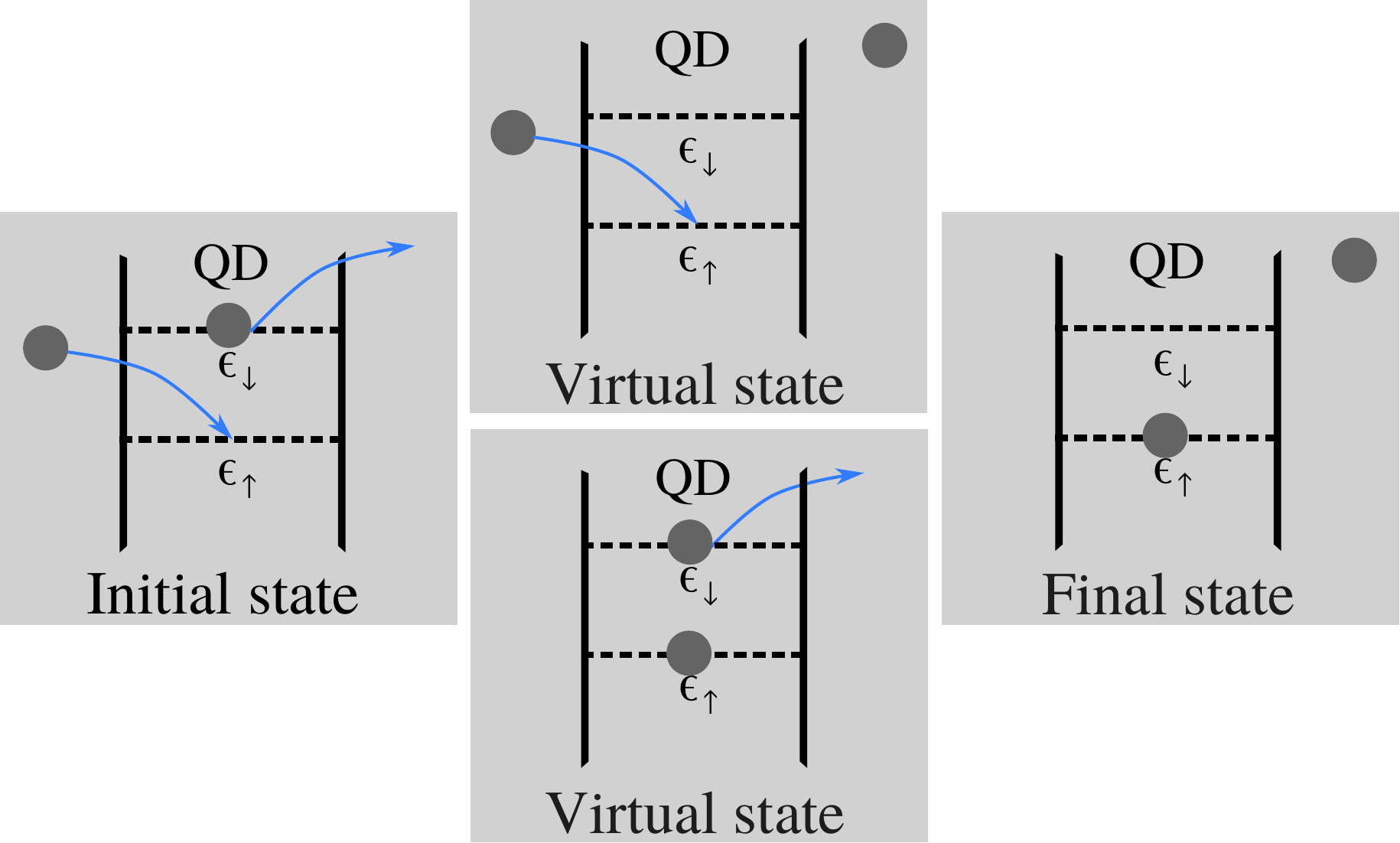}
	\caption{Diagram of the inelastic cotunneling process $L\rightarrow R$ with incoming electron withh spin $\uparrow$ and QD initially in $(0,1)$.}
	\label{fig:ineLR}
\end{figure}
In case (i), see Fig.~\ref{fig:ineLR}, there are two possible intermediate states (QD fully occupied and QD empty), depending on the order of the two tunneling processes.
We obtain the following cotunneling rates
\begin{equation}\label{Cotunneling_ine:1TLR}
\begin{split}
		&T_{L\bar{\sigma}\rightarrow R\sigma}=\frac{\hbar}{2\pi}\Gamma_L\Gamma_R\int f_L(E)f_R^-(E+\epsilon_{\bar{\sigma}}-\epsilon_\sigma)
		\\
		&\times\left|\frac{1}{\epsilon_\sigma-E+i\eta} + \frac{1}{E-\epsilon_{\sigma}-E_C+i\eta}\right|^2 dE.
\end{split}
\end{equation}
These processes transfer both charge and energy, so that the current flowing out of the left lead is given by
\begin{equation}\label{Cotunneling_ine:1currLR}
\begin{split}
		&\mathcal{J}^{c/u}_{L\bar{\sigma}\rightarrow R\sigma}=\frac{\hbar}{2\pi}\Gamma_L\Gamma_R\int f_L(E)f_R^-(E+\epsilon_{\bar{\sigma}}-\epsilon_\sigma)
		\\
		&\times\left|\frac{1}{\epsilon_\sigma-E+i\eta} + \frac{1}{E-\epsilon_{\sigma}-E_C+i\eta}\right|^2 KdE,
\end{split}
\end{equation}
where $K=-e$ for the charge current (superscript $c$) and $K=E$ for the energy current (superscript $u$).

\begin{figure}[!htb]
	\centering	\includegraphics[width=0.9 \columnwidth]{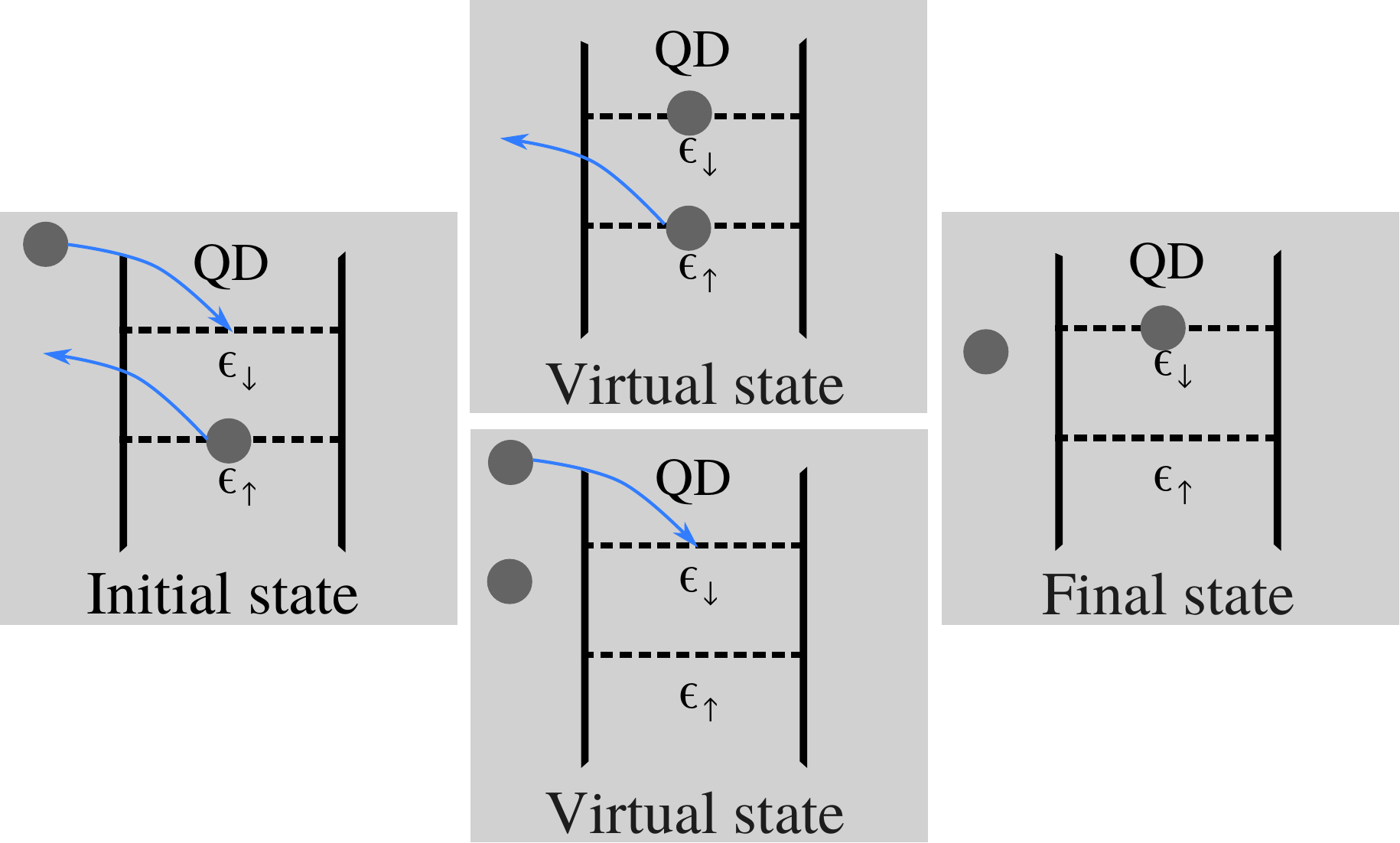}
	\caption{Diagram of the inelastic cotunneling process $L\rightarrow L$ with incoming electron of spin $\uparrow$ and QD initially in $(0,1)$.}
	\label{fig:ineLL}
\end{figure}
Also in case (ii), see Fig.~\ref{fig:ineLL}, there are two possible intermediate states (QD fully occupied and QD empty), depending on the order of the two tunneling processes.
We obtain the following cotunneling rates
\begin{equation}\label{Cotunneling_ine:1TLL}
\begin{split}
		&T_{L\bar{\sigma}\rightarrow L\sigma}=\frac{\hbar}{2\pi}\Gamma_L^2\int f_L(E)f_L^-(E+\epsilon_{\bar{\sigma}}-\epsilon_\sigma)
		\\
		&\times\left|\frac{1}{\epsilon_\sigma-E+i\eta} + \frac{1}{E-\epsilon_{\sigma}-E_C+i\eta}\right|^2 dE.
\end{split}
\end{equation}
These processes transfer only energy, with a current flowing out of the left lead given by
\begin{equation}\label{Cotunneling_ine:1currLL}
\begin{split}
		&\mathcal{J}^u_{L\bar{\sigma}\rightarrow L\sigma}=\frac{\hbar}{2\pi}\Gamma_L\Gamma_R\int f_L(E)f_L^-(E+\epsilon_{\bar{\sigma}}-\epsilon_\sigma)
		\\
		&\times\left|\frac{1}{\epsilon_\sigma-E+i\eta} + \frac{1}{E-\epsilon_{\sigma}-E_C+i\eta}\right|^2 KdE,
\end{split}
\end{equation}
where $K=\epsilon_\sigma-\epsilon_{\bar{\sigma}}$.

When one of the electron is initially in the right lead, analogous calculations lead to the following cotunneling rates
\begin{equation}\label{Cotunneling_ine:1TRL0}
\begin{split}
		&T_{R\bar{\sigma}\rightarrow L\sigma}=\frac{\hbar}{2\pi}\Gamma_R\Gamma_L\int f_R(E)f_L^-(E+\epsilon_{\bar{\sigma}}-\epsilon_\sigma)
		\\
		&\times\left|\frac{1}{\epsilon_\sigma-E+i\eta} + \frac{1}{E-\epsilon_{\sigma}-E_C+i\eta}\right|^2 dE,
\end{split}
\end{equation}
\begin{equation}\label{Cotunneling_ine:1TRR}
\begin{split}
		&T_{R\bar{\sigma}\rightarrow R\sigma}=\frac{\hbar}{2\pi}\Gamma_R^2\int f_R(E)f_R^-(E+\epsilon_{\bar{\sigma}}-\epsilon_\sigma)
		\\
		&\times\left|\frac{1}{\epsilon_\sigma-E+i\eta} + \frac{1}{E-\epsilon_{\sigma}-E_C+i\eta}\right|^2 dE,
\end{split}
\end{equation}
and expressions of the current flowing out of the left lead (only for the former processes, since the latter processes do not produce any current in the left lead)
\begin{equation}\label{Cotunneling_ine:1currRL}
\begin{split}
		&\mathcal{J}^{c/u}_{R\bar{\sigma}\rightarrow L\sigma}=\frac{\hbar}{2\pi}\Gamma_L\Gamma_R\int f^-_L(E)f_R(E+\epsilon_\sigma-\epsilon_{\bar{\sigma}})
		\\
		&\times\left|\frac{1}{\epsilon_{\bar{\sigma}}-E+i\eta} + \frac{1}{E-\epsilon_{\bar{\sigma}}-E_C+i\eta}\right|^2 KdE,
\end{split}
\end{equation}
where $K=-e$ for the charge current (superscript $c$) and $K=E$ for the energy current (superscript $u$).

Finally we notice that, in the present case, cotunneling inelastic processes occur via two intermediate states (contrary to elastic processes, which occur via a single intermediate state), thus giving rise to quantum interference effects.

\subsection{Master equations and cotunnelling rates}
\label{AppQDME}
When inelastic co-tunnelling processes are included, the MEs need to be modified and can be written as follows
\begin{equation}
	\label{meIN}
  \begin{aligned}
	\frac{d}{dt}P_0=\left[\Sigma_\uparrow^-P_\uparrow+\Sigma_\downarrow^-P_\downarrow-(\Sigma_\uparrow^++\Sigma_\downarrow^+)P_0\right] \\
	+ \left[T_{\rm out}P_2-T_{\rm in}P_0\right] , \\
	\frac{d}{dt}P_2 = \left[P_\uparrow\Theta^+_\downarrow+P_\downarrow\Theta^+_\uparrow-P_2(\Theta^-_\uparrow+\Theta^-_\downarrow)\right] \\
	+\left[P_0T_{\rm in}-P_2T_{\rm out}\right], \\
	\frac{d}{dt}P_\uparrow=\left[P_0\Sigma^+_\uparrow+P_2\Theta^-_\downarrow-P_\uparrow(\Sigma_\uparrow^-+\Theta^+_\downarrow)\right]\\
	+\left[P_\downarrow T^{01\rightarrow10}-P_\uparrow T^{10\rightarrow01}\right],
  \end{aligned}
\end{equation}
where 
\begin{equation}\label{Cotunneling_meq:Tout}
	T_{\rm out}=T^{11\rightarrow00}_{LL}+T^{11\rightarrow00}_{LR,\uparrow}+T^{11\rightarrow00}_{LR,\downarrow}+T^{11\rightarrow00}_{RR}
\end{equation}
is the sum of the inelastic cotunneling transition rates relative to the processes that empty the QD, and
\begin{equation}\label{Cotunneling_meq:Tin}
		T_{\rm in}=T^{00\rightarrow11}_{LL}+T^{00\rightarrow11}_{LR,\uparrow}+T^{00\rightarrow11}_{LR,\downarrow}+T^{00\rightarrow11}_{RR}
\end{equation}
is the sum of the inelastic cotunneling rates of the processes that fill the QD.
Similarly, $T^{01\rightarrow10}$ and $T^{10\rightarrow01}$ are defined as the sums of the inelastic cotunneling rates that exchange the level in the QD which is occupied.
In Eqs.~(\ref{meIN}), $P_0$ represents the probability for the QD to be unoccupied, $P_2$ represents the probability for the QD to be doubly occupied, while $P_{\uparrow}$ ($P_{\downarrow}$) is the probability for the lower $\epsilon_\uparrow$ (upper $\epsilon_\downarrow$) level of the QD to be occupied.
Moreover, we have defined
\begin{equation}\label{_1}
	\Theta_\sigma^+=\Gamma_Lf_L(\epsilon_\sigma+E_C)+\Gamma_Rf_R(\epsilon_\sigma+E_C)
\end{equation}
and
\begin{equation}\label{_2}
	\Theta_\sigma^-=\Gamma_Lf_L^-(\epsilon_\sigma+E_C)+\Gamma_Rf_R^-(\epsilon_\sigma+E_C),
\end{equation}
where $\Theta_\sigma^+$ describes the total rate of the sequential processes that move an electron with spin $\sigma$ from the leads to the QD that has already one electron inside, while $\Theta_\sigma^-$ represents the inverse process, in which the electron with spin $\sigma$ tunnels from the doubly occupied QD to the leads.
Notice that, in Eqs.~(\ref{meIN}), the first square brackets contain the sequential tunneling contribution, see Eq.~\eqref{Sequential:mastereq3}, whereas, the second square brackets refer to the cotunneling corrections.
The equation relative to $P_\downarrow$ is obtained by exchanging the labels $\uparrow$ and $\downarrow$ in the last line of Eqs.~(\ref{meIN}).

In the four-terminal device, the MEs which allow to determine the probabilities $P_0$, $P_\uparrow$, $P_\downarrow$ and $P_2$ in the sequential tunneling regime are formally equal to Eqs.~(\ref{meIN}), where we consider only the first square brackets in each line.
In this case, however, the quantities $\Theta_\sigma^\pm$ and $\Sigma_\sigma^\pm$ are defined as follows
\begin{equation}
  \begin{aligned}
	\Theta_\uparrow^+=\Gamma_{L1}f_{L1}(\epsilon_\uparrow+E_C)+\Gamma_{L2}f_{L2}(\epsilon_\uparrow+E_C), \\
	\Theta_\uparrow^-=\Gamma_{L1}f_{L1}^-(\epsilon_\uparrow+E_C)+\Gamma_{L2}f_{L2}^-(\epsilon_\uparrow+E_C), \\
	\Theta_\downarrow^+=\Gamma_{R1}f_{R1}(\epsilon_\downarrow+E_C)+\Gamma_{R2}f_{R2}(\epsilon_\downarrow+E_C), \\
	\Theta_\downarrow^-=\Gamma_{R1}f_{R1}^-(\epsilon_\downarrow+E_C)+\Gamma_{R2}f_{R2}^-(\epsilon_\downarrow+E_C)
  \end{aligned}
\end{equation}
and
\begin{equation}
  \begin{aligned}
	\Sigma_\uparrow^+=\Gamma_{L1}f_{L1}(\epsilon_\uparrow)+\Gamma_{L2}f_{L2}(\epsilon_\uparrow), \\
	\Sigma_\uparrow^-=\Gamma_{L1}f_{L1}^-(\epsilon_\uparrow)+\Gamma_{L2}f_{L2}^-(\epsilon_\uparrow), \\
	\Sigma_\downarrow^+=\Gamma_{R1}f_{R1}(\epsilon_\downarrow)+\Gamma_{R2}f_{R2}(\epsilon_\downarrow), \\
	\Sigma_\downarrow^-=\Gamma_{R1}f_{R1}^-(\epsilon_\downarrow)+\Gamma_{R2}f_{R2}^-(\epsilon_\downarrow)
  \end{aligned}
\end{equation}
and the tunneling constants $\Gamma_\alpha$ are defined by Eq.~(\ref{tunc}), with $\alpha={L1, L2, R1, R2}$.

\subsection{Expressions for the currents}
\label{cot-cur}
In the case of a QD with two levels one finds
\begin{equation}\label{Cotunneling_c&r:Jcseq}
\begin{split}
	I^c_{\rm seq}&=-e\Gamma_L\left(P_0\left[f_{L\uparrow}+f_{L\downarrow}\right]-P_2\left[F^-_{L\uparrow}+F^-_{L\downarrow}\right]+\right.
	\\
	&\left. +P_\uparrow\left[F_{L\downarrow}-f^-_{L\uparrow}\right]+P_\downarrow\left[F_{L\uparrow}-f^-_{L\downarrow}\right]\right),
\end{split}
\end{equation}
and
\begin{equation}\label{Cotunneling_c&r:Juseq}
\begin{split}
	I_{\rm seq}&=P_0\Gamma_L[\epsilon_\uparrow f_{L\uparrow}+\epsilon_\downarrow f_{L\downarrow}]-
	\\
	&-P_2\Gamma_L[(E_C+\epsilon_\uparrow)F^-_{L\uparrow}+(E_C+\epsilon_\downarrow)F^-_{L\downarrow}]+
	\\
	&+P_\uparrow\Gamma_L[(E_C+\epsilon_\downarrow)F_{L\downarrow}-\epsilon_\uparrow f^-_{L\uparrow}]+
	\\
	&+P_\downarrow\Gamma_L[(E_C+\epsilon_\uparrow)F_{L\uparrow}-\epsilon_\downarrow f^-_{L\downarrow}],
\end{split}
\end{equation}
where $F_{L\sigma}=f_{L}(\epsilon_\sigma+E_C)$ and $F_{L\sigma}^-=1-f_{L}(\epsilon_\sigma+E_C)$.
Notice that Eqs.~(\ref{Cotunneling_c&r:Jcseq}) and (\ref{Cotunneling_c&r:Juseq}) reduce to Eqs.~(\ref{Sequential:J^cP}) and (\ref{Sequential:J^uP}), respectively, when $E_C$ diverges, thus causing $F_{L\sigma}$ and $P_2$ to vanish.

In Eq.~(\ref{eq:j_cot}) the cotunneling contributions contain the sum of elastic and inelastic processes, namely
\begin{equation}
	I^{c}_\text{cot}(\Delta T) = I^{c}_\text{el}(\Delta T)+ I^{c}_\text{in}(\Delta T),
\end{equation}
and
\begin{equation}
	I_\text{cot}(\Delta T) = I_\text{el}(\Delta T)+ I_\text{in}(\Delta T).
	\label{eq:j_cot-el}
\end{equation}
The elastic components are given by
\begin{equation}\label{Cotunneling_c&r:Jcela}
\begin{split}
	I^{c}_{\rm el}=&({\cal J}^{00,c}_{\uparrow}+{\cal J}^{00,c}_{\downarrow})P_0+({\cal J}^{10,c}_{\uparrow}+{\cal J}^{10,c}_{\downarrow})P_\uparrow+
	\\
	&({\cal J}^{01,c}_{\uparrow}+{\cal J}^{01,c}_{\downarrow})P_\downarrow+({\cal J}^{11,c}_{\uparrow}+{\cal J}^{11,c}_{\downarrow})P_2 
\end{split}
\end{equation}
and
\begin{equation}\label{Cotunneling_c&r:Juela}
\begin{split}
	I_{\rm el}=&({\cal J}^{00,u}_{\uparrow}+{\cal J}^{00,u}_{\downarrow})P_0+({\cal J}^{10,u}_{\uparrow}+{\cal J}^{10,u}_{\downarrow})P_\uparrow+
	\\
	&+({\cal J}^{01,u}_{\uparrow}+{\cal J}^{01,u}_{\downarrow})P_\downarrow+({\cal J}^{11,u}_{\uparrow}+{\cal J}^{11,u}_{\downarrow})P_2,
\end{split}
\end{equation}
while the inelastic components are
\begin{equation}\label{Cotunneling_c&r:Jcine}
\begin{split}
	I^c_{\rm in}&=({\cal J}^{00\rightarrow11,c}_{LL}+{\cal J}^{00\rightarrow11,c}_{LR\uparrow}+{\cal J}^{00\rightarrow11,c}_{LR\downarrow})P_0+
	\\
	&+({\cal J}^c_{L\downarrow\rightarrow R\uparrow}-{\cal J}^c_{R\downarrow\rightarrow L\uparrow})P_\uparrow+({\cal J}^c_{L\uparrow\rightarrow R\downarrow}-{\cal J}^c_{R\uparrow\rightarrow L\downarrow})P_\downarrow-
	\\
	&-({\cal J}^{11\rightarrow00,c}_{LL}+{\cal J}^{11\rightarrow00,c}_{LR\uparrow}+{\cal J}^{11\rightarrow00,c}_{LR\downarrow})P_2
\end{split}
\end{equation}
and
\begin{equation}\label{Cotunneling_c&r:Juine}
\begin{split}
	I_{\rm in}&=({\cal J}^{00\rightarrow11,u}_{LL}+{\cal J}^{00\rightarrow11,u}_{LR\uparrow}+{\cal J}^{00\rightarrow11,u}_{LR\downarrow})P_0+
	\\
	&+({\cal J}^u_{L\downarrow\rightarrow R\uparrow}-{\cal J}^u_{R\downarrow\rightarrow L\uparrow}+{\cal J}^u_{L\downarrow\rightarrow L\uparrow})P_\uparrow+
	\\
	&+({\cal J}^u_{L\uparrow\rightarrow R\downarrow}-{\cal J}^u_{R\uparrow\rightarrow L\downarrow}+{\cal J}^u_{L\uparrow\rightarrow L\downarrow})P_\downarrow-
	\\
	&-({\cal J}^{11\rightarrow00,u}_{LL}+{\cal J}^{11\rightarrow00,u}_{LR\uparrow}+{\cal J}^{11\rightarrow00,u}_{LR\downarrow})P_2 .
\end{split}
\end{equation}
Such probabilities are calculated through a ME which also account for inelastic co-tunnelling processes (see App.~\ref{AppQDME}).
The elastic co-tunnelling {\it single-process} currents (${\cal J}^{ij,c/u}_{\sigma}$) appearing in Eqs.~(\ref{Cotunneling_c&r:Jcela}) and (\ref{Cotunneling_c&r:Juela}) are defined in App.~\ref{AppCotEl}, while the inelastic co-tunnelling {\it single-process} currents ($J^{kk\rightarrow ll,c/u}_{LL}$, $J^{kk\rightarrow ll,c/u}_{LR\sigma}$ and $J^{c/u}_{L/R\bar{\sigma}\rightarrow R/L\sigma}$) appearing in Eqs.~(\ref{Cotunneling_c&r:Jcine}) and (\ref{Cotunneling_c&r:Juine}) are defined in App.~\ref{AppCotInel}.
It is worth noticing that the elastic cotunneling contributions to the heat current can give rise to rectification, despite the fact that the quantities ${\cal J}^{ij,u}_{\sigma}$ in Eq.~(\ref{Cotunneling_c&r:Juela}) depend (under an energy integration) on the difference between the Fermi functions of the two leads at the same energy. Indeed, the probabilities $P_0$, $P_2$, $P_{\uparrow}$  and $P_{\downarrow}$ actually depend on the sign of the temperature bias, in a more noticeable way for non-degenerate levels.

\begin{figure}[!htb]
	\centering	\includegraphics[width=0.9 \columnwidth]{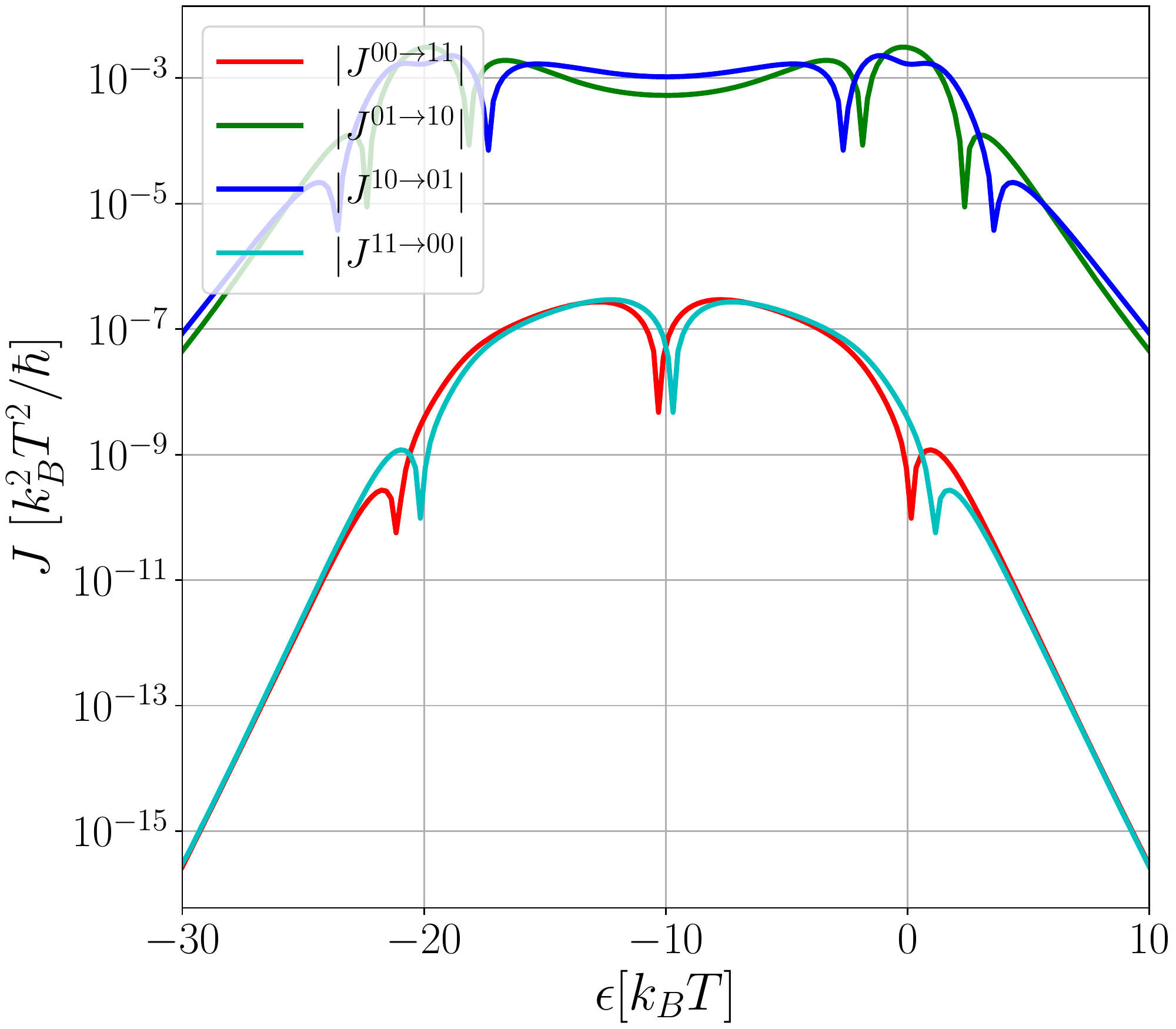}
	\caption{Closed-circuit setup. Heat current contributions (in units of $k_B^2T^2/\hbar$) for the case of two non-degenerate levels as functions of the average energy of the  levels $\epsilon$. All parameters are the same used for Fig.~\ref{Sequential:1:grafici}, while $\Delta\epsilon=2k_BT$ and $E_C=20k_BT$.}
	\label{fig:ineCurr}
\end{figure}
Furthermore, we can identify each line of Eq.~\eqref{Cotunneling_c&r:Juine} with the contribution to the energy current of the corresponding change of QD state, namely
\begin{equation}
I_\text{in} = J^{00\rightarrow11} + J^{10\rightarrow01} + J^{01\rightarrow10} + J^{11\rightarrow00}.
\end{equation}
The contributions $J^{00\rightarrow11}$ and $J^{11\rightarrow00}$ are suppressed because the respective \textit{single-process} currents and  QD occupation probabilities decrease exponentially in different energy regions.
Indeed, for the process $(0,0)\rightarrow(1,1)$ [$(1,1)\rightarrow(0,0)$], the probability $P_0$ ($P_2$) decreases as the electrons (holes) from the leads can occupy the QD levels, just through sequential tunneling, when $\epsilon\lesssim 0$ ($\epsilon\gtrsim -E_C$).
On the other hand, the \textit{single-process} currents decreases as the leads electrons (holes) do not have enough energy to overcome the charging energy $E_C$ when $\epsilon\gtrsim-E_C/2$ ($\epsilon\lesssim -E_C/2$).
Instead, the cotunneling processes $(1,0)\leftrightarrow(0,1)$ are not much suppressed because the combined energy of the involved electrons does not have to overcome the charging energy $E_C$.
In Fig.~\ref{fig:ineCurr} these inelastic cotunneling contributions to the heat currents are compared: $(1,0)\leftrightarrow(0,1)$ and $(0,0)\leftrightarrow(1,1)$ contributions differ in order of magnitudes.

\section{Renormalization of cotunneling integrals}
\label{AppIntegrals}
To calculate the transition rates and the currents of the cotunneling processes, we have to integrate over all the initial and final states.
We note that all these integrals diverge when $\eta\rightarrow 0$, so we have to separate and remove the divergent part from the rest.
As shown below, such divergent part is associated with the sequential tunneling \cite{turek2002}.
A generic integral is in the following form:
\begin{equation}
	{\cal I}=\int\left|\frac{1}{E-A+i\eta}+\frac{1}{E-B+i\eta}\right|^2g(E)dE,
\end{equation}
where $A$ and $B$ are some constant energies and $g(E)$ is a smooth function of $E$.
Expanding the squared modulus, we can separate the integral in three parts
\begin{equation}
	\begin{split}
		&{\cal I}=\int\left[\frac{1}{(E-A)^2+\eta^2}+\frac{1}{(E-B)^2+\eta^2}+\right.
		\\
		&\left.+2\Re{\left(\frac{1}{(E-A+i\eta)(E-B-i\eta)}\right)}\right]g(E)dE,
		\\
		&{\cal I}= \mathcal{I}_{1}(A)+\mathcal{I}_{1}(B)+\mathcal{I}_{2},
	\end{split}
\end{equation}
where the integrals $\mathcal{I}_1$ contain the squared moduli of the fractions, while the integral $\mathcal{I}_2$ contains the real part of the product between the fractions.

Let's see how to regularize the integral $\mathcal{I}_1$.
Expanding the squared modulus and changing the integration variable, we obtain
\begin{equation}
	\mathcal{I}_1(A)=\int \frac{g(E)}{(E-A)^2+\eta^2}dE = \int\frac{g(E+A)}{E^2+\eta^2}dE.
\end{equation}
Then, we can sum and subtract the quantity $g(A)$ at the numerator of the integral, obtaining
\begin{equation}
	\mathcal{I}_1(A)=\int \frac{g(A)}{E^2+\eta^2}dE+\int \frac{g(E+A)-g(A)}{E^2+\eta^2}dE.
\end{equation}
The first integral can be calculated exactly and is proportional to $1/\eta$, namely
\begin{equation}\label{integrals:divergence}
	\mathcal{I}_1(A)=\pi\frac{g(A)}{\eta}+\int  \frac{g(E+A)-g(A)}{E^2+\eta^2}dE.
\end{equation}
The first term is divergent in the limit $\eta\rightarrow0$, so we remove it.
Whereas the second term converges at vanishing $\eta$.
Then, the integral $\mathcal{I}_1$ is reduced to
\begin{equation}
		\mathcal{I}_1(A)\rightarrow \int\frac{g(E+A)-g(A)}{E^2}dE.
\end{equation}
The integrand diverges at $E=0$, so, we need to eliminate this divergence to compute numerically the integrals.
We note that, changing the sign of the integration variable, it yields 
\begin{equation}
	\int\frac{g(E+A)-g(A)}{E^2+\eta^2}dE=\int \frac{g(-E+A)-g(A)}{E^2+\eta^2}dE.
\end{equation}
Therefore, summing the integrals, we obtain
\begin{equation}\label{integrals:I1}
	\mathcal{I}_1(A)\rightarrow \int \frac{g(E+A)+g(-E+A)-2g(A)}{2E^2},
\end{equation}
in which the integrand does not diverge at $E=0$ and can be computed numerically.
We note that after we remove the divergence proportional to $1/\eta$, the sign of the integral $\mathcal{I}_1(A)$ is no longer guaranteed to be positive.
Therefore, the transition rates of the cotunneling processes can have negative values.
However, the total transition rate of the processes that move the system from the initial state to the final state, namely the sequential tunneling processes and the cotunneling processes, is always positive, as it is given by Eq.~\eqref{integrals:divergence}.

Now, let's regularize the integral $\mathcal{I}_2$.
First, we expand the real part of the product of the fractions, obtaining
\begin{equation}
	\begin{split}
		\mathcal{I}_2&=\int 2\Re{\left[\frac{1}{(E-A+i\eta)(E-B-i\eta)}\right]}g(E)dE,
		\\	
		&=\int 2\frac{(E-A)(E-B)+\eta^2}{[(E-A)(E-B)+\eta^2]^2+\eta^2(A-B)^2}g(E)dE.
	\end{split}
\end{equation}
In the limit $\eta\rightarrow0$, the term proportional to $(E-A)(E-B)$ becomes the principal values of the integral, while the term proportional to $\eta^2$ vanishes.
Therefore, the integral $\mathcal{I}_2$ is reduced to
\begin{equation}
		\mathcal{I}_2\rightarrow \int \frac{2g(E)}{(E-A)(E-B)}dE,
\end{equation}
where the integral is performed on the principal value.
To compute the integral numerically, we can split the denominator in the following way:
\begin{equation}
	\frac{1}{(E-A)(E-B)}=\frac{1}{A-B}\left(\frac{1}{E-A}-\frac{1}{E-B}\right).
\end{equation}
Then, we can split the integral and change variables to move the singularity in $E=0$, namely
\begin{equation}
		\mathcal{I}_2\rightarrow\int \frac{2g(E+A)-2g(E+B)}{(A-B)E}dE.
\end{equation}
Finally, we can change the integration variable from $E$ to $-E$ and obtain a similar integral with opposite sign.
Summing these integrals together, we obtain 
\begin{equation}\label{integrals:I2}
\begin{split}
	\mathcal{I}_2\rightarrow\int \left[\frac{g(E+A)-g(-E+A)}{(A-B)E}+\right. \\
		\left. +\frac{-g(E+B)+g(-E+B)}{(A-B)E}\right]dE,
\end{split}
\end{equation}
which is computable numerically because the integrand has no divergences.

We remove the divergent part because it is generated by the sequential processes.
Indeed, in the term proportional to $1/\eta$ there is energy conservation between the initial, the intermediate, and the final state.
Therefore, the system can arrive in the final state through two sequential tunneling processes.
Moreover, the imaginary parameter $i\eta$ is associated with the rate of leaving the intermediate state~\cite{averin1994}.
Indeed, imaginary energies describe metastable states and decay processes.
For instance, suppose to initialize the system in a state $|\alpha\rangle$, which has energy $E-i\eta$.
Then, after a time $t$, the state has evolved into
\begin{equation}
		|\alpha\rangle_t=e^{-i(E-i\eta)t/\hbar}|\alpha\rangle,
\end{equation}
and the probability of finding the system in the state $|\alpha\rangle$ becomes $e^{-2\eta t/\hbar}$.
In the same way, $\eta$ is due to the processes that move the system out of the state $|\alpha\rangle$.

For example, let us consider a QD with two non-degenerate levels of energies $\epsilon\pm\Delta\epsilon/2$ and let us calculate the divergent part of the cotunneling processes that have as the intermediate state the state in which the QD has one electron of spin $\sigma$.
Using the currents calculated in section \ref{AppCotEl}, and considering the divergent part that arises from Eq.~\eqref{integrals:divergence}, we can calculate the divergent part of the elastic cotunneling charge current that leaves the left lead, namely
\begin{equation}
	I^\sigma_{\rm ela}=\frac{-e\hbar}{2\eta}\Gamma_L\Gamma_R\left[P_0(f_{L\sigma}-f_{R\sigma})+P_2(F_{L\bar{\sigma}}-F_{R\bar{\sigma}})\right].
\end{equation}
Then, we obtain the divergence of the inelastic cotunneling processes using the currents calculated in section \ref{AppCotInel}, namely
\begin{equation}
	\begin{split}
	I^\sigma_{\rm ine}&=\frac{-e\hbar}{2\eta}\Gamma_L^2\left[2P_0f_{L\sigma}F_{L\bar{\sigma}}-2P_2f^-_{L\sigma}F^-_{L\bar{\sigma}}\right]+
	\\
	&+\frac{-e\hbar}{2\eta}\Gamma_L\Gamma_R\left[P_0(f_{L\sigma}F_{R\bar{\sigma}}+f_{R\sigma}F_{L\bar{\sigma}}) \right.
	\\
	&-\left. P_2(f^-_{L\sigma}F^-_{R\bar{\sigma}}+f^-_{R\sigma}F^-_{L\bar{\sigma}}) \right] .
	\end{split}
\end{equation}
Summing together such currents, we get 
\begin{equation}\label{integrals:Icot}
		\begin{split}
		&I^\sigma_{\rm cot}=\frac{-e\hbar}{2\eta}\Gamma_L\left[\Gamma_RP_0(f_{L\sigma}-f_{R\sigma})+P_0(f_{L\sigma}\Theta^+_{\bar{\sigma}}+\right.
		\\
		&\left. +\Sigma^+_\sigma F_{L\bar{\sigma}})+\Gamma_R P_2(F_{L\bar{\sigma}}-F_{R\bar{\sigma}})-P_2(f^-_{L\sigma}\Theta^-_{\bar{\sigma}}+\Sigma^-_\sigma F^-_{L\bar{\sigma}})\right].
		\end{split}
\end{equation}

Now, we remember that the rate equation of the probability $P_\sigma$ in the sequential tunneling regime is
\begin{equation}\label{integrals:rateeqPs}
	\frac{d}{dt}P_\sigma=-P_\sigma(\Sigma_\sigma^-+\Theta^+_{\bar{\sigma}})+P_0\Sigma^+_\sigma+P_2\Theta^-_{\bar{\sigma}},
\end{equation}
and the left charge current transferred by the processes that enter such a rate equation is
\begin{equation}\label{integrals:currPs}
	I^\sigma_{\rm{seq},L}=-e\Gamma_L\left[P_0f_{L\sigma}+P_\sigma(F_{L\bar{\sigma}}-f^-_{L\sigma})-P_2F^-_{L\bar{\sigma}}\right].
\end{equation}
In the stationary condition, the charge current becomes
\begin{equation}\label{integrals:IF}
	\begin{split}
		&I^\sigma_{\rm{seq}, L}=\frac{-e\Gamma_L}{\Theta^+_{\bar{\sigma}}+\Sigma^-_\sigma}\left[\Gamma_RP_0(f_{L\sigma}-f_{R\sigma})+P_0(f_{L\sigma}\Theta^+_{\bar{\sigma}}+\right.
		\\
		&\left. +\Sigma^+_\sigma F_{L\bar{\sigma}})+\Gamma_R P_2(F_{L\bar{\sigma}}-F_{R\bar{\sigma}})-P_2(f^-_{L\sigma}\Theta^-_{\bar{\sigma}}+\Sigma^-_\sigma F^-_{L\bar{\sigma}})\right].
	\end{split}
\end{equation}
Comparing Eqs.~\eqref{integrals:Icot} and \eqref{integrals:IF}, we can find that $\eta$ satisfies
\begin{equation}\label{_}
		\eta=\hbar\frac{\Theta^+_{\bar{\sigma}}+\Sigma^-_\sigma}{2},
\end{equation}
which is half the rate of the sequential tunneling processes that move the QD out of the state with one electron of spin $\sigma$. This state is the intermediate state of the cotunneling processes considered.

\end{appendix}


\begin{thebibliography}{100}
\bibliographystyle{unsrt}

\bibitem{giazotto2006}
F. Giazotto, T. T. Heikkil{\"a}, A. Luukanen, A. M. Savin, and J. P. Pekola, \href{https://doi.org/10.1103/RevModPhys.78.217}{Rev. Mod. Phys. {\bf 78}, 217 (2006).}

\bibitem{giazotto2012}
F. Giazotto and M. J. Mart{\'i}nez-P{\'e}rez, \href{https://doi.org/10.1038/nature11702}{Nature {\bf 492}, 401 (2012).}

\bibitem{pekola2015}
J. P. Pekola, \href{https://doi.org/10.1038/nphys3169}{Nat. Phys. {\bf 11}, 118 (2015).}

\bibitem{ronzani2018}
A. Ronzani, B. Karimi, J. Senior, Y. -C. Chang, J. T. Peltonen, C. -D. Chen, and J. P. Pekola, \href{https://doi.org/10.1038/s41567-018-0199-4}{Nat. Phys. {\bf 14}, 991 (2018).}

\bibitem{maillet2019}
O. Maillet, P. A. Erdman, V. Cavina, B. Bhandari, E. T. Mannila, J. T. Peltonen, A. Mari, F. Taddei, C. Jarzynski, V. Giovannetti, and J. P. Pekola,
\href{https://doi.org/10.1103/PhysRevLett.122.150604}{Phys. Rev. Lett. {\bf 122}, 150604 (2019).}

\bibitem{maillet2020}
O. Maillet, D. A. S. Rengel, J. T. Peltonen, D. S. Golubev, and J. P. Pekola, \href{https://doi.org/10.1038/s41467-020-18163-8}{Nat. Commun. {\bf 11}, 4326 (2020).}

\bibitem{starr1935}
C. Starr, \href{https://doi.org/10.1063/1.1745338}{Physics {\bf 7}, 15 (1936).}

\bibitem{terraneo2002}
M. Terraneo, M. Peyrarad, and G. Casati, \href{https://doi.org/10.1103/PhysRevLett.88.094302}{Phys. Rev. Lett. {\bf 88}, 094302 (2002).}

\bibitem{li2004} B. Li, L. Wang, and G. Casati, \href{https://doi.org/10.1103/PhysRevLett.93.184301}{Phys. Rev. Lett. {\bf 93}, 184301 (2004).}

\bibitem{segal2005}
D. Segal, and A. Nitzan, \href{https://doi.org/10.1103/PhysRevLett.94.034301}{Phys. Rev. Lett. {\bf 94}, 034301 (2005).}

\bibitem{eckmann2006}J.-P. Eckmann, and C. M. Monasterio, \href{https://doi.org/10.1103/PhysRevLett.97.094301}{Phys. Rev. Lett. {\bf 97}, 094301 (2006).}

\bibitem{zeng2008} 
N. Zeng, and J. -S. Wang, \href{https://doi.org/10.1103/PhysRevB.78.024305}{Phys. Rev. B {\bf 78}, 024305 (2008).}

\bibitem{ojanen2009}
T. Ojanen, \href{https://doi.org/10.1103/PhysRevB.80.180301}{Phys. Rev. B {\bf 80}, 180301 (R) (2009).}

\bibitem{ruokola2009}
T. Ruokola, T. Ojanen, and A. -P. Jauho, \href{https://doi.org/10.1103/PhysRevB.79.144306}{Phys. Rev. B {\bf 79}, 144306 (2009).}

\bibitem{wu2009}
L. -A. Wu, and D. Segal, \href{https://doi.org/10.1103/PhysRevLett.102.095503}{Phys. Rev. Lett. {\bf 102}, 095503 (2009).}

\bibitem{wu2009b}
L. -A. Wu, C. X. Yu, and D. Segal, \href{https://doi.org/10.1103/PhysRevE.80.041103}{Phys. Rev. E {\bf 80}, 041103 (2009).}

\bibitem{otey2010}
C. R. Otey, W. T. Lau, and S. Fan, \href{https://doi.org/10.1103/PhysRevLett.104.154301}{Phys. Rev. Lett. {\bf104}, 154301 (2010).}

\bibitem{kuo2010} D. M.-T. Kuo, and Y. Chang, \href{https://doi.org/10.1103/PhysRevB.81.205321}{Phys. Rev. B {\bf 81}, 205321 (2010).}

\bibitem{kuo2010b} David M.-T. Kuo,  \href{https://doi.org/10.1143/JJAP.49.105202}{Jpn. J. Appl. Phys. {\bf 49}, 105202 (2010).}

\bibitem{ruokola2011}  
T. Ruokola, and T. Ojanen, \href{https://doi.org/10.1103/PhysRevB.83.241404}{ Phys. Rev. B \textbf{83}, 241404(R) (2011)}.

\bibitem{gunawardana2012}
K. G. S. H. Gunawardana, K. Mullen, J. Hu, Y. P. Chen, and X. Ruan, \href{https://doi.org/10.1103/PhysRevB.85.245417}{Phys. Rev. B {\bf 85}, 245417 (2012).}

\bibitem{martinez2013}
M. J. Mart\'inez-P\'erez, and F. Giazotto, \href{https://doi.org/10.1063/1.4804550}{Appl. Phys. Lett. {\bf 102}, 182602 (2013).}

\bibitem{giazotto2013}
F. Giazotto, and F. S. Bergeret,  \href{https://doi.org/10.1063/1.4846375}{Appl. Phys. Lett. {\bf 103}, 242602 (2013).}

\bibitem{landi2014} 
G. T. Landi, E. Novais, M. J. de Oliveira, and D. Karevski, \href{https://doi.org/10.1103/PhysRevE.90.042142}{Phys. Rev. E. {\bf 90}, 042142 (2014).}

\bibitem{liu2014}
Y. -Y. Liu, W. -X. Zhou, L. -M. Tang, K. -Q. Chen, \href{https://doi.org/10.1063/1.4902427}{Appl. Phys. Lett. {\bf 105}, 203111 (2014).}

\bibitem{jiang2015}J. -H. Jiang, M. Kulkarni, D. Segal, and Y. Imry, \href{https://doi.org/10.1103/PhysRevB.92.045309}{Phys. Rev. B {\bf 92}, 045309 (2015).}

\bibitem{sanchez2015}
R. S\'anchez, B. Sothmann, and A. N. Jordan, \href{https://doi.org/10.1088/1367-2630/17/7/075006}{New J. Phys. {\bf 17}, 075006 (2015).}

\bibitem{joulain2016}
K. Joulain, J. Drevillon, Y. Ezzahari, and J. Ordonez-Miranda, \href{https://doi.org/10.1103/PhysRevLett.116.200601}{Phys. Rev. Lett. {\bf 116}, 200601 (2016).}

\bibitem{agarwalla2017}
B. K. Agarwalla, D. Segal, \href{https://doi.org/10.1088/1367-2630/aa6657}{New J. Phys. {\bf 19}, 043030 (2017).}

\bibitem{vicioso2018} A. Marcos-Vicioso, C. L\'opez-Jurado, M. Ruiz-Garcia, and R. S\'anchez, \href{https://doi.org/10.1103/PhysRevB.98.035414}{Phys. Rev. B {\bf 98}, 035414 (2018).}

\bibitem{goury2019} D. Goury, and R. S\'anchez, \href{https://doi.org/10.1063/1.5109100}{Appl. Phys. Lett. {\bf 115}, 092601 (2019).}

\bibitem{giazotto2020}
F. Giazotto, and F. S. Bergeret, \href{https://doi.org/10.1063/5.0010148}{Appl. Phys. Lett. {\bf 116}, 192601 (2020).}

\bibitem{bhandari2021} B. Bhandari, P. A. Erdman, R. Fazio, E. Paladino, and F. Taddei, \href{https://doi.org/10.1103/PhysRevB.103.155434}{Phys. Rev. B {\bf 103}, 155434 (2021).}

\bibitem{iorio2021} A. Iorio, E. Strambini, G. Haack, M. Campisi, and F. Giazotto,
\href{https://doi.org/10.1103/PhysRevApplied.15.054050}{Phys. Rev. Applied {\bf 15}, 054050 (2021).}

\bibitem{upadhyay2021} V. Upadhyay, M. T. Naseem, R. Marathe, and O. E. M\"ustecapl\i o\u{g}lu, \href{https://arxiv.org/abs/2106.02891}{arXiv preprint arXiv:2106.02891.}

\bibitem{xueou2008} C. Xue-Ou, D. Bing, and L. Xiao-Lin, \href{https://iopscience.iop.org/article/10.1088/0256-307X/25/8/080}{Chinese Phys. Lett. {\bf 25} 3032 (2008).}

\bibitem{lopez2013} R. L\'opez, and D. S\'anchez, \href{https://doi.org/10.1103/PhysRevB.88.045129}{Phys. Rev. B {\bf 88}, 045129 (2013).}
 
\bibitem{kuo2020} D. M. T. Kuo, \href{https://doi.org/10.1063/1.5123403}{AIP Advances {\bf 10}, 045222 (2020).}

\bibitem{aligia2020} A. A. Aligia, D. P\'erez Daroca, Liliana Arrachea, and P. Roura-Bas,
\href{https://doi.org/10.1103/PhysRevB.101.075417}{Phys. Rev. B {\bf 101}, 075417 (2020).}

\bibitem{chang2006}
C. W. Chang, D. Okawa, A. Majumdar, A. Zettl, \href{https://doi.org/10.1126/science.1132898}{Science {\bf 314}, 1121 (2006).}

\bibitem{scheibner2008}
R. Scheibner, M. K\"onig, D. Reuter, A. D. Wieck, C. Gould,
H. Buhmann, and L. W. Molenkamp, \href{https://doi.org/10.1088/1367-2630/10/8/083016}{New J. Phys. {\bf 10}, 083016 (2008).}

\bibitem{schmotz2011}
M. Schmotz, J. Maier, E. Scheer and P. Leiderer, \href{https://doi.org/10.1088/1367-2630/13/11/113027}{New J. Phys. {\bf 13}, 113027 (2011).}

\bibitem{martinez2015}
M. J. Mart\'inez-P\'erez, A. Fornieri, and F. Giazotto, \href{https://doi.org/10.1038/nnano.2015.11}{Nat. Nanotechnol. {\bf 10}, 303 (2015).}

\bibitem{senior2019}
J. Senior, A. Gubaydullin, B. Karimi, J. T. Peltonen, J. Ankerhold, and J. P. Pekola, \href{https://doi.org/10.1038/s42005-020-0307-5}{Commun. Phys. {\bf 3}, 40 (2020).}

\bibitem{mcclure2007}
D. T. McClure, L. DiCarlo, Y. Zhang, H.-A. Engel, C. M. Marcus, M. P. Hanson, and A. C. Gossard,
\href{https://doi.org/10.1103/PhysRevLett.98.056801}{Phys. Rev. Lett. {\bf 98}, 056801 (2007).}

\bibitem{shinkai2009}
G. Shinkai, T. Hayashi, T. Ota, K. Muraki, and T. Fujisawa,
\href{https://doi.org/10.1143/APEX.2.081101}{Appl. Phys. Express {\bf 2}, 081101 (2009).}

\bibitem{shinkai2009b}
G. Shinkai, T. Hayashi, T. Ota, and T. Fujisawa,
\href{https://doi.org/10.1103/PhysRevLett.103.056802}{Phys. Rev. Lett. {\bf 103}, 056802 (2009).}

\bibitem{bischoff2015}
D. Bischoff, M. Eich, O. Zilberberg, C. R\"ossler, T. Ihn, and K. Ensslin,
\href{https://doi.org/10.1021/acs.nanolett.5b02167}{Nano Lett. {\bf 15}, 6003 (2015).}

\bibitem{hartmann2015}
F. Hartmann, P. Pfeffer, S. H\"ofling, M. Kamp, and L. Worschech,
\href{https://doi.org/10.1103/PhysRevLett.114.146805}{Phys. Rev. Lett. {\bf 114}, 146805 (2015).}

\bibitem{thierschmann2015}
H. Thierschmann, R. S\'anchez, B. Sothmann, F. Arnold, C. Heyn, W. Hansen, H. Buhmann, and L. W. Molenkamp, 
\href{https://doi.org/10.1038/nnano.2015.176}{Nat. Nanotech. {\bf 10}, 854 (2015).}

\bibitem{volk2015}
C. Volk, S. Engels, C. Neumann, and C. Stampfer,
\href{https://doi.org/10.1002/pssb.201552445}{Phys. Status Solidi B {\bf 252}, 2461 (2015).}

\bibitem{koski2015}
J. V. Koski, A. Kutvonen, I. M. Khaymovich, T. Ala-Nissila, and J. P. Pekola,
\href{https://doi.org/10.1103/PhysRevLett.115.260602}{Phys. Rev. Lett. {\bf 115}, 260602 (2015).}

\bibitem{keller2016}
A. J. Keller, J. S. Lim, D. S\'anchez, R. L\'opez, S. Amasha, J. A. Katine, H. Shtrikman, and D. Goldhaber-Gordon, 
\href{https://doi.org/10.1103/PhysRevLett.117.066602}{Phys. Rev. Lett. {\bf 117}, 066602 (2016).}

\bibitem{singh2019}
S. Singh, \'E. Rold\'an, I. Neri, I. M. Khaymovich, D. S. Golubev, V. F. Maisi, J. T. Peltonen, F. J\"ulicher, and J. P. Pekola,
\href{https://doi.org/10.1103/PhysRevB.99.115422}{Phys. Rev. B {\bf 99}, 115422 (2019).}

\bibitem{mu2021}
J. Mu, W. Li, S. Huang, D. Pan, Y. Chen, J.-Y. Wang, J. Zhao, and H. Q. Xu,
\href{https://arxiv.org/abs/2105.11162}{arXiv preprint arXiv:2105.11162.}
 
\bibitem{Zianni2007}
X. Zianni,
\href{https://doi.org/10.1103/PhysRevB.75.045344}{Phys. Rev. B {\bf 75}, 045344 (2007).}

\bibitem{Erdman2017}
P. A. Erdman, F. Mazza, R. Bosisio, G. Benenti, R. Fazio, and F. Taddei,
\href{https://doi.org/10.1103/PhysRevB.95.245432}{Phys. Rev. B {\bf 95}, 245432 (2017).}
 
\bibitem{nota2}
Whereas the sequential heat current, as we just argued, is nearly proportional to the sequential charge current.

\bibitem{bhandari2018}
B. Bhandari, G. Chiriac\`o, P. A. Erdman, R. Fazio, and F. Taddei, \href{https://doi.org/10.1103/PhysRevB.98.035415}{Phys. Rev. B {\bf 98}, 035415 (2018).}

\bibitem{cooling}
For suitable choice of parameters, in the drag circuit heat can be extracted from one of the lead and deposited to the other one producing a non-local cooling effect.

\bibitem{Erdman2018}
P. A. Erdman, B. Bhandari, R. Fazio, J. P. Pekola, and F. Taddei, \href{https://doi.org/10.1103/PhysRevB.98.045433}{Phys. Rev. B {\bf 98}, 045433 (2018).}

\bibitem{nota1}
In the closed-circuit setup $\mu_{\rm L}=\mu_{\rm R}=0$.

\bibitem{nazarov2009} Y. V. Nazarov and Y. M. Blanter, {\it Quantum transport: introduction to nanoscience} (Cambridge University Press, 2009).

\bibitem{schon1997}
T. Dittrich, P. H\"anggi, G.-L. Ingold, B. Kramer, G. Sch\"on, and W. Zwerger, {\it Quantum Transport and Dissipation.} (Wiley-VCH, 1998).

\bibitem{turek2002}
M. Turek and K. A. Matveev, \href{https://doi.org/10.1103/PhysRevB.65.115332}{Phys. Rev. B \textbf{65}, 115332 (2002)}.

\bibitem{averin1994}
D. V. Averin,  \href{https://doi.org/10.1016/0921-4526(94)90819-2}{Physica B: Condensed Matter {\bf 194-196}, 979 (1994).}

\end{thebibliography}
\end{document}